\documentclass[fleqn,usenatbib]{mnras}
\pdfoutput=1
\usepackage{times}
\usepackage{amsmath}
\usepackage{amssymb}
\usepackage{booktabs}
\usepackage[usenames,dvipsnames]{color}
\usepackage{dsfont}
\usepackage{graphicx}
\usepackage{url}
\usepackage{listings} 
\usepackage[point]{rccol}
\usepackage{soul}
\usepackage{subfig}
\usepackage{verbatim}
\usepackage{natbib}
\usepackage{hyperref}
\usepackage{pythonhighlight}
\usepackage{ulem}

\usepackage{newtxtext,newtxmath}
\usepackage[T1]{fontenc}

\allowdisplaybreaks[1]

\newcommand{\App}[1]{Appendix~\ref{#1}}

\newcommand{\Eq}[1]{Eq.~(\ref{#1})}

\newcommand{\Fig}[1]{Fig.~\ref{#1}}
\newcommand{\Sec}[1]{\S\ref{#1}}
\newcommand{\Tab}[1]{Table~\ref{#1}}

\def\adaptahop    {{\sc AdaptaHOP}}

\def\bpass    {{\sc Bpass}}

\def\music    {{\sc Music}}
\def\ramses    {{\sc Ramses}}
\def\ramsesrt  {{\sc Ramses-RT}}
\def\rascas    {{\sc Rascas}}
\def\sphinx  {{\sc Sphinx}}
\def\sphinxtw  {{\sc Sphinx$^{20}$}}
\def\sphinxten  {{\sc Sphinx$^{10}$}}
\def\sphinxtenf  {{\sc Sphinx$^{10}_{\rm early}$}}

\def\PI        {{R18}}

\def\he    {{\rm{He}}}
\def\hi    {{\rm{H\textsc{i}}}}
\def\hii   {\rm{H\textsc{ii}}}
\def\hei    {{\rm{He\textsc{i}}}}
\def\heii    {{\rm{He\textsc{ii}}}}
\def\heiii    {{\rm{He\textsc{iii}}}}
\def\htwo    {${\rm H_2}$}
\def\nh {n_{\rm{H}}}

\def\Nhi {N_{\rm{H\textsc{i}}}}

\def\cci {{\rm{cm}}^{-3}}   
\def\cs {\rm{cm}^{2}}       
\def\Msun {{\rm{M}_{\odot}}}
\def\sm {\rm{s}^{-1}}       

\def\csavg {\sigma}           

\def\Dxmin {\Delta x_{\rm min}}
\def\eavg {\bar{\epsilon}}   

\def\fesc {f_{\rm esc}}
\def\feschundred {f_{\rm esc,100}}

\def\Gammahi {\Gamma_{\rm HI}}
\def\Gyrm {\rm Gyr^{-1}}

\def\Lbox {L_{\rm box}}
\def\Lesc {L_{\rm esc}}
\def\Lintr {L_{\rm int}}

\def\Lyc {\rm LyC}

\def\mDM {m_{\rm DM}}

\def\MDMcell {M_{\rm DM, cell}}

\def\Mbaryonscell {M_{\rm b, cell}}

\def\MFifteen {{M_{1500}}}

\def\Mstar {M_{*}}

\def\Mvir {M_{\rm vir}}

\def\Qhi {Q_{\rm HI}}

\def\Rvir {R_{\rm vir}}

\def\Sigsfr {\Sigma_{\rm SFR}}

\def\sSFR {\rm sSFR}
\def\sSFRc {\rm sSFR_{100}}

\def\sSFRmax {\rm sSFR_{10,max}}
\def\sSFRten {\rm sSFR_{10}}

\def\Zhalo {Z_{\rm halo}}           
\def\Zsun {Z_{\odot}}           

\mathchardef\mhyphen="2D

\long\def\symbolfootnote[#1]#2{\begingroup%
\def\thefootnote{\fnsymbol{footnote}}\footnote[#1]{#2}\endgroup}

\title[Escape fractions in the EoR]{LyC escape from SPHINX galaxies in
  the Epoch of Reionization}

\author[Rosdahl et al.] {Joakim Rosdahl$^{1}$\thanks{E-mail:
    karl-joakim.rosdahl@univ-lyon1.fr},
  J\'er\'emy Blaizot$^{1}$,
  Harley Katz$^{2}$,
  Taysun Kimm$^{3}$,
  Thibault Garel$^{4}$,
  \newauthor Martin Haehnelt$^{5}$,
  Laura C. Keating$^{6}$,
  Sergio Martin-Alvarez$^{5}$,
  L\'eo Michel-Dansac$^{1}$,
  \newauthor and Pierre Ocvirk$^{7}$ \\
  $^1$Univ Lyon, Univ Lyon1, Ens de Lyon, CNRS, Centre de Recherche
  Astrophysique de Lyon UMR5574, F-69230, Saint-Genis-Laval, France \\
  $^2$Sub-department of Astrophysics, University of Oxford,
  Keble Road, Oxford OX1 3RH, UK \\
  $^3$Department of Astronomy, Yonsei University, 50 Yonsei-ro,
  Seodaemun-gu, Seoul 03722, Republic of Korea \\
  $^4$Observatoire de Gen\`eve, Universite\'e de Gen\`eve,
  51 Ch. des Maillettes, 1290 Versoix, Switzerland \\
  $^5$Kavli Institute for Cosmology and Institute of Astronomy,
  Madingley Road, Cambridge CB3 0HA, UK \\
  $^6$Leibniz-Institute for Astrophysics Potsdam (AIP), An der Sternwarte
  16, 14482 Potsdam, Germany \\
  $^7$Observatoire Astronomique de Strasbourg, Universit\'e de
  Strasbourg, CNRS UMR 7550, 11 rue de l'Universit\'e, 67000
  Strasbourg, France
}

\date{Accepted XXX. Received YYY; in original form ZZZ}

\pubyear{2015}

\begin{document}
\label{firstpage}
\pagerange{\pageref{firstpage}--\pageref{lastpage}}
\maketitle

\begin{abstract}
  We measure escape fractions, $\fesc$, of ionizing radiation from
  galaxies in the \sphinx{} suite of cosmological
  radiation-hydrodynamical simulations of reionization, resolving
  halos with $\Mvir\ga7.5 \times 10^7 \ \Msun$ with a minimum cell
  width of $\approx 10$ pc. Our new and largest $20$ co-moving Mpc
  wide volume contains tens of thousands of star-forming galaxies with
  halo masses up to a few times $10^{11} \ \Msun$.  The simulated
  galaxies agree well with observational constraints of the UV
  luminosity function in the Epoch of Reionization. The escape
  fraction fluctuates strongly in individual galaxies over timescales
  of a few Myrs, due to its regulation by supernova and radiation
  feedback, and at any given time a tiny fraction of star-forming
  galaxies emits a large fraction of the ionizing radiation escaping
  into the inter-galactic medium. Statistically, $\fesc$ peaks in
  intermediate-mass, intermediate-brightness, and low-metallicity
  galaxies ($\Mstar \approx 10^7 \ \Msun$, $\MFifteen\approx -17$,
  $Z\lesssim 5 \times 10^{-3} \ \Zsun$), dropping strongly for lower and
  higher masses, brighter and dimmer galaxies, and more metal-rich
  galaxies. The escape fraction correlates positively with both the
  short-term and long-term specific star formation rate. According to
  \sphinx{}, galaxies too dim to be yet observed, with
  $\MFifteen \ga -17$, provide about $55$ percent of the photons
  contributing to reionization. The global averaged $\fesc$ naturally
  decreases with decreasing redshift, as predicted by UV background
  models and low-redshift observations. This evolution is driven by
  decreasing specific star formation rates over cosmic time.
\end{abstract}

\section{Introduction} \label{Intro.sec}

Following the Big Bang, the Universe underwent rapid expansion and
cooling. When reaching a temperature of a few thousand Kelvin, protons
and electrons recombined, forming atoms. The moment of this transition
from an ionized to neutral Universe is perpetually observable in the
Cosmic Microwave Background \citep[CMB;][]{Penzias1965}. However these
neutral Dark Ages did not last forever: about a billion years later,
the reverse transition took place and the inter-galactic medium (IGM)
became reionized, a state in which it remains today.

There is much still to be understood about
reionization. Observationally it is known from the Thomson optical
depth for CMB radiation \citep[e.g.][]{PlanckCollaboration2018}, the
Gunn-Peterson trough \citep{Gunn1965}, and observations of
extreme-redshift Lyman-alpha emitters \citep[e.g.][]{Inoue2018}, that
the Epoch of Reionization (EoR) lasted for several hundred million
years and most likely ended between redshifts $6$ and $5$
\citep{Keating2020, Becker2021}.

It is commonly assumed that reionization was driven by ionizing
Lyman-continuum (\Lyc) radiation from massive stars in the first
galaxies. In this scenario, the redshift-evolution of the ionized
hydrogen fraction in the IGM can be modelled with four parameters that
describe the competition between star-powered photo-ionization and
recombination of atoms \citep{Madau1999}: for photo-ionization i) the
star formation rate density (SFRD) in the early Universe, ii) the
number of \Lyc{} photons produced per unit stellar mass formed, and
iii) the fraction -- $\fesc$ -- of the \Lyc{} photons produced by
stars that make it out of the inter-stellar medium (ISM) of their
source galaxies and into the IGM -- and for recombination iv) the
clumping factor of the IGM which sets the average recombination
rate. Out of those four parameters, the escape fraction of \Lyc{}
photons from galaxies is the most poorly understood.

Observationally, direct measurements of $\fesc$ are not feasible at
$z\ga4$ because almost all the \Lyc{} radiation is absorbed by the
intervening IGM on its way to the observer
\citep[][]{Inoue2008}. Indirect measurements are made using metal-line
ratios, but these ratios are still not well understood
\citep{Katz2020a, Katz2021c} and hence the derived escape fractions
are poorly constrained. At lower redshift direct and indirect
measurements can be made of escape fractions from individual galaxies
but these are difficult and highly uncertain and they tend to vary a
lot both from study to study and from galaxy to galaxy. Typically
upper limits of a few percent are estimated for the mean escape
fraction at $z\lesssim4$ \citep[see][and references therein]{Dayal2019,
  Robertson2021}, with the notable exception of \cite{Steidel2018} who
estimate an average $\fesc=9 \%$ at $z=3$ (though the same sample of
124 galaxies was later estimated by \citealt{Pahl2021} to have an
average $\fesc=6 \%$ ). Most observational studies find little
indication of a dependence on galaxy properties \citep{Saxena2021} or
redshift evolution \citep{Mestric2021} though there are exceptions
such as \cite{Faisst2016}, who used measurements of oxygen line
ratios to argue for both a galaxy mass dependency and a redshift
evolution.

Indirectly, one can also constrain the global escape fraction,
i.e. the total number of escaping over emitted \Lyc{} photons in the
Universe, via the HI photo-ionization rate in the IGM, derived from
the observed Lya forest. By comparing this to the estimated intrinsic
emission of \Lyc{} photons per volume from galaxies (and quasars,
typically assuming unity escape fraction), UV background models derive
a small $z\sim3$ escape fraction of one or two percent
\citep{Haardt2012, Khaire2015, Puchwein2019, FaucherGiguere2020,
  Yung2020}. These same models however require at least a ten times
higher global $\fesc$ for $z\ga6$ in order to reionize the
Universe. Therefore they postulate an evolving global escape fraction
which is high during the EoR and decreases with redshift \citep[see
also][making similar predictions but with different
methodology]{Price2016}. Such a varying escape fraction with time is
difficult to verify and to explain, though various authors have
suggested that if it exists, it may be due to a redshift evolution
towards more massive galaxies with lower escape fractions
\citep{Alvarez2012, Ferrara2012, Sun2015}, towards more metal-rich
galaxies with lower $\fesc$ \citep{Yoo2020}, or a decreasing
efficiency of stellar feedback in the expanding Universe in blowing
out escape routes for the radiation \citep{Sharma2017,
  FaucherGiguere2020, Ma2020}.

The best theoretical approach we rely on to understand and predict
$\fesc$ is simulations. Most cosmological simulations of reionization
aim for large volumes approaching cosmologically homogeneous scales on
the order of hundreds of Mpc \cite[e.g.][]{Iliev2013, Gnedin2014,
  Ocvirk2018b, Kannan2021}. These are very useful for understanding
the overall process of reionization and make predictions for
observable signals for different reionization scenarios. However,
large volume comes at the cost of resolution and these simulations are
far from resolving the propagation of radiation through the ISM, so
$\fesc$ is an input parameter rather than a prediction. Many works
have used the zoom-simulation technique, where the evolution of one
or a few galaxies is targeted in a cosmological environment, allowing
for high enough resolution that the propagation of radiation through
the ISM, and hence $\fesc$, can be predicted \cite[e.g.][]{Gnedin2008,
  Yajima2011, Wise2014, Ma2015, Ma2016, Kimm2017,
  Trebitsch2017}. Other works forego the cosmological environment to
gain even better resolution in idealised galaxies
\cite[e.g.][]{Wise2009, Yoo2020}. A lot of discrepancies and
disagreements remain between simulation works that predict $\fesc$,
but the emerging picture is that escape fractions appear highly
fluctuating but overall they decrease with increasing galaxy mass and
metallicity, suggesting that reionization is disproportionally driven
by low-mass metal-poor galaxies.

Except for a few existing works that either use large \citep{Kimm2014,
  Paardekooper2015, Xu2016, Trebitsch2021} or many \citep{Ma2020} zoom
regions, these zoom or idealised approaches are limited by the small
number of simulated galaxies, which prevents a systematic study of how
escape fractions vary with galaxy properties such as mass,
metallicity, and star formation activity. This limitation is amplified
by the fluctuating nature of $\fesc$ for individual galaxies which
enhances the need for many objects in a statistical
study. Additionally, the zoom and idealised galaxies approaches have
the limitation that the reionization process is not modelled, making
it difficult to tell how the predicted escape fractions translate to
the reionization of the Universe. Also, the feedback effect of
reionization on $\fesc$ is not captured \citep[e.g.][]{Katz2020,
  Ocvirk2021}.

Armed with the \sphinx{} suite of simulations \citep{Rosdahl2018} we
aim for the best of both worlds: a large sample of tens of thousands
of resolved galaxies evolving in non-zoomed cosmological
simulations. The methods we have developed for \sphinx{}
\citep{Rosdahl2013, Rosdahl2015a, Katz2017, Rosdahl2018} unlock the
goal of this paper: to statistically predict the escape of \Lyc{}
radiation from EoR galaxies while self-consistently modelling the
reionization history and high-z luminosity function, with unparalleled
sample size and resolution. The simulation data provide us a mock
Universe which reionizes at a reasonable redshift and in which we can
measure the actual escape fractions, how they scale with various
galaxy properties, which galaxies predominantly power reionization,
and if -- and if so, why -- the global escape fraction evolves with
time.

The setup of this paper is as follows. In Section
\ref{simulations.sec} we introduce the \sphinx{} simulations used for
the current analysis and recap their main characteristics. In Section
\ref{results.sec} we present our results: we first compare to
observational constraints for the luminosity function at various
redshifts during the EoR (and hence the luminosity budget) and the
reionization history. We present the redshift-evolution of the global
escape fraction of \Lyc{} photons in \sphinx{} and show how $\fesc$
scales with halo mass, galaxy mass, metallicity, magnitude, and
specific star formation rate. We then show how galaxies with
different properties contribute to the \Lyc{} luminosity budget during
reionization and finally we probe what effects drive the
redshift-evolution of $\fesc$ in \sphinx{}. We discuss our results in
Section \ref{Discussion.sec} and present our conclusions in Section
\ref{Conclusions.sec}.
 
\section{Methods} \label{simulations.sec}

The three cosmological radiation-hydrodynamical (RHD) simulations we
use in this paper are part of the \sphinx{} project and extend the
series presented in R18. With a few exceptions, the code and setup is
identical here to that described in \citet[][hereafter
\PI]{Rosdahl2018}. We refer the reader to \PI{} for a full description
of the methods and parameters and highlight below the main
characteristics of the runs used here.

\if
We study \Lyc{} luminosities and escape fractions in three
radiation-hydrodynamical (RHD) cosmological simulations of galaxy
evolution in the epoch of reionization. The simulations are part of
the \sphinx{} simulation suite first described in \citet[][which we
will in this section refer to as \PI]{Rosdahl2018}. Our main
simulation is a 20 cMpc (co-moving Mpc) wide volume with stellar
population \Lyc{} luminosities described by the BPASS spectral energy
distribution (SED) model version 2.2.1. The other two simulations are
smaller 10 cMpc volumes which we study for completeness, one otherwise
identical to the fiducial simulation, and the other with a slightly
different SED model.

With a few exceptions, the code and setup are identical to what we
described in \PI{}. We refer to that pilot \sphinx{} paper for
details, but recap here the main ingredients of the simulation setup
and clarify all differences from the original simulations.
\fi

\subsection{Simulation code and setup} \label{setup.sec}

We use the \ramsesrt{} code described in \cite{Rosdahl2013} and
\cite{Rosdahl2015a}. This is an RHD extension of the \ramses{}
parallel adaptive mesh refinement code \citep{Teyssier2002}, adding
radiative transfer and interactions of the radiation with gas to the
cosmological hydrodynamics.

The initial conditions (ICs) for our periodic cosmological volumes are
generated with \music{} \citep{Hahn2011} using $\Lambda$CDM
cosmological parameters compatible with the \cite{Ade2014} results.
We minimise the effect of cosmic variance on the \Lyc{} luminosity
budget by selecting the most ``average'' initial conditions for each
of our volumes from a set of $60$ ICs generated with \music{}. We
assume constant and homogeneous H and He mass fractions of $X=0.76$
and $Y=0.24$ respectively. We use an initial homogeneous metal mass
fraction of
$Z_{\rm init}=6.4 \times 10^{-6} ( = 3.2 \times 10^{-4} \ \Zsun$,
assuming a Solar metal mass fraction of $\Zsun = 2 \times 10^{-2}$
throughout this work). This artificial and unrealistically non-zero
initial metal mass fraction is necessary to allow gas to cool below
$\approx 10^4$ Kelvin and collapse in the absence of \htwo{} formation
from primordial gas, which is not modelled.

Our main simulation is a (20 cMpc)$^3$ (co-moving Mpc) volume, 8 times
larger than the largest volume in \PI. The other two simulations are
smaller (10 cMpc)$^3$ volumes -- using the exact same ICs as in \PI{}
-- which we study for completeness, one otherwise identical in setup
to the main simulation, and one with a slightly different spectral
energy distribution (SED) model (see below).  All three simulations
have the same resolution and refinement strategy. Dark matter (DM)
particles have a mass of $\mDM=2.5 \times 10^5 \ \Msun$, with $512^3$
particles in the smaller volume simulations and $1024^3$ in the large
one. We resolve halos at the atomic cooling mass of
$3 \times 10^7 \ \Msun$ \citep[][]{Wise2014} with 120 DM
particles. The simulations have a minimum cell width of $10.9$ pc at
$z=6$, having a co-moving finest cell width of $76.3$ cpc, and a
maximum cell width in the diffuse IGM of 2.8 kpc (19.6 ckpc) at
z=6. Note that fixed co-moving maximum resolution means that the
maximum physical resolution degrades linearly with the cosmological
expansion factor $a$ and the finest cell width doubles as $a$ doubles,
e.g. from $z=11$ to $z=5$.

We use adaptive refinement criteria to resolve dense and
Jeans-unstable regions. A parent cell is split into 8 equal-size
children cells if: i) $\MDMcell + \Mbaryonscell/f_b > 8 \ \mDM$, where
$\MDMcell$ and $\Mbaryonscell$ are the total DM and baryonic (gas plus
stars) masses in the cell and $f_b=0.154$ is the baryon mass fraction;
or ii) the local Jeans length is smaller than four local cell widths.

For the M1 radiative transfer \citep[see][]{Rosdahl2013}, we use the
variable speed of light approximation described in \cite{Katz2017} to
speed up the calculation, such that the speed of light goes from
$1.25\%$ of the real speed of light in the highest resolution regions
in the ISM to $20\%$ in the coarsest cells representing IGM voids. We
subcycle the radiative transfer within the hydro timestep to reduce
the computational cost. Whereas we used three radiation groups in
\PI{}, for \hi- \hei- and \heii-ionizing photons, we merge the two
more energetic groups into one in the current simulations. This is
done to reduce the number of radiation variables and hence mitigate
the memory cost of the simulation. In \Tab{groups.tbl} we show the
energy ranges for the two photon groups as well as their typical
energies and cross sections, which are updated regularly to reflect
the ``mean'' stellar particles. Merging the two high energy (\hei{}
and \heii{}) groups from \PI{} into one is acceptable because only a
tiny fraction of the \Lyc{} photons emitted by stellar populations are
in the highest energy (\heii-ionizing) group, whereas the two
lower-energy groups both contribute strongly in \Lyc{} (see Appendix D
in \PI{} for details). It does not matter for \hi{} and \hei{}
reionization whether one merges this highest energy group or simply
omits it, and we have verified in post-processing that escape
fractions are insensitive to whether the HeII cross section is set to
zero or our non-zero SED-derived value.

\begin{table*}
  \begin{center}
  \caption
  {Photon group energy (frequency) intervals and properties. The
    energy intervals are indicated in units of eV by $\epsilon_0$ and
    $\epsilon_1$ and in units of \AA{}ngstr\"om by $\lambda_0$ and
    $\lambda_1$. The last four columns show photon properties derived
    every $10$ coarse time-steps from the stellar luminosity weighted
    SED model. These properties evolve over time as the stellar
    populations age, and the approximate variation is indicated in the
    column headers (the difference being similar in the two SED models
    we use). $\eavg$ denotes photon energies while
    $\csavg_{\hi}$, $\csavg_{\hei}$, and $\csavg_{\heii}$ denote 
    cross sections for ionization of hydrogen and helium,
    respectively.}
  \label{groups.tbl}
  \begin{tabular}{l|R{2}{2}R{2}{2}rr|rrrr}
    \toprule
    Photon & \multicolumn{1}{c}{$\epsilon_0$ [eV]} 
    & \multicolumn{1}{c}{$\epsilon_1$ [eV]}
    & $\lambda_0$ [\AA{}] & $\lambda_1$ [\AA{}]
    & $\eavg$ [eV] & $\csavg_{\hi} \, [\cs]$ 
    & $\csavg_{\hei} \, [\cs]$ & $\csavg_{\heii} \, [\cs]$ \\
    group & \multicolumn{1}{c}{} 
    & \multicolumn{1}{c}{}
    &  & 
    & $\pm 20 \%$ & $\pm 10 \%$ 
    & $\pm 10 \%$ & $\pm 25 \%$ \\
    \midrule
    UV$_{\hi}$ & 13.6 & 24.59 & $9.1 \times 10^{2}$ & $5.0 \times 10^{2}$ & 18.3 
            & $3.2 \times 10^{-18}$ & 0 & 0 \\
    UV$_{\he}$    & 24.59 & \multicolumn{1}{c}{$\infty$} & $5.0 \times 10^{2}$ 
            & 0 & 33.9 
            & $6.2 \times 10^{-19}$& $4.7 \times 10^{-18}$ & $1.5 \times 10^{-21}$ \\
    \bottomrule
  \end{tabular}
  \end{center}
\end{table*}

Another memory-saving change we make from \PI{} is to change the
precision of all RHD cell variables from double to single precision,
reducing the memory cost by almost half. We have performed idealised
tests and smaller volume \sphinx{} runs to confirm that the reduction
in precision has no effect on our results. We note that for all
calculations, such as the radiation-hydro-gravity solver and
thermochemistry, we cast the cell variables to double precision and
then back to single precision when updating the cell state, to
minimise the impact of numerical underflow and overflow.

\begin{figure}
  \centering
  \includegraphics[width=0.45\textwidth]
    {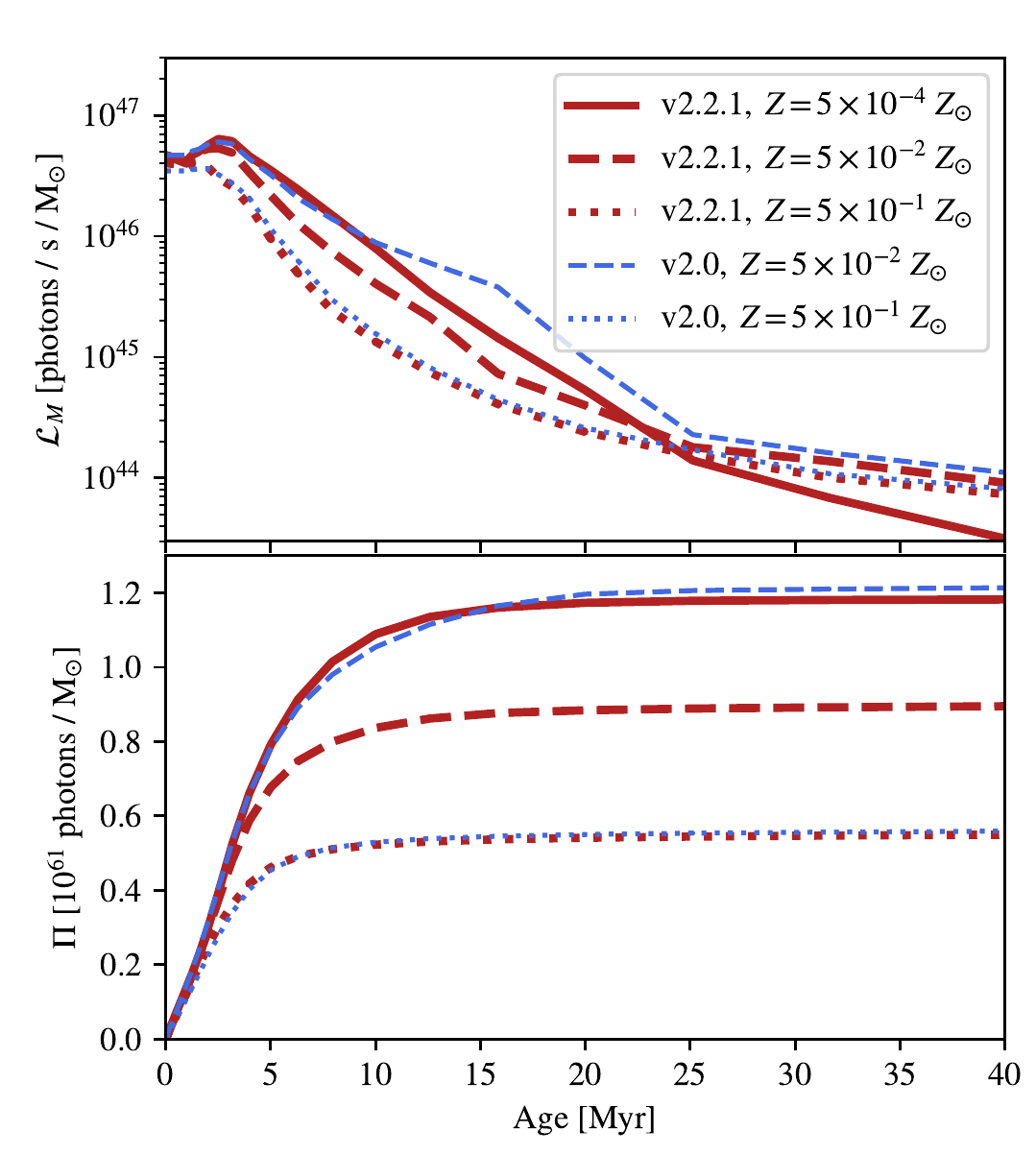}
  \caption
  {\label{SEDs.fig}Ionising number luminosity (top) and cumulative
    number of ionising photons (bottom) per Solar mass with BPASS
    versions 2.2.1 and 2.0. Note that the very lowest metallicity,
    $5 \times 10^{-4} \ \Zsun$, only exists in the newer model, while
    the older model starts at $5 \times 10^{-2} \ \Zsun$. The newer
    model has significantly lower late-time luminosities for
    $Z \ga 5 \times 10^{-2} \ \Zsun$ than the older model at the same
    metallicities.}
\end{figure}

For stellar population \Lyc{} luminosities as a function of age and
metallicity, we use the Binary Population and Spectral Synthesis
model\footnote{\protect\url{https://bpass.auckland.ac.nz}}
\citep[\bpass:{}][]{Eldridge2007, Stanway2016, Stanway2018}, assuming
a \cite{Kroupa2001}-like initial mass function (IMF) with a slope of
$-1.3$ from $0.1$ to $0.5 \ \Msun{}$ and $-2.35$ from $0.5$ to
$100 \ \Msun$. For the fiducial run, which we name \sphinxtw{} (the
superscript referring to the volume width) we use BPASS version
2.2.1. This is an update from \PI{} where we used the older BPASS
version 2.0. The newer version is extended to metallicities two orders
of magnitude lower than in the older version, which is more
representative for the metallicities of stellar populations formed in
\sphinx{}. As shown in \Fig{SEDs.fig}, the older version has
significantly higher \Lyc{} luminosities than the new one for stellar
populations older than a few million years and with low metallicities
of $10^{-3} \Zsun \lesssim Z << \Zsun$.

As shown in \PI{}, the escape of \Lyc{} radiation from galaxies is
highly sensitive to subtle variations in the assumed SED model,
especially at advanced stellar population ages and low
metallicities. Indeed, as demonstrated later in this paper, using the
two BPASS versions in otherwise identical simulations leads to very
different reionization histories that bracket observational
constraints, with the older version giving much earlier reionization
than the newer one. Therefore we analyse simulations using each
version. Performing two runs with the volume size of \sphinxtw{},
however, is prohibitively expensive, so we report on two
smaller-volume simulations: \sphinxtenf{} (because it reionizes early)
is a ($10$ cMpc)$^3$ volume where we use the older BPASS version 2.0, and
\sphinxten{} is the same volume, but with BPASS 2.2.1, to provide a
clean comparison to \sphinxtenf{} (i.e. separating the effects of
changing the SED model and the cosmological volume). The simulations
are listed in \Tab{sims.tbl}.

\begin{table}
  \begin{center}
  \caption
  {Simulations used in this work. From left to right: simulation name,
    volume width, and SED model used for stellar population \Lyc{}
    luminosities. }
  \label{sims.tbl}
  \begin{tabular}{l|l|l|}
    \toprule
    Name & $\Lbox$ [cMpc] & SED model \\
    \midrule    
    \sphinxtw   & $20$ & BPASS v2.2.1 \\
    \sphinxten & $10$ & BPASS v2.2.1 \\
    \sphinxtenf   & $10$ & BPASS v2.0 \\
    \bottomrule
  \end{tabular}
  \end{center}
\end{table}

A stellar \Lyc{} luminosity factor is often applied in RHD simulations
of reionization, to account for either unresolved absorption of
radiation (factor <1) or unresolved channels through which the
radiation can escape (factor > 1), and calibrated to match
observational constraints of the reionization history. This factor is
unity in all \sphinx{} simulations presented here, i.e. stellar
particles inject into their host cells the luminosities given by the
assumed SED model, and our escape fractions emerge as the ratio of the
rate of photons escaping a halo versus its total \Lyc{}
luminosity. This does not mean, however, that our simulations resolve
the transport of radiation on the smallest scales and we indeed
demonstrate later in the paper that our overall escape fractions are
not converged with resolution and are likely somewhat under-predicted
compared to reality.  It simply happens, as we will show, that our
combination of resolution, sub-grid models, and assumed stellar
evolution models, produces a fairly realistic model of the early
Universe and its reionization and hence presents a plausible model to
probe how radiation escapes from galaxies with different properties
and which galaxies predominantly contribute to reionization.

Gas cooling and heating is described in detail in
\cite{Rosdahl2013}. The non-equilibrium hydrogen and helium
thermochemistry is coupled with the local ionizing radiation and
evolved semi-implicitly via collisional ionisation, collisional
excitation, photo-ionization, recombination, bremsstrahlung, Compton
cooling off cosmic microwave background radiation, and di-electric
recombination. The non-equilibrium abundances of \hi{}, \hii{},
\hei{}, \heii{}, and \heiii{} are tracked and stored in each gas
cell. For $T>10^4$ K, additional cooling rates from heavier elements
are pre-calculated from Cloudy \citep[][]{Ferland1998}, assuming
photo-ionization equilibrium with a redshift-evolving UV
background. For $T\le10^4$ K we use fine structure metal cooling rates
from \cite{Rosen1995}, allowing the gas to cool to a temperature floor
of $15$ K.

Star formation is performed with the thermo-turbulent model described
in \PI{}. Gas cells are eligible for star formation only if the local
hydrogen density $\nh>10 \ \cci$, the local turbulent Jeans length is
smaller than the finest cell width, and the gas is locally
convergent. If a cell is elegible for star formation under these
conditions, its gas is converted stochastically into stellar particles
as described by \cite{Rasera2006}, and its star formation efficiency
is a non-linear function of the local virial parameter and turbulence
\citep[][]{Federrath2012}. The stellar particles, representing coeval
stellar populations, have an initial mass of $400 \ \Msun$, somewhat
smaller than the $1000 \ \Msun$ particles in \PI{}.

Supernova (SN) feedback is also unchanged from \PI{}.  Stellar
particles undergo individual $10^{51}$ erg explosions from an age of 3
to 50 Myrs. Injection of energy and momentum from those individual SN
explosions is performed with the ``mechanical'' feedback model
described in e.g. \cite{Kimm2015}, where the energy is injected
thermally if the resolution is sufficient to correctly capture the
Sedov-Taylor expansion. Otherwise, to overcome direct numerical
overcooling, a physically motivated density- and metallicity-dependent
amount of momentum is injected, which has been derived from
high-resolution numerical experiments. SN explosions from a stellar
particle are sampled in time to produce an average mass return of $20$
percent of the original stellar mass back into the ISM, a metal yield
of 0.075, and an average of 4 SN explosions per $100 \ \Msun$. The
mass return is close to that of a \cite{Kroupa2001} IMF, and therefore
consistent with the BPASS model. However, the number of SN explosions
per Solar mass is boosted roughly four-fold compared to
\cite{Kroupa2001}. This calibrated and artificial boost in the number
and total energy of SN explosions is necessary to suppress star
formation enough to produce a realistic high-redshift luminosity
function. The necessity of such a boost in SN feedback is still not
fully understood and reflects an important unsolved question in galaxy
formation, where simulated galaxies tend to form stars too efficiently
compared to observations. It may be due to a lack of additional
ill-understood feedback channels such as cosmic rays
\citep[e.g.][]{Farcy2022}, or to insufficient resolution to
self-consistently model a multi-phase turbulent ISM.

To recap, the differences in setup from \PI{} are less
massive stellar particles, single precision RHD, two radiation groups
(reduced from three), and, most significantly, an updated BPASS
version for stellar \Lyc{} luminosities in our fiducial \sphinxtw{}
run.

At the end of the run at $z=4.64$, \sphinxtw{} contains about
$5 \times 10^9$ cells and a billion stellar particles. The run,
performed on $10,080$ cores, required about $50$ million core-hours.
We write simulation outputs every $5$ Myrs, giving a total of 216
outputs between $z=20$ and the end at $z=4.64$. The size per output
increases significantly from start to finish due to structure
formation and increasing AMR refinement, being about $350$ GB at the
beginning and $1.2$ TB at the end, and the total size of all output is
$105$ TB. The two smaller volume simulations have been performed on
$2,880$ cores and require roughly a tenth of the memory and CPU-time
compared to \sphinxtw{}.

\subsection{Halos and escape fractions} \label{fesc_rascas.sec}

To identify dark matter halos, we use the \adaptahop{} halofinder
\citep{Aubert2004,Tweed2009} on the dark matter particles, using the
same parameters as described in \PI{}.

We associate stars to halos as follows. Each stellar particle is
assigned to the closest (sub-) halo, using the weighted distance
measurement $d=r/\Rvir$, where $r$ is the distance between the
particle and the halo centre and $\Rvir$ is the virial radius of the
halo. A stellar particle with $d>\Rvir$ for any halo is not assigned
at all. Such unassigned stellar particles do exist, but they are
negligible in both number and ionizing luminosities, even if we assume
all their radiation escapes into the IGM. We do not assign stars to
sub-halos fully enclosed within $\Rvir$ of their parent halo
(i.e. stars within the sub-halo are assigned to its parent
halo). 

We compute \Lyc{} and $1500$ \AA{} escape fractions from halos in
post-processing using the Monte-Carlo radiative transfer code
\rascas{} \citep[][]{Michel-Dansac2020}. We prefer to use ray-tracing
to measure escape fractions rather than using the M1 radiation fluxes
in the simulation outputs directly, as our variable speed of light
generates significant and variable delays between the emission of
radiation and escape at $\Rvir$ that are difficult to
track. Furthermore, ray-tracing allows us to individually measure
escape fractions from each stellar particle. \cite{Trebitsch2017}
showed, using the same methods but with a constant speed of light,
that \rascas{} yields almost identical escape fractions as using the
M1 flux directly. Contrary to \PI{} where we obtained galaxy escape
fractions by computing the optical depth along 500 rays from each star
particle, here we use a Monte Carlo sampling technique, where we cast
photon packets from star particles with a probability proportional to
their LyC luminosity. These photon packets are propagated with
\rascas{} until either they escape the virial radius of the host halo
or are absorbed by an H or He atom. The escape fraction is then the
fraction of photon packets which reached $\Rvir$ without being
absorbed. We set the number of photon packets per halo to be 100 times
the number of star particles, with a minimun of $10^4$ and a maximum
of $10^7$. We have checked that the resulting escape fractions are
converged and accurate to sub-percent relative precision regarding the
number of photon packets. Our results are also insensitive to our
choice of $\Rvir$ as the escape distance, which is rather arbitrary
but commonly used in the simulation literature to measure escape
fractions. We have tested using instead both $2 \Rvir$ and 10 kpc from
each photon source (the latter corresponding to the virial radius for
$\Mvir\approx10^{10} \ \Msun$ halos at $z\sim 6$) and found that the
measured escape fractions are also sub-percent converged. Note that
for consistency, we repeat here the \ramsesrt{} approximation and bin
radiation in two frequency bins with mean cross sections, and the
total escape fraction of a galaxy is the \Lyc -luminosity-weighted
mean of that of the two radiation groups.  For reference,
\cite{Mauerhofer2021} compute escape fractions by propagating the full
spectra with RASCAS, which gives similar results within a few percent
relative precision (Valentin Mauerhofer, private comm.).

Throughout the paper, the sample of halos considered is the full
population of galaxies, i.e. any halo hosting stellar particles,
without any threshold on its star formation rate. The \Lyc{} escape
fraction considered is always the intrinsic-\Lyc -luminosity weighted
(not number-weighted) mean escape fraction of all halos or halos with
a given property, or equivalently the fraction of \Lyc{} photons
emitted that escapes into the IGM.

For attenuated $1500$ \AA{} (or UV) luminosities of galaxies embedded
in DM halos, we use the same Monte-Carlo routine to derive UV escape
fractions, but here with absorption by dust instead of hydrogen and
helium. We use the formulation of \cite{Laursen2009}, assuming that
the dust absorption coefficient scales linearly with the local metal
mass fraction of the gas and the neutral hydrogen density, as also
described in \cite{Garel2021}. The UV escape fraction is then found as
described above, but using $1500$ \AA{} luminosities for the stellar
particles, as provided by the assumed SED model.

For the $\Lyc$ radiation we ignore absorption by dust. This is for
consistency with the on-the-fly absorption in the simulation itself,
where dust is also ignored.  Dust absorption of \Lyc{} radiation is
expected to be negligible \citep{Yoo2020} because the dust optical
depth, while not necessarily always small, is always much smaller than
the photo-ionization optical depth \citep{Mauerhofer2021}. We have
verified this by calculating escape fractions with dust absorption in
post-processing for a few of the last snapshots in \sphinxtw{}, using
the same formulation for dust as described above. These tests indeed
reveal that the largest effect of dust on $\fesc$ is in UV-bright
galaxies that already have very low escape fractions via atomic
absorption, and that the effect on mean escape fractions, at any UV
brightness, is negligible. A deeper analysis on the effect of dust on
\Lyc{} escape fractions in \sphinx{} will be presented in Katz et
al. in prep.

\section{Results}\label{results.sec}

\begin{figure*}
  \centering
  \includegraphics[width=1.0\textwidth]
    {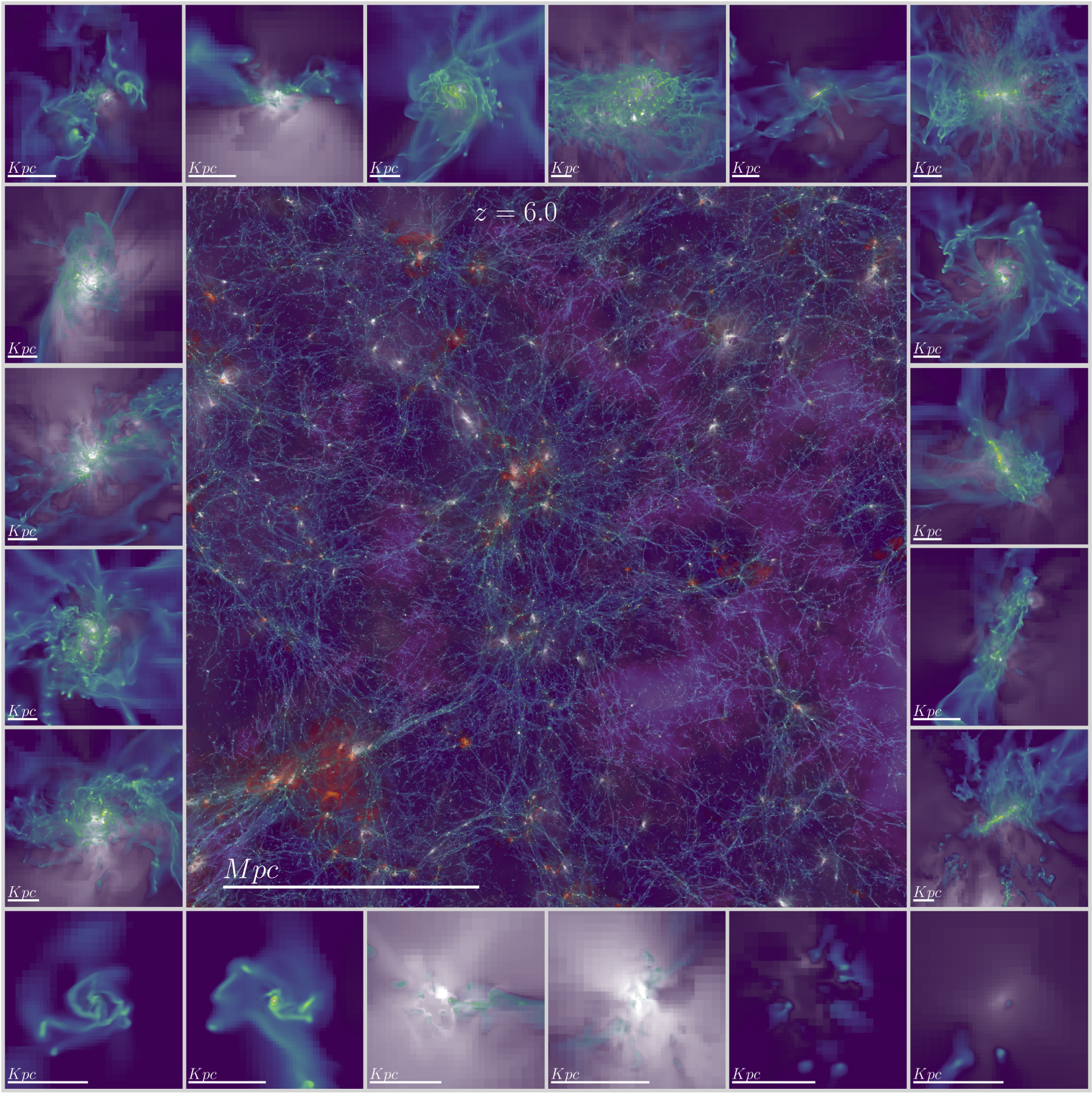}
  \caption
  {\label{map_fullbox.fig}Galaxies from \sphinxtw{} at $z=6$. The
    large central panel shows a projection of the full simulation
    volume, with green filaments and clumps showing gas density, red
    denoting hot outflows ($T \ga 10^5 K$) driven by SN feedback,
    white denoting intense hydrogen photo-ionization rates $\Gammahi$,
    and purple denoting islands of still-neutral hydrogen. The smaller
    panels show individual galaxies, with green denoting neutral
    hydrogen column density $\Nhi$ and white denoting $\Gammahi$, with
    the same ranges in all the small panels of
    $\Nhi=10^{20.5}-10^{24.5} \ \cci$ and
    $\Gammahi=10^{-13}-10^{-7} \ \sm$. The length scale is shown in
    the bottom left corner of each panel. The bottom row of panels
    shows halos with virial mass $\Mvir\approx 10^9 \ \Msun$, while
    other panels show some of the most massive halos at $z=6$, with
    $\Mvir=10^{10}-10^{11} \ \Msun$. \sphinx{} galaxies have a wide
    range of morphologies and properties, some emitting a lot of
    \Lyc{} photons into the IGM and others not.
  }
\end{figure*}

The central panel of \Fig{map_fullbox.fig} shows a projection of the
full \sphinxtw{} volume at $z=6$ in gas density (green), temperature
(red), photo-ionization rate (white), and neutral fraction
(purple). Smaller panels in the same figure show zoomed-in projections
of neutral hydrogen column density and photo-ionization rate for some
of the individual galaxies, with panel widths corresponding to 20
percent of the parent halo radius. The bottom-row of panels shows
different low-mass galaxies at $z=6$, residing in halos with
$\Mvir \approx 10^9 \ \Msun$, while the other zoom-in panels show some
of the most massive galaxies at the same redshift, with halo masses in
the range $10^{10}-10^{11} \ \Msun$. The simulation produces a wide
range of galaxy masses and morphologies due to different environments
and accretion histories. At $z=6$, the simulation contains
$\approx 32$ thousand star-forming halos, i.e. halos containing one or
more stellar particles, $\approx 57$ thousand resolved halos with
$\Mvir>7.5 \times 10^7 \ \Msun$, and $\approx 454$ thousand halos
identified by the halofinder with $10$ DM particles or more.

The bottom row galaxies represent four stages typically found for
low-mass galaxies in \sphinx{}, in a looping sequence going from left
to right: i) gas accumulation with inefficient star formation (SF) --
and hence low \Lyc{} luminosity -- and low $\fesc$ on the far left,
ii) onset of star formation, with efficient SF and low $\fesc$ in the
second panel from left, iii) disruptive feedback with still-high SF
and high $\fesc$ in third and fourth panels from left, and iv) total
disruption, with high $\fesc$ but inefficient SF in the two rightmost
panels. A low-mass galaxy can remain in that last stage for tens of
millions of years, but eventually starts accumulating gas again and
repeats the cycle, starting again at stage i).

We split our results into several sections. We first demonstrate that
the simulated galaxies realistically represent actual high-z galaxies
via comparing their properties to available observational constraints
and show that they power reionization in a reasonable timeline. We
then show the evolution of the global escape fraction with
redshift. We go on to investigate trends of ionizing escape fractions
with halo properties and which halos dominate the budget of ionising
radiation reaching the IGM. Finally we investigate the evolution of
the global escape fraction with time.

\subsection{UV luminosity function}\label{galprops.sec}

\begin{figure}
  \centering
  \includegraphics[width=0.47\textwidth]
    {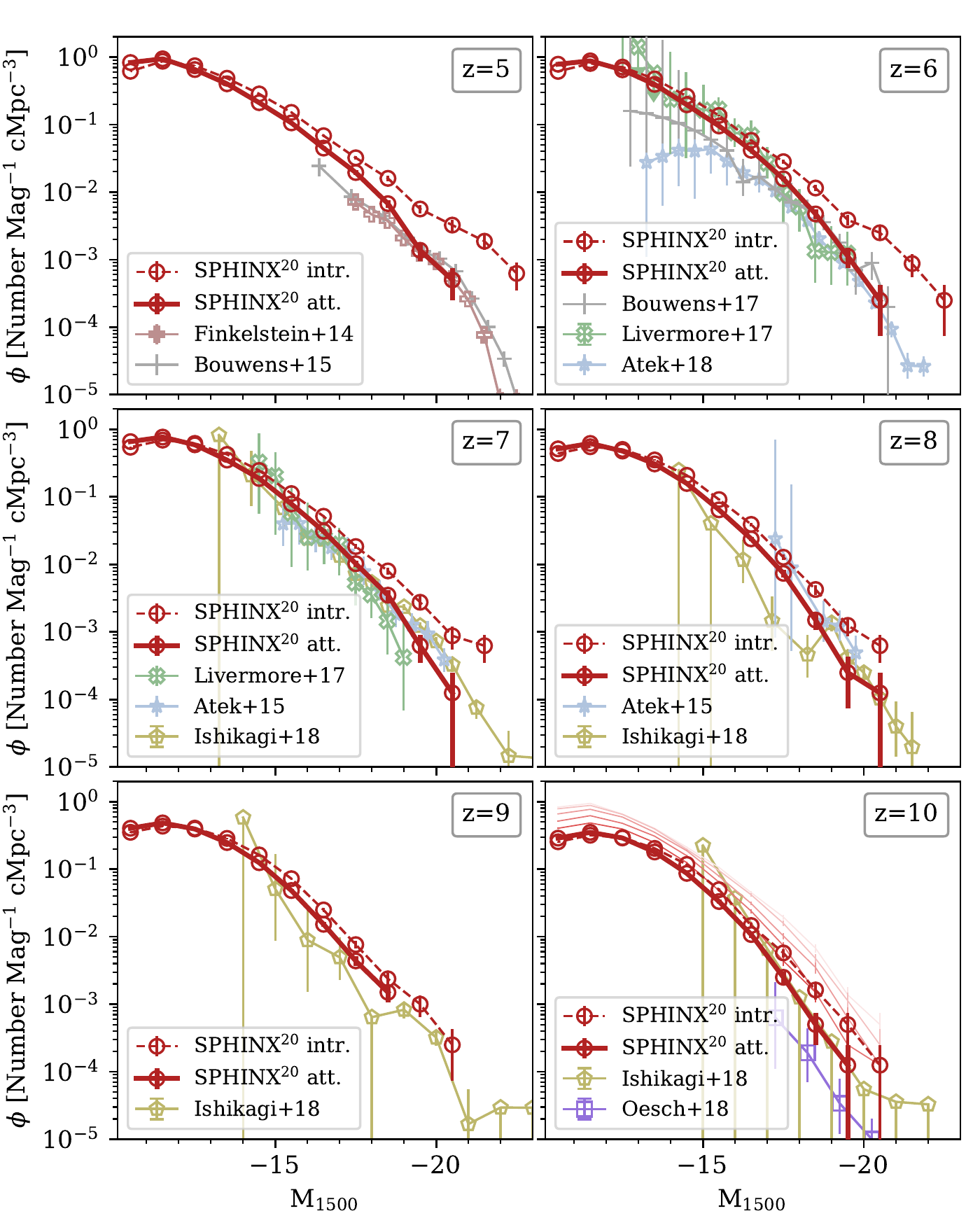}
  \caption
  {\label{flum.fig}1500 \AA{} UV luminosity function at different
    redshifts, as indicated in the top right corner of each panel,
    from \sphinxtw{} and observations, as indicated in the
    legends. For \sphinxtw{}, intrinsic (dust-attenuated) luminosities
    are shown in dashed (thick solid) red curves. To clearly show the
    evolution of the simulated LF, we re-plot the attenuated
    \sphinxtw{} LF from other panels in the bottom-right z=10 panel,
    with decreasing redshift represented by increasingly light red
    color.  We use Poissonian error-bars for the \sphinx{} data (but
    note that the error from cosmic variance is typically larger). The
    simulated LF agrees well with observational limits at all
    redshifts shown, implying a realistic intrinsic \Lyc{}
    volume-emissivity during the simulated EoR. }
\end{figure}

To establish that our \sphinxtw{} simulation forms stars at
approximately the correct rate, we start with the UV luminosity
function at $1500$ \AA{} in \sphinxtw{}, shown in
\Fig{flum.fig} at integer redshifts from $z=5$ to $z=10$. The dashed
and solid red curves show the simulated intrinsic and dust-attenuated
LFs, respectively, with dust-attenuation taken into account using the
method described in \Sec{fesc_rascas.sec}. Compared to the intrinsic
LF, the attenuated LF is shifted to dimmer magnitudes primarily at the
bright end, by up to two dex. This strong attenuation at the bright
end is expected, since the brightest galaxies tend also to be the most
metal-enriched and hence dusty ones.

We include at each redshift a sub-set of published observational
limits, as indicated by the legends. At $z=5$ we show results from
\cite{Bouwens2015, Finkelstein2015} using different combinations of
Hubble Space Telescope (HST) surveys. At higher redshifts we show
results from works using the Hubble Frontier Fields (HFF) clusters,
which magnify the high-z sources via lensing: \citet[][two HFF
clusters]{Livermore2017}, \citet[][four HFF clusters]{Bouwens2017},
\citet[][6 HFF clusters]{Atek2018}, \citet[][three HFF
clusters]{Atek2015}; \citet[][six HFF clusters]{Ishigaki2017}, and
\citet[][six HFF clusters, plus legacy HST datasets]{Oesch2018}. All
these observational results have large uncertainties due to lensing
model assumptions, cosmic variance, and small sample sizes. This can
be seen from the error bars in \Fig{flum.fig} as well as in the
difference between individual observational analyses, which use
different models and/or model parameters. The attenuated \sphinxtw{}
luminosity function mostly falls within the range of observational
data, though it does tend to be on the high side of that range, and
therefore is in good agreement with observational constraints. The
same is true for the smaller volume \sphinx{} simulations we report
on. They have almost identical UV luminosity functions as shown here
for \sphinxtw{}, except that they are cut off at dimmer magnitudes,
being smaller volumes, and hence agree equally well with
observations. This similarity in the UV luminosity function between
simulations with different BPASS versions is because star formation is
insensitive to the assumed SED model and the two SED models have very
similar 1500 \AA{} intensities, even if \Lyc{} intensities vary.  The
simulations thus provide a good working model of the high-redshift
Universe through which we can explore the production and escape of
ionizing radiation from galaxies\footnote{We note also that
  \citet{Garel2021} have shown good agreement of \sphinx{} with the
  extreme-z Lyman-alpha luminosity function, though here there is 
  little overlap between the low-luminosity range of simulations and
  high-luminosity range of observations.}.

\iftrue
\begin{figure}
  \centering
  \includegraphics[width=0.4\textwidth]
    {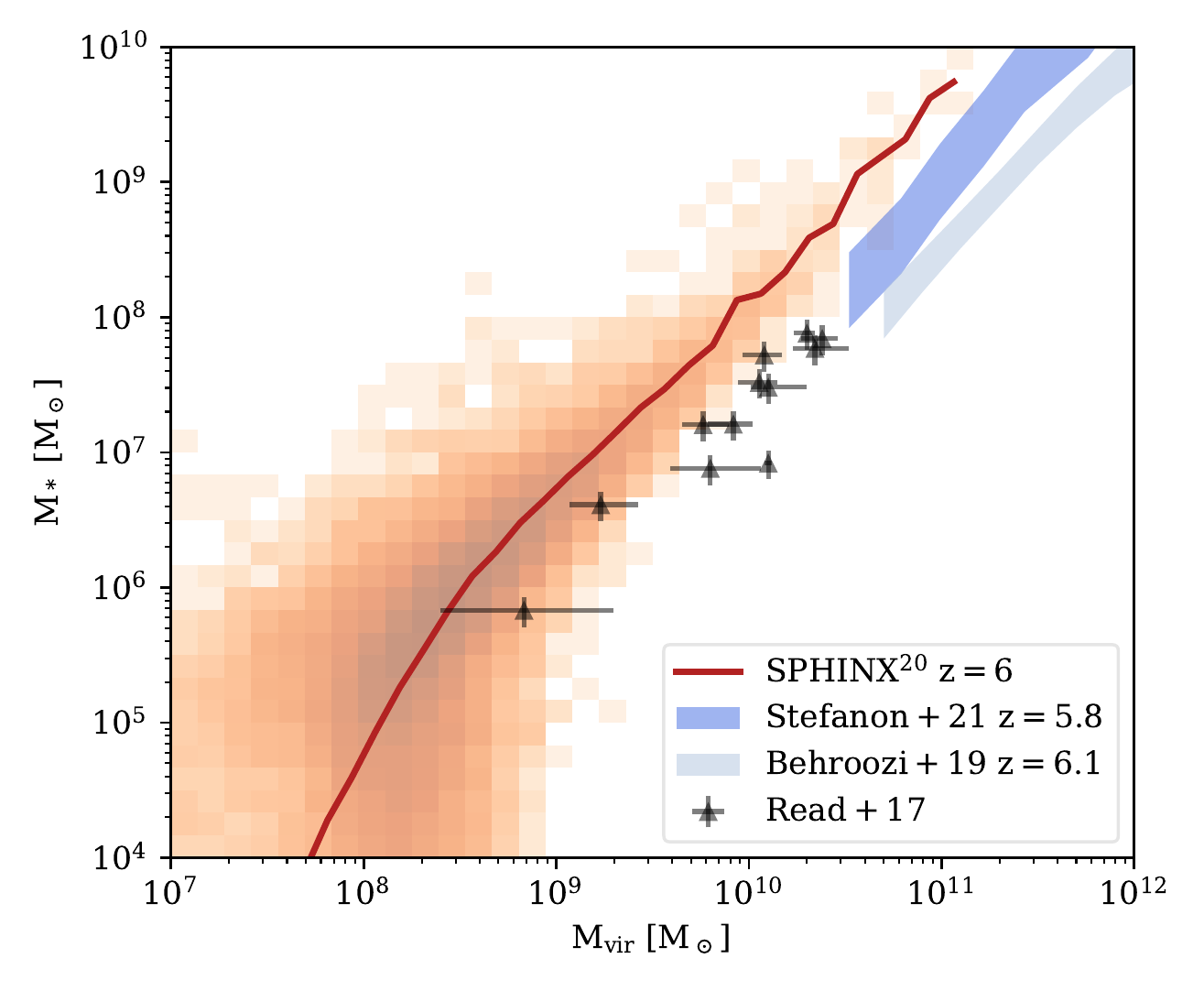}
  \caption
  {\label{smhm.fig}Stellar mass to halo mass relation in \sphinxtw{}
    at $z=6$. The red solid curve shows mean stellar mass per halo
    mass bin. The orange pixels show the distribution in stellar mass
    on a logarithmic color scale, with the lightest pixels containing
    single halos and the darkest containing about a thousand. The
    shaded areas in the upper right corner show $z\approx6$ abundance
    matching constraints from \citet{Stefanon2021a} and
    \citet{Behroozi2019}, and black triangles show observational
    constraints of $z = 0$ isolated dwarf galaxies \citep{Read2017}.}
\end{figure}

\subsection{Stellar mass to halo mass}\label{smhm.sec}

We consider in \Fig{smhm.fig} the stellar mass to halo mass (SMHM)
relation of \sphinxtw{} galaxies at $z=6$, compared to the abundance
matching constraints of \citet{Stefanon2021a} and \citet{Behroozi2019}
at similar redshifts and observational estimates for local dwarf
galaxies from \citet{Read2017}. We do not show the SMHM relations for
the smaller \sphinx{} volumes here, but they are very similar except
for not having as massive halos.  The \sphinxtw{} galaxies show a wide
scatter in stellar mass for the lower-mass halos, but this scatter
shrinks with increasing halo mass. Where there is overlap in halo
mass, the \sphinx{} stellar masses are a factor of a few larger than
those derived from abundance matching by \citet{Stefanon2021a}, and a
bit further away from the results of \citet{Behroozi2019}. This
suggests that \sphinx{} galaxies may be somewhat too star-forming,
although it is non-trivial to align this discrepancy with the good
agreement shown in \Fig{flum.fig} between \sphinx{} and the observed
luminosity function at various redshifts.

We note that the stellar masses reported here for \sphinxtw{} are a
few tens of percent higher than for the smaller volume \sphinx{}
simulations described in \PI{}. This is due to the more inclusive way
we assign stars to halos in the current paper: here we assign all
stars within a halo, whereas in \PI{} we used a galaxy finder
algorithm on the stellar particles to identify galaxies, which tends
to exclude satellites and diffuse stellar distributions in the
halo. The latter method is closer to the spirit of observations, but
we prefer the former here, since it is how we assign stars to halos
throughout the paper to compute escape fractions.

\fi

\subsection{Reionization}\label{reionization.sec}

\begin{figure}
  \centering
  \includegraphics[width=0.49\textwidth]
    {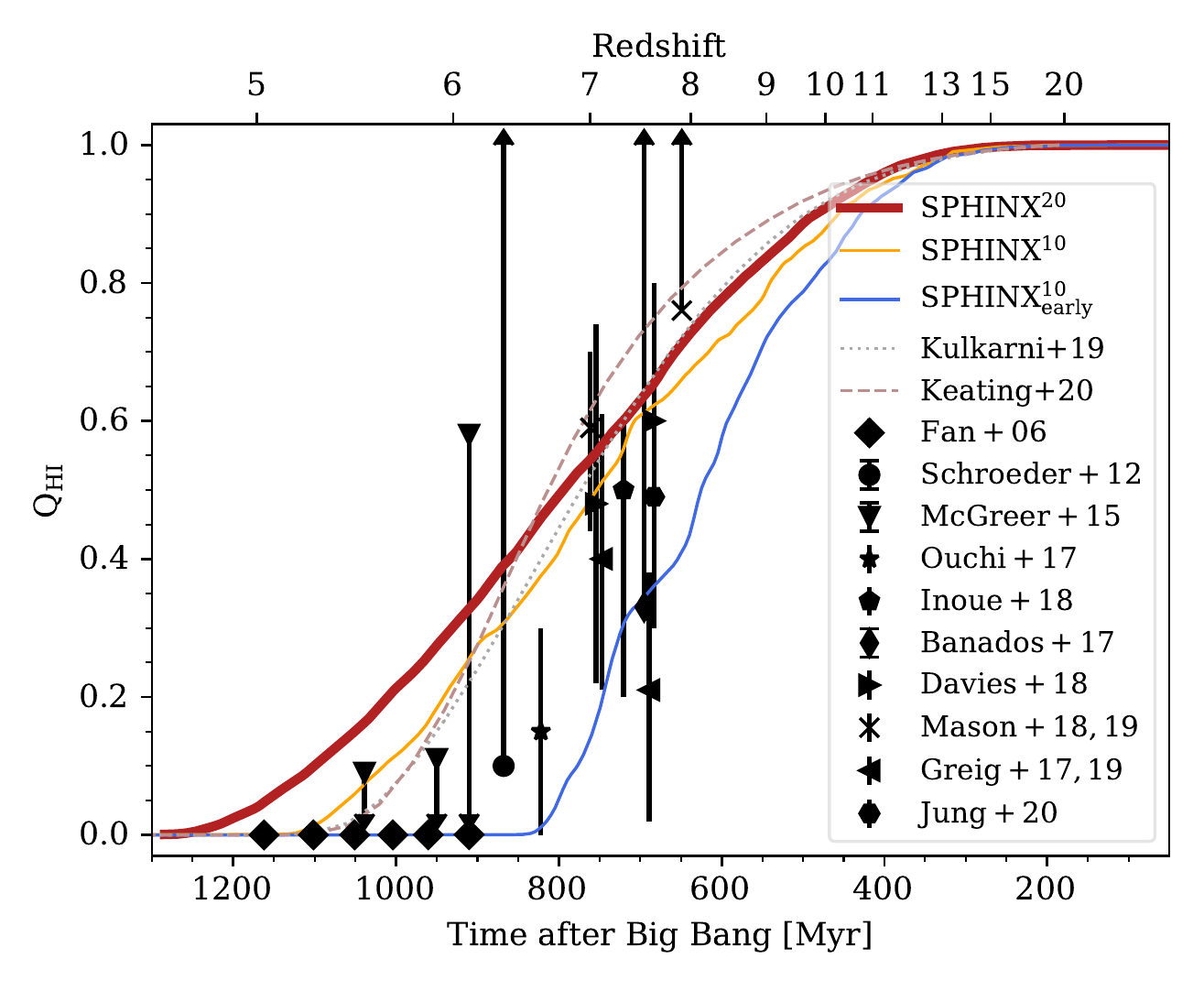}
  \caption
  {\label{xhi.fig}Evolution of the volume-filling neutral hydrogen
    fraction in three \sphinx{} runs (solid curves). The thick
    solid red curve represents the fiducial \sphinxtw{} run we
    focus on in this work. For comparison we show observational
    estimates from \citet{Fan2006a}, \citet{Schroeder2012},
    \citet{McGreer2015}, \citet{Ouchi2017}, \citet{Inoue2018},
    \citet{Banados2017}, \citet{Davies2018}, \citet{Mason2018},
    \citet{Mason2019}, \citet{Greig2017}, \citet{Greig2019}, and
    \citet{Jung2020}, as indicated in the legend. We also show the
    late reionization scenarios from \citet{Kulkarni2019} and
    \citet{Keating2020}. The timing of reionization is very sensitive
    to small variations in the SED model (solid blue vs. solid orange)
    and somewhat sensitive to the cosmological ICs (solid red
    vs. solid orange).}
\end{figure}

We now compare the simulated reionization histories against
observational constraints. We show in \Fig{xhi.fig} the reionization
history of our fiducial \sphinxtw{} simulation (thick solid red curve)
in the form of the volume-weighted neutral fraction, $\Qhi{}$, versus
redshift. We include the same for the two smaller volume simulations,
\sphinxten{} (same SED model) and \sphinxtenf{} (brighter SED
model). The updated BPASS version 2.2.1 we use in this work for
\sphinxtw{} and \sphinxten{} has about $30$ percent less total
ionising radiation emitted over the lifetime of an extremely
metal-poor stellar population than the older version 2.0 used in
\sphinxtenf{} and a factor $\approx4$ lower ionising luminosity at a
stellar population age of $\approx 15$ Myr (see \Fig{SEDs.fig}).

The \sphinxtw{} simulation achieves full reionization ($\Qhi=10^{-3}$)
at $z=4.64$. This is quite late compared to the various existing
observational estimates shown with black symbols in \Fig{xhi.fig}, and
well beyond the standard scenario of reionization by z=6
\citep{Fan2006}. However, large variations are expected in volumes as
small as ours and besides, \sphinxtw{} reionization is not very far
from \textit{late reionization} scenarios proposed recently by
\cite{Kulkarni2019} and \cite{Keating2020}, shown in dotted curves.

Switching to an 8 times smaller volume than \sphinxtw{} but otherwise
identical setup, \sphinxten{} is fully reionized about 150 Myrs
earlier (orange solid curve), or at $z=5.1$. This is presumably
somewhat random and due to cosmic variance, i.e. slightly different
halo mass functions between the two simulations. Now switching from
\sphinxten{} to \sphinxtenf{}, which uses BPASS version 2.0 but is
otherwise identical, we find a large effect on reionization, which is
achieved about 300 Myrs earlier with the older SED version, or at
$z=6.5$. Such sensitivity to the assumed SED model has already been
demonstrated in \cite{Ma2016} and \cite{Rosdahl2018}.

The latest constraints of the Thomson optical depth
$\tau=[0.0471, 0.0617]$, reported in \citet{PlanckCollaboration2018},
almost precisely bracket the three \sphinx{} simulations, with
$\tau=0.049, 0.052$, and $0.061$ in \sphinxtw, \sphinxten, and
\sphinxtenf{} respectively (calculated from the redshift evolution of
the hydrogen and helium ionization fractions using eq. 6 in
\citealt{Zaroubi2012}).

The three \sphinx{} simulations used here produce reionization
histories that bracket most observational constraints and recent
models. Even though the \sphinxtw{} simulation pushes a bit even the
recent late-reionization models, we still prefer to focus our analysis
in the current paper on this simulation, since it has almost an order
of magnitude better statistics -- i.e. more galaxies -- and an order
of magnitude more massive galaxies than the smaller volumes. Both
these aspects are very important for a meaningful statistical study of
the escape of ionising radiation from galaxies, due to its enormous
variability over time and from one galaxy to another. We show in
\App{bpass.app} how our main results vary with different volume size
and BPASS version, and to summarise we find that the correlations we
study as well as our conclusions are insensitive to those factors.

Any reionization model has degeneracies in $\fesc$, star formation,
and the amount of \Lyc{} photons produced per stellar mass, and none
of those factors are very well constrained (although the uncertainties
on $\fesc$ are probably the largest). Models with different
combinations of these three factors, within reasonable limits, may be
equally successful in reproducing observational constraints of the
reionization history and high-z luminosity functions.  Indeed we find
different \sphinx{} simulations can have different $\fesc$ and yet
bracket the aforementioned observational constraints. Therefore this
work is not very predictive on the overall magnitudes of escape
fractions, which can plausibly be larger than what we find in this
work, combined with somewhat lower star formation rates and/or more
conservative SED models. Our goal is not to predict the magnitudes but
rather how escape fractions correlate with galaxy properties and how
they may evolve with redshift. These correlations, which we will
demonstrate to be well converged with resolution, volume size, and SED
models considered, should be the take-away message of this paper.

\if
, except that on the whole,
escape fractions are somewhat higher in the smaller volume (because it
contains less massive and more metal-poor galaxies) and significantly
higher with the older version of BPASS (because it has much higher
ionising luminosities at moderately high stellar population ages).
\fi

\subsection{Global LyC escape fraction}\label{fesc.sec}

We now examine the global \Lyc -luminosity weighted escape fraction
$\fesc$ of \Lyc{} photons in the simulated volumes. We show the
redshift-evolution of $\fesc$ in the three \sphinx{} simulations in
\Fig{fesc_global.fig}. Thin solid transparent curves show
instantaneous $\fesc$ while the corresponding thick opaque curves show
their sliding luminosity-weighted average over the last 100 Myrs, or
$\feschundred$. We note the following observations from the figure:

\begin{figure}
  \centering
  \includegraphics[width=0.49\textwidth]
    {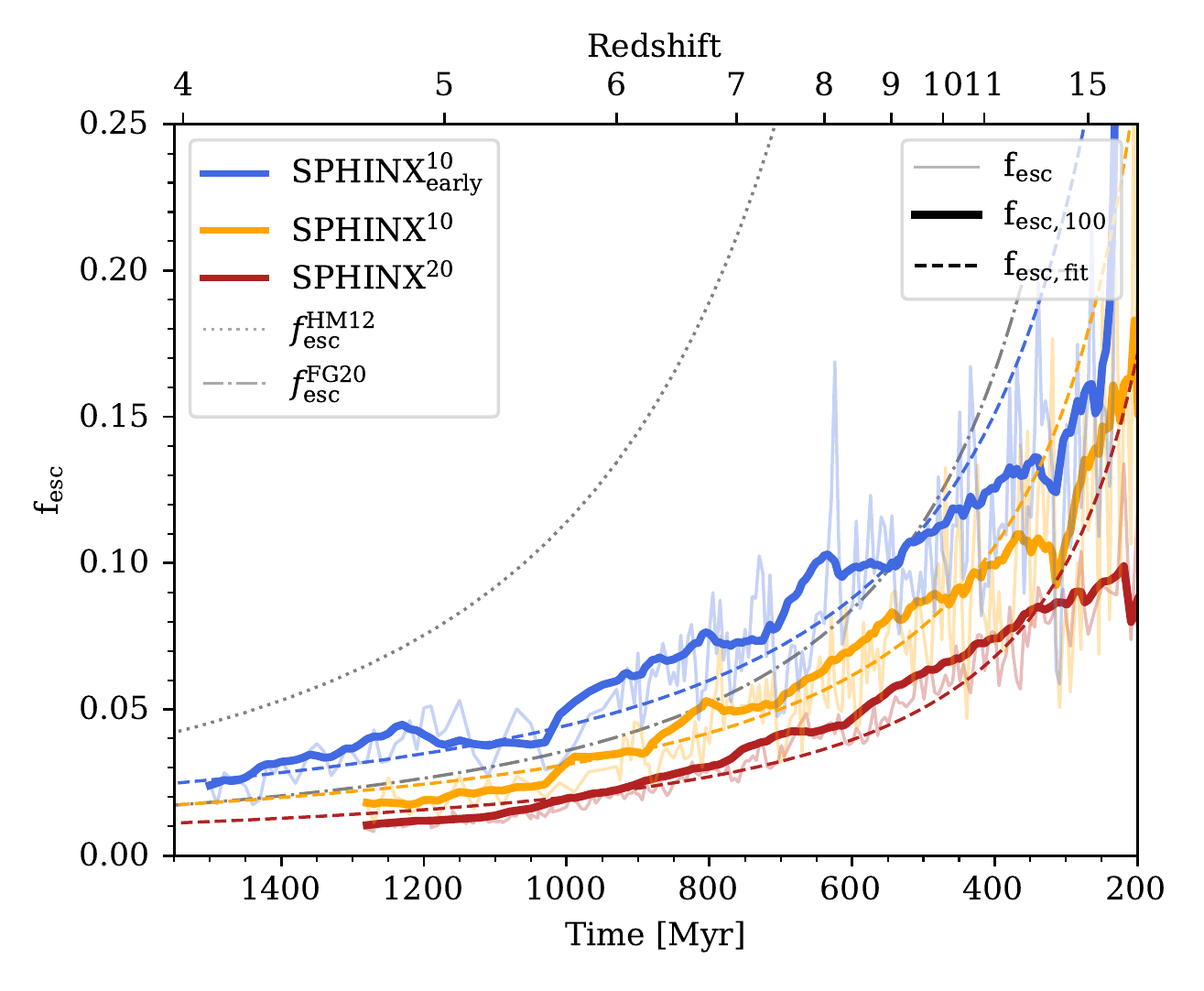}
  \caption
  {\label{fesc_global.fig}Time-evolution of the global
    (luminosity-weighted) escape fraction $\fesc$ in the three
    \sphinx{} simulations. The transparent highly fluctuating $\fesc$
    curves are instantaneous escape fractions, while the opaque thick
    curves show sliding luminosity-weighted averages over the previous
    100 Myrs. Dotted and dash-dotted grey curves show the models from
    \citet{Haardt2012} and \citet{FaucherGiguere2020}, respectively,
    to demonstrate how a trend of decreasing $\fesc$ with redshift is
    typically assumed in reionization models. The thin dashed curves
    show our fits to the global $\fesc$ evolution in the three
    \sphinx{} simulations to the functional form used by
    \citet{Haardt2012}, $\fesc(z)=a (1+z)^b$, with
    $a=4.5 \times 10^{-4}, 7 \times 10^{-4}$, and $10^{-3}$ for
    \sphinxtw, \sphinxten, and \sphinxtenf{} respectively, and $b=2$
    in all cases.  The global escape fraction in \sphinx{} decreases
    with time, as required to reionize the high-redshift Universe
    while simultaneously reproducing the observational constraint of
    very low escape fractions at low redshift.}
\end{figure}

\begin{enumerate}
\item Even if averaged over up to two billion stellar particles and
  tens of thousands of halos in \sphinxtw{}, the instantaneous escape
  fraction fluctuates strongly over time, and even more so in the
  smaller volume simulations. We focus for now on $\feschundred$,
  where the difference between simulations is easier to assess.
\item The escape fraction is significantly ($\approx 50-100 \%$)
  higher for the older and brighter BPASS version 2.0 SED model (blue)
  than for the identical simulation volume with the fiducial version
  2.2.1 (orange). This partly explains the earlier reionization with
  the older version shown in \Fig{xhi.fig} (complementary to the
  higher intrinsic stellar luminosities in the older version). We
  address in \Sec{sed_model.sec} why the more luminous SED model
  results in higher escape fractions.
\item The escape fraction is lower for the larger ($20$ cMpc)$^3$
  volume (red) than for the smaller ($10$ cMpc)$^3$ volume (orange),
  even if both use the same SED model. We address this difference in
  \Sec{volume_sed.sec}.
\item The global escape fraction decreases with time. For the
  \sphinxtw{} volume, it goes from just under $10$ percent at $z=15$
  to about one percent at $z=5$, and a qualitatively similar evolution
  is seen in the other \sphinx{} simulations.  A drop in $\fesc$ with
  time is almost always assumed by empirical reionization models, two
  of which are shown in dotted and dashed curves in
  \Fig{fesc_global.fig} for reference (eq. 56 in \citealt{Haardt2012}
  and eq. 12 in \citealt{FaucherGiguere2020}, though note these
  empirical models can vary greatly in the overall magnitudes due to
  aforementioned degeneracies with star formation rates and stellar
  \Lyc{} luminosities as well as uncertainties in the timing and
  duration of reionization). Different physical reasons for what
  drives such a drop have been suggested, but none confirmed. The
  \sphinx{} simulations naturally produce a drop in $\fesc$ with
  redshift, in qualitative agreement with empirical reionization
  models, and we address in \Sec{fesc_regulation.sec} what drives this
  evolution.
\item The escape fraction we find is low, well below $10\%$ during
  most of the EoR. While this is at odds with some values used in the
  literature (e.g. $20\%$ in \citealt{Ouchi2009a} or the
  \citealt{Haardt2012} model shown in \Fig{fesc_global.fig}), it is
  very consistent with the recent model of \citet{FaucherGiguere2020}
  included in \Fig{fesc_global.fig}, as well as with
  \citet{Finkelstein2019} in whose model an average $\fesc<5\%$ is
  sufficient to reionize the Universe.
\end{enumerate}

The average $\fesc$ at a given time as shown in \Fig{fesc_global.fig}
does not represent the escape fraction in a typical halo in the
simulation. Rather, most halos have almost zero $\fesc$ and/or
negligible intrinsic luminosities, while a tiny fraction of halos have
high $\fesc$ and non-negligible luminosities simultaneously. At a
given time, the escaping ionising emissivity is hence typically
dominated by a small fraction of halos, which explains the large
fluctuations in the global $\fesc$ in \Fig{fesc_global.fig}. In
addition, due to the stochasticity of both star formation and escape
fractions for individual halos, this situation changes rapidly, with
the halos that dominate the escaping emissivity at a given time
becoming sub-dominant only a few million years later. Reionization of
the \sphinx{} volumes is ``disco-like'', as coined by
\cite{Matthee2021}, with galaxies emitting brief flashes of \Lyc{}
radiation into the IGM and then going dark.\footnote{Readers are
  encouraged to experience disco-reionization animations on the
  \sphinx{} website: \url{http://sphinx.univ-lyon1.fr}. Disco music is
  not provided, so please bring your own.}

\begin{figure}
  \centering
  \includegraphics[width=0.45\textwidth]
    {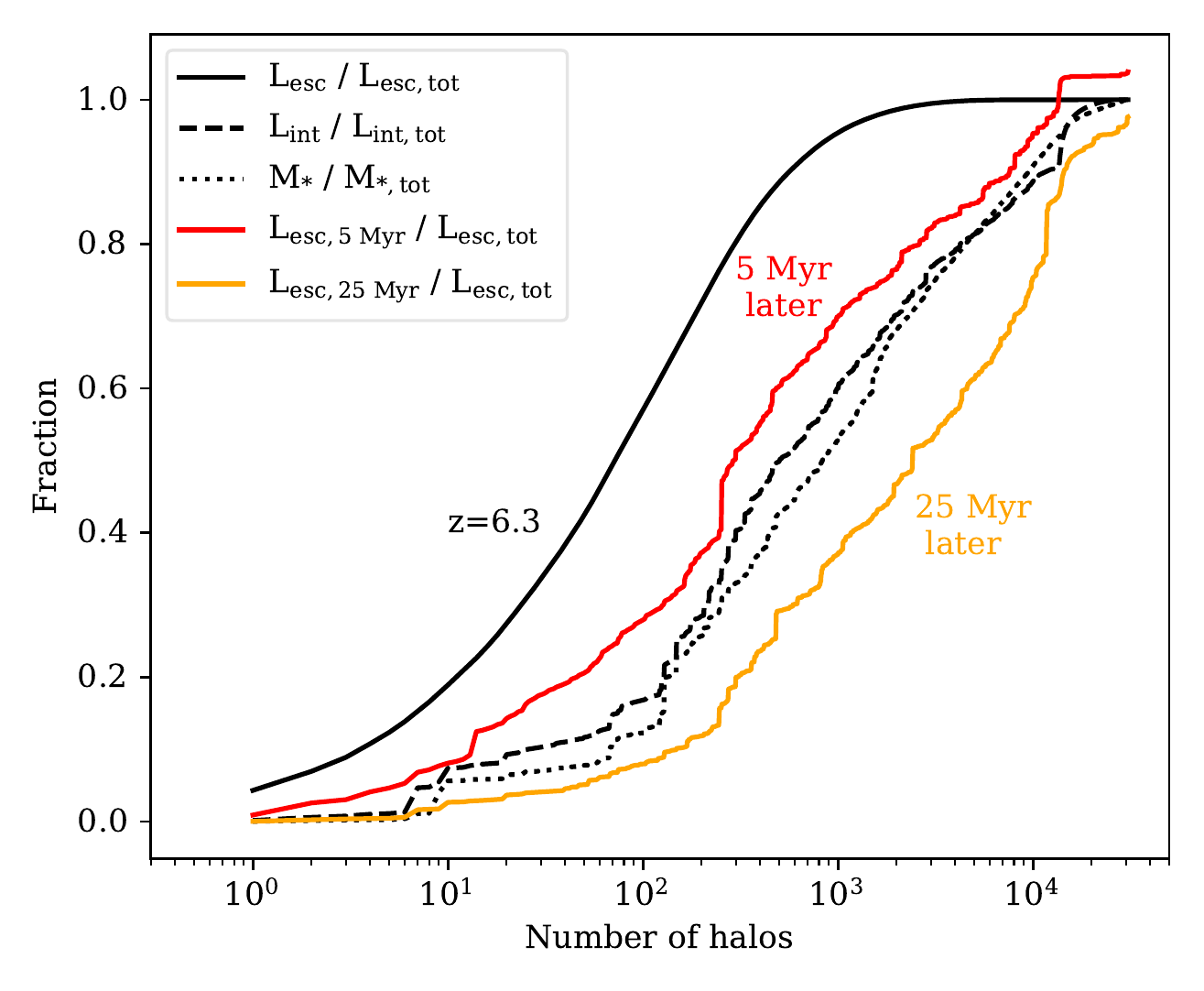}
  \caption
  {\label{christmas.fig} Fraction of total escaping and intrinsic
    \Lyc{} luminosities, in solid and dashed black respectively, from
    individual halos in a single snapshot at $z=6.3$ in \sphinxtw{},
    ordered by decreasing escaping halo luminosity. The dotted black
    curve shows the cumulative stellar mass fraction of the same halos
    and in the same order. The escaping \Lyc{} emissivity in the
    simulated volume is dominated by a relatively few halos, with only
    about $100$ halos out of $\approx 3 \times 10^4$ being responsible
    for almost $60$ percent of the escaping emissivity, and about $95$
    percent of the escaping emissivity coming from a thousand
    halos. The red and orange curves show the cumulative escaping
    emissivity from the same halos in the same order $5$ and $25$ Myrs
    later. The escaping emissivities of individual galaxies change
    rapidly and dramatically, with the $100$ previously most dominant
    halos accounting for only about a quarter and the tenth of the
    escaping photons $5$ and $25$ Myrs later. Note that the red and
    orange curves are normalised to the total escaping luminosity at
    $z=6.3$, which is why they do not converge to unity. In other
    words, the total escaping emissivity changes over time.}
\end{figure}

We demonstrate this in \Fig{christmas.fig}. Here we show the fractions
of the total escaping emissivities in \sphinxtw{} in solid black at
$z=6.3$, ordering the halos by decreasing escaping luminosity,
i.e. with the \Lyc -brightest halos on the left. Out of the
$\approx 3 \times 10^4$ halos containing stars, a single halo
contributes almost $5$ percent of the total escaping emissivity
$\Lesc$, $10$ halos contribute to $\approx 20$ percent of it, and
$100$ halos to $\approx 60$ percent. These contributions are quite
different from the \textit{intrinsic} emissivities, $\Lintr$, shown in
dashed black, with e.g. the $100$ $\Lesc$-brightest halos contributing
to $\lesssim 20$ percent of $\Lintr$. This is because halos with
similar intrinsic luminosities can exhibit very different escape
fractions. Both the escape fractions and luminosities vary
dramatically over time. The solid red and orange curves show the
fraction of $\Lesc$ for the same halos and in the same order, $5$ and
$25$ Myrs later, respectively. The $10$ halos that contributed to
$\approx 20$ percent of the total $\Lesc$ at $z=6.3$ contribute $25$
Myrs later to about $2-3$ percent of the total, and the $100$ halos
that contributed $\approx 60 \%$ earlier now contribute
$\approx 10\%$.

Given the large time-variability in escape fractions and \Lyc{}
emissivities, we find it both reasonable and necessary to stack
simulation snapshots in most of the forthcoming analysis of escape
fractions and relative contributions of halos to the escaping
emissivity. For each redshift interval, we collect data from all
snapshots within that interval, always having $5$ Myrs between
individual snapshots. We therefore treat a single halo, evolving over
time, as many. Without such stacking, the combination of small
statistics and large scatter leads to excessive noise when binning
escape fractions by halo (or galaxy) properties, as we do in the next
sections.

\subsection{Escape of LyC photons from halos}\label{halos_fesc.sec}

We now assess how \Lyc{} escape fractions correlate with several halo
properties. We consider halo mass ($\Mvir$), stellar mass ($\Mstar$),
metallicity ($\Zhalo$) and specific star formation rate ($\sSFR$),
each of which has been predicted by different theoretical works to
have an effect on the escape fraction, and also UV magnitude
($\MFifteen$), in order to estimate the fraction of escaping \Lyc{}
photons which are accounted for in the bright and well-constrained
part of the high-z luminosity function.

\subsubsection{Halo mass}
\begin{figure}
  \centering
  \includegraphics[width=0.49\textwidth]
    {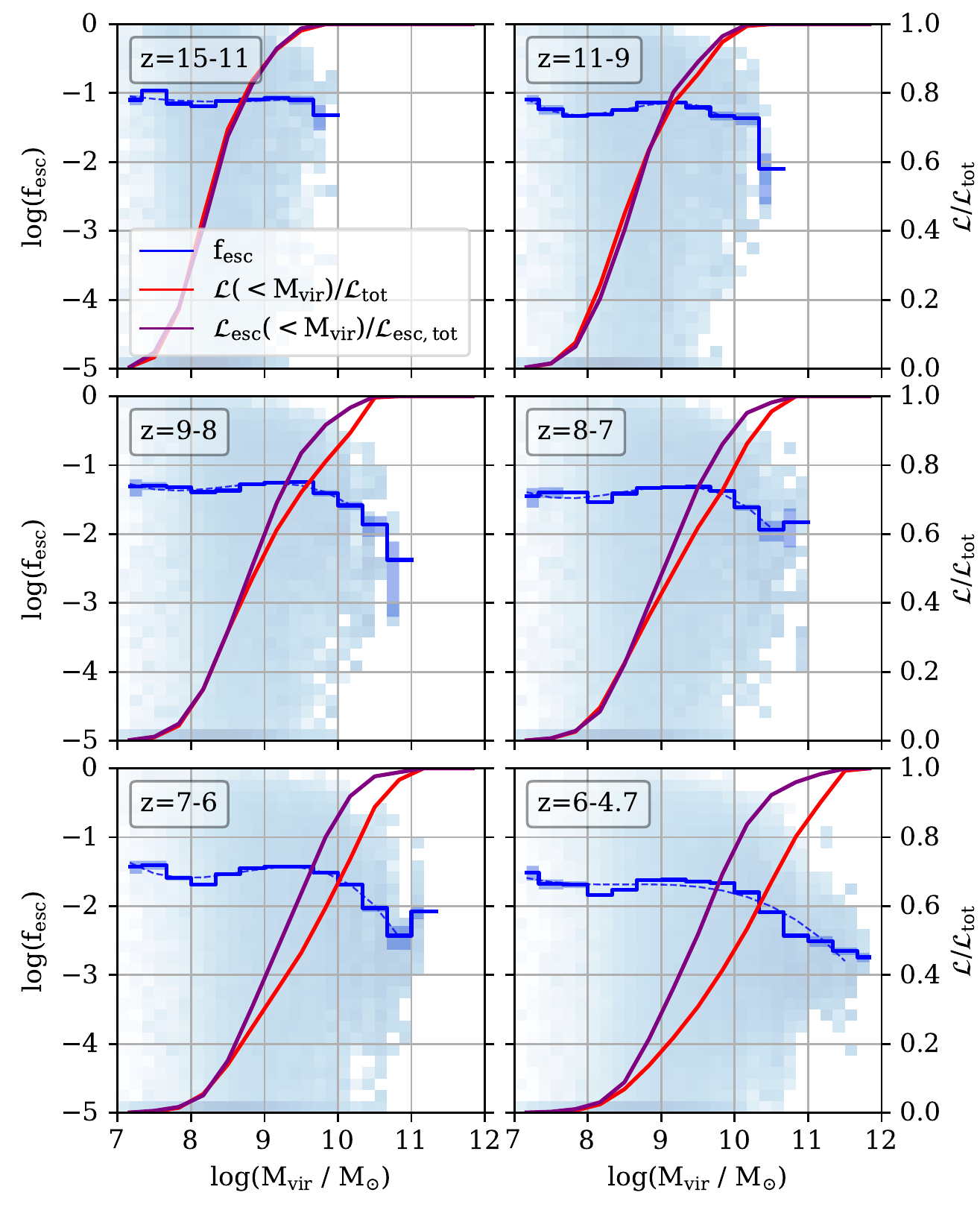}
    \caption
    {\label{fesc_hist_Mvir.fig} Intrinsic-\Lyc{}-luminosity-weighted
      mean escape fraction in \sphinxtw versus halo mass ($\Mvir$), in
      solid blue, stacked over snapshots with 5 Myr intervals at
      different redshift ranges, as indicated by legends in each
      panel. We also show the fraction of the total intrinsic
      (escaping) \Lyc{} luminosity in red (purple) emitted by halos
      with mass $<\Mvir$, to be read against the right-side axis. To
      illustrate the scatter in $\fesc$ for individual halos, we show
      in the background of each panel a 2D histogram of the normalised
      intrinsic \Lyc{} luminosity in bins of $\fesc$ (floored at the
      lower y-limit) and halo mass, and the standard error of $\fesc$
      with a shaded region around the solid blue curve. The dashed
      blue curves show our best third-degree polynomial fits for
      $\log(\fesc)$, which we describe in detail in \App{fits.app}. At
      any given redshift, $\fesc$ is fairly flat for
      $\Mvir\lesssim 3\times10^9 \ \Msun$, above which it decreases
      strongly with increasing halo mass. As seen by the red curve
      drifting towards more massive halos with decreasing redshift,
      the \Lyc{} photons are produced by increasingly massive halos,
      with decreasing $\fesc$, over time.}
\end{figure}

We show in \Fig{fesc_hist_Mvir.fig} how \Lyc{} escape fractions
correlate with halo mass $\Mvir$ at different redshifts. Each panel
covers a redshift interval, with the highest redshifts in the top left
panel and the lowest redshifts in the bottom right one. For each
redshift interval, we plot the \Lyc{}-luminosity-weighted mean escape
fraction per halo mass in blue, and we show the
intrinsic-luminosity-weighted probability distribution of $\fesc$ in
the background to illustrate the scatter for individual halos. We also
show in red the fraction of total intrinsic \Lyc{} luminosity produced
by halos below a given mass and in purple the corresponding
\textit{escaping} luminosity (which is just the product of the blue
and red curves). In each panel we stack all snapshots belonging to the
relevant redshift range in order to reduce the noise.

Several insights can be drawn from \Fig{fesc_hist_Mvir.fig}. Firstly,
at all simulated redshifts, we find a fairly flat $\fesc$ for all
halos with $\Mvir \lesssim 3 \times 10^{9} \ \Msun$ and a drop in $\fesc$
for more massive halos. This drop for high halo (or galaxy) masses is
not particularly surprising and a similar feature has been reported in
several works \citep[][]{Razoumov2010, Kimm2014, Paardekooper2015,
  Xu2016, Yajima2020, Ma2020, Lewis2020}, though there is not full
agreement on this point and some have predicted the opposite trend
\citep{Gnedin2008, Wise2009, Naidu2019}. The reason for the drop in
$\fesc$ with increasing halo mass is that stellar feedback gradually
becomes less efficient with mass in clearing out gas and allowing
radiation to escape. While most recent works agree that the escape
fraction decreases at high halo masses, there is no agreement on the
magnitudes of escape fractions or the halo mass limit at which the
escape fraction starts to drop significantly. This is because both are
sensitive to the efficiency and timing of stellar feedback
\citep{Rosdahl2018} and likely to resolution as well \citep{Ma2020}.

A second feature in \Fig{fesc_hist_Mvir.fig} is the drift of the
production of \Lyc{} photons to more and more massive halos with
decreasing redshift (the red curve moves right). At the highest
redshift range of $z=15-11$, almost all the \Lyc{} photons are emitted
in halos with $\Mvir\lesssim 3 \times 10^{9} \ \Msun$. Since $\fesc$ is
fairly flat at these low halo masses, the halos that produce \Lyc{}
photons are also the ones from which \Lyc{} photons escape, and the
red and purple curves are overlaid. As halos become more massive, the
intrinsic luminosity drifts with decreasing redshift to higher
masses. In the final redshift range of $z=6-4.7$, \Lyc{} photons are
predominantly emitted in massive halos, with only about $30\%$ of them
emitted in halos with $\Mvir\lesssim 3 \times 10^{9} \ \Msun$. However,
because of the shape of the $\fesc$ versus $\Mvir$ relation, these low
mass halos contributing to $30$ percent of the intrinsic \Lyc{}
emissivity still contribute to about $50$ percent of the
\textit{escaping} emissivity, shown in purple.

\subsubsection{Galaxy mass} \label{galmass.sec}

\begin{figure}
  \centering
  \includegraphics[width=0.49\textwidth]
    {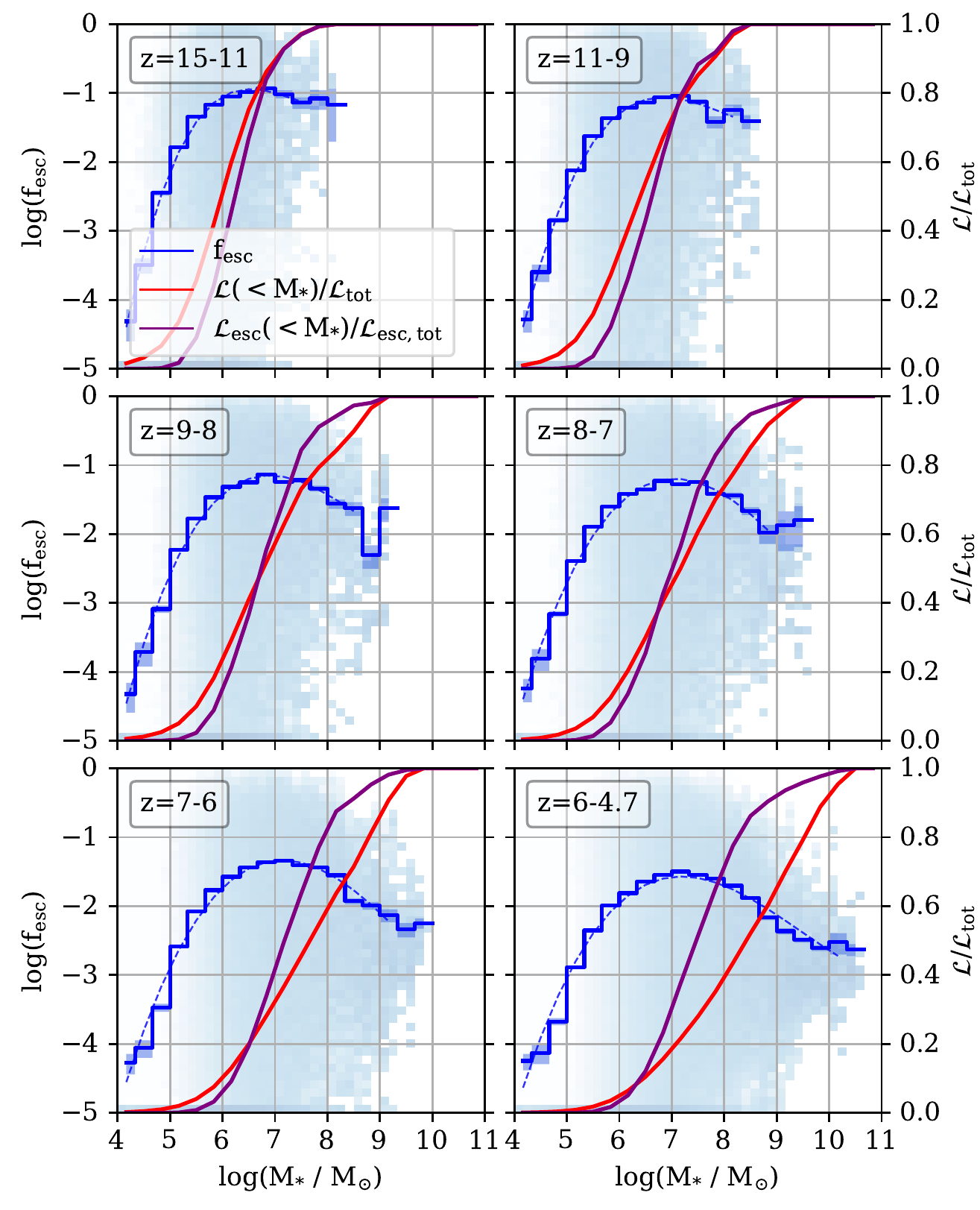}
  \caption
  {\label{fesc_hist_Mstar_all.fig}Correlation of
    intrinsic-luminosity-weighted mean escape fraction with galaxy
    stellar mass in \sphinxtw at different redshifts. See the caption
    of \Fig{fesc_hist_Mvir.fig} for a detailed explanation of the
    panels. At all redshifts, the escape fraction peaks at
    $\Mstar \approx 10^7 \ \Msun$ and decreases for both lower and
    higher masses. As seen by the red curve drifting towards higher
    galaxy mass with redshift, \Lyc{} photons are produced in
    increasingly massive galaxies over time, which have decreasing
    $\fesc$.}
\end{figure}

We now look at $\fesc$ versus galaxy stellar mass, $\Mstar$, in
\Fig{fesc_hist_Mstar_all.fig}. At all redshifts we find a strong
dependence of $\fesc$ on $\Mstar$, with the highest $\fesc$ found at
intermediate masses of $\approx 10^7 \ \Msun$ and declining
$\fesc$ for both increasing and decreasing $\Mstar$.

The drop in $\fesc$ for $\Mstar \ga 10^7 \ \Msun$ is due to the
decreasing efficiency of stellar feedback with mass as already
discussed and simply reflects the drop in $\fesc$ for the parent halos
with $\Mvir \ga 3 \times 10^9 \ \Msun$ as seen in
\Fig{fesc_hist_Mvir.fig}. The strong drop with $\fesc$ towards the
lowest galaxy masses, however, is more surprising and is not reflected
in the $\fesc$ versus halo mass relation. This drop appears to be due
to star formation being unsustained over time in the lowest-mass
galaxies. For the more massive galaxies with
$\Mstar \ga 10^6 \ \Msun$, star formation happens in bursts sustained
over a few Myrs, leading to a) a collective burst of strong SN
feedback and b) ``support'' by the SN feedback of the first stars born
in clearing channels for \Lyc{} radiation to escape from successively
formed ones. For the lowest-mass galaxies, however, the bursts are
more discrete and less sustained, with typically one or very few
stellar particles formed ($400 \ \Msun$ in mass each), making their SN
feedback less disruptive and weaker in facilitating the \Lyc{}
radiation to escape. This is likely due to radiation feedback: as
shown in e.g. \cite{Agertz2020, Smith2020}, ionizing radiation
feedback suppresses the clustering of star formation in low-mass
galaxies, leading to a negative form of feedback where the efficiency
of SN explosions is weakened. As low-mass galaxies grow, star
formation becomes more clustered, whereby SN explosions become more
efficient and increasingly the first stars born in a starburst can
help clear away gas from their younger siblings, increasing the escape
fraction.

At first glance, the correlation between $\fesc$ and galaxy mass
appears inconsistent with that for halo mass
(\Fig{fesc_hist_Mvir.fig}), which does not show a decline in $\fesc$
for the lowest halo masses. However, the unvarying escape fraction for
the lowest halo masses is just a consequence of the large scatter in
the stellar mass to halo mass at those low masses, as seen in
\Fig{smhm.fig}. For the lowest-mass halos, the mean escape fraction
(which, we remind, is luminosity weighted) is dominated by the most
luminous galaxies in the bin, which also have the highest stellar mass
and the highest escape fractions. This leads to a fairly flat $\fesc$
with halo mass.

A similar peak at a certain galaxy mass and turndown for both lower
and higher galaxy masses was found by \cite{Ma2020}, though their peak
is at $\Mstar \approx 10^8 \ \Msun$, about ten times higher than in
our case. The reason for their peak being at a higher stellar mass is
likely that their sub-grid (radiation and SN) stellar feedback is
significantly more efficient in suppressing star formation than in
\sphinx{}, leading to a stronger negative radiation feedback at low
masses and stronger positive SN feedback at higher masses. As do we,
\cite{Ma2020} find their escape fraction to be fairly insensitive to
\textit{halo} mass at the low-mass end, again likely due to the
scatter in the stellar to halo mass at the low-mass end.

However, for these lowest-mass galaxies, we are in discord with other
works using extreme-resolution simulations. \cite{Kimm2017} studied
the escape of \Lyc{} radiation from mini-halos
($\Mvir \sim 10^8 \ \Msun$) in the EoR using cosmological RHD zoom
simulations of extremely high resolution ($\Dxmin \lesssim 1$
pc). They used the same code as we do in this paper and similar
methods for star formation and feedback, but additionally included
PopIII stars and molecular hydrogen formation. Contrary to the current
paper, they found high (luminosity-weighted) mean escape fractions on
the order of $\approx 40$ percent for halos with $\Mvir \lesssim 10^7$
and a significant drop in $\fesc$ with increasing mass for
$10^7 \Msun \lesssim \Mvir \lesssim 10^8 \ \Msun$. Similar results
were found, using a different code and methods but comparable
resolution and physics, by \cite{Wise2014, Xu2016}. These authors
attribute these very high escape fractions in mini-halos to radiation
feedback, i.e. the \Lyc{} radiation breaking itself out in these
low-mass halos. Although this can in part be due to their inclusion of
PopIII stars, which are very luminous, we find it likely that the lack
of strong enough \Lyc{} feedback in low-mass \sphinx{} galaxies is a
numerical limitation due to our finite resolution \citep[$\approx 10$
pc versus $\approx 1$ pc in][]{Wise2014, Kimm2017}. Presumably, higher
resolution would produce high escape fractions in galaxies with
$\Mstar \lesssim 10^5 \ \Msun$, corresponding to the halo masses where
escape fractions are high in the aforementioned works, and in
accordance with those high-resolution results, the escape fractions
would drop and converge with what we find for SN feedback regulation
at $\Mstar \ga 10^5 \ \Msun$. Reassuringly, \cite{Kimm2017} find that
these low mass halos with $\Mvir \lesssim 10^8 \ \Msun$ (or
$\Mstar \lesssim 10^5 \ \Msun$) contribute very little to reionization
despite their high escape fractions, because they are very susceptible
to feedback and hence are very dim. To summarise, the drop we see in
$\fesc$ for $\Mstar \lesssim 10^7 \ \Msun$ is probably real, due to
negative radiation feedback, but in reality the escape fraction rises
again for $\Mstar \lesssim 10^5 \ \Msun$ where radiation feedback can
do serious damage.

We now consider the redshift-evolution of the fraction of intrinsic
\Lyc{} luminosities of galaxies versus mass, shown in red in the
panels of \Fig{fesc_hist_Mstar_all.fig}. At the highest redshift range
of $z=15-11$, about $90 \%$ of the \Lyc{} photons are coming from
galaxies below the peak-$\fesc$ mass of $10^7 \ \Msun$. Since the
escape fraction is rising with increasing galaxy mass in these
galaxies, the escaping \Lyc{} luminosity, in purple, is shifted
somewhat towards more massive galaxies compared to the intrinsic
one. As galaxies become more massive, the intrinsic luminosity drifts
to the right with decreasing redshift, and at the end of reionization
(rightmost bottom panel), $\approx 80\%$ of the emitted \Lyc{} photons
are coming from galaxies above the peak-$\fesc$ mass (i.e.
$\Mstar>10^7 \Msun$). Because of the shape of the $\fesc$ versus
galaxy mass relation, the escaping \Lyc{} emissivity is now shifted to
lower galaxy masses compared to the intrinsic emissivity, with e.g.
$\Mstar<10^8 \ \Msun$ galaxies accounting for $\approx70\%$ of the
escaping luminosity but only $40\%$ of the intrinsic luminosity.

\subsubsection{Metallicity}
\begin{figure}
  \centering
  \includegraphics[width=0.49\textwidth]
    {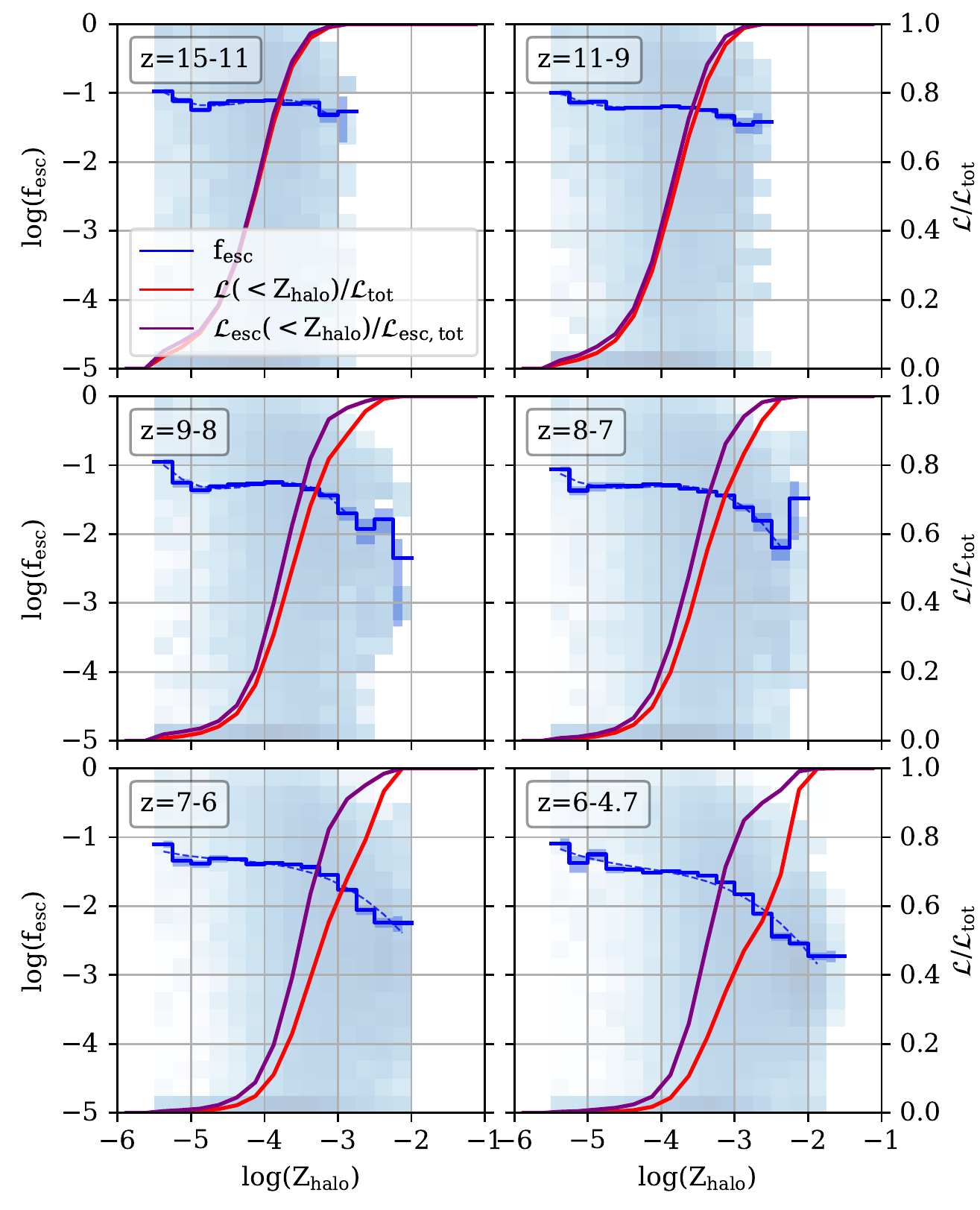}
  \caption
  {\label{fesc_hist_Z.fig} Correlation of
    intrinsic-luminosity-weighted mean escape fraction with
    luminosity-weighted metal mass fraction, $\Zhalo$, in
      \sphinxtw at different redshifts. See the caption of
    \Fig{fesc_hist_Mvir.fig} for a detailed explanation of the
    panels. Galaxies with metallicities $\Zhalo \ga 3 \times 10^{-4}$
    (more than about one percent Solar) tend to have low escape
    fractions, due to efficient cooling and because they are
    massive. The global \Lyc{} emissivity drifts with decreasing
    redshift towards more metal-rich galaxies, which have low
    $\fesc$.}
\end{figure}
We examine $\fesc$ versus \Lyc{}-luminosity-weighted stellar
metallicities of galaxies, or $\Zhalo$, in \Fig{fesc_hist_Z.fig}. We
find a fairly flat escape fraction for
$\sim 10^{-5} \lesssim \Zhalo \lesssim 3\times 10^{-4}$, a drop in $\fesc$ for
higher metallicities, and a spike in $\fesc$ for the floor metallicity
of $\Zhalo=6.4 \times 10^{-6}$. The decrease in $\fesc$ with ``high''
metallicities has also been found by \cite{Yoo2020} and is due to
enhanced cooling in metal-rich gas which weakens the efficiency of SN
explosions in clearing away dense \Lyc{}-absorbing gas.

We note that we find more or less the same correlation and redshift
evolution for \textit{mass-weighted} stellar metallicities, which is
less biased towards the youngest (and most luminous) stars in each
galaxy.

A comparison of the different panels in \Fig{fesc_hist_Z.fig} shows
that \Lyc{} photons are produced in increasingly metal-rich galaxies
with decreasing redshift (red curve drifting towards right with
redshift). The drift for \textit{escaping} radiation with redshift
(purple curves), however, is slower, due to those metal-rich galaxies
having very low escape fractions.

\subsubsection{Specific star formation rate} \label{ssfr.sec} Several
works have argued that if the escape of \Lyc{} is regulated by
feedback, it should be the highest in galaxies experiencing the
strongest starbursts. \cite{Heckman2011} proposed that intense star
formation, followed by extreme feedback, is a driver of high
$\fesc$. \cite{Sharma2017} similarly argued that galaxies with high
star formation rate surface densities have high $\fesc$, as these same
galaxies are observed to be able to generate galactic winds, which
should clear a way for \Lyc{} radiation escape.
\cite{FaucherGiguere2020} took the argument a step further, suggesting
that a decreasing intensity of star formation and, hence, feedback in
the expanding Universe could drive the global escape fraction to
decrease with cosmic time.

We therefore explore the correlation between $\fesc$ and star
formation in \sphinx{}. Using the star formation rate (SFR) of
galaxies as a proxy for extreme star formation is not the best
approach, since it is very biased towards the most massive
galaxies. Instead we quantify the extremeness of star formation via
the specific star formation rate, $\sSFR_{\tau}$, which is calculated
for each galaxy as

\begin{align} \label{ssfr.eq}
  \sSFR_{\tau}=\frac{SFR_{\tau}}{\Mstar}
  = \frac{\frac{\Mstar(age<\tau)}{\tau}}{\Mstar},
\end{align}
where $\Mstar$ is the stellar mass of the galaxy and $\tau$ is the
lookback time, in Myr, over which we average the star formation rate
(SFR). In other words, $\sSFR_{\tau}$ is the ratio of the mass of
stars formed in the previous $\tau$ Myrs versus the total stellar mass
of the galaxy\footnote{Note that for simplicity we use the
  \textit{initial} or formed stellar mass in both the numerator and
  denominator of \Eq{ssfr.eq}, whereas it is more traditional to use
  the \textit{current} mass (accounting for stellar mass loss) in the
  denominator.}.  We consider specific star formation rates with two
lookback times, short-term with $\sSFRten$ and long-term with
$\sSFRc$. The former roughly corresponds to star formation rates
observationally determined via H$\alpha$ and the latter via FUV
\citep[e.g.][]{Kennicutt2012}. By our \Eq{ssfr.eq} definition,
$\sSFRten$ can be in the range $[0,100]$ Gyr$^{-1}$ and $\sSFRc$ can
be in the range $[0,10]$ Gyr$^{-1}$, the upper limits meaning that all
the stars in a galaxy have been formed in the last $\tau$ Myrs. Since
$\sSFRten$ is measured on a shorter timescale, it is a better
indicator of the extremeness of star formation than $\sSFRc$, but is
restricted to either ongoing or very recent starbursts. The
longer-term $\sSFRc$ can either indicate intense starbursts in the
last $100$ Myrs or alternatively a steady rate of star formation over
the same period, and is hence not a perfect indicator of the most
intense starbursts.

\begin{figure}
  \centering
  \includegraphics[width=0.49\textwidth]
    {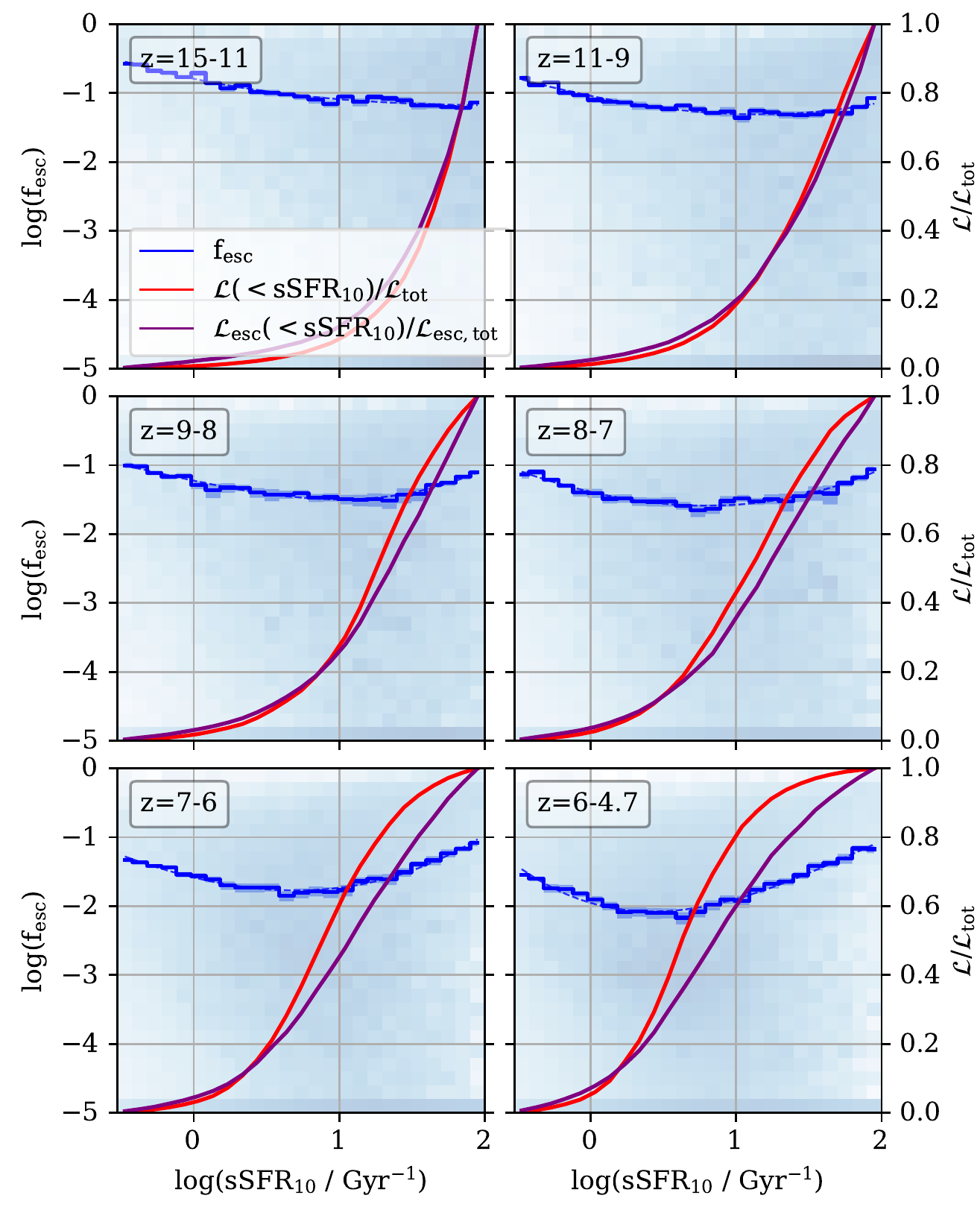}
  \caption
  {\label{fesc_halo_hist_sSFR10.fig} Correlation of
    intrinsic-luminosity-weighted mean escape fraction with short-term
    specific star formation rate, $\sSFRten$, in \sphinxtw at
    different redshifts. See the caption of \Fig{fesc_hist_Mvir.fig}
    for a detailed explanation of the panels. Note that we have
    floored galaxies with $\sSFRten$ below the range to the leftmost
    $\sSFRten$ bin, so they are counted in the plots.  The galaxies
    experiencing the highest $\sSFRten$ have higher mean escape
    fractions than those with moderate $\sSFRten$. Galaxies with very
    low $\sSFR$s also have high mean escape fractions, especially at
    high z, though they do not contribute much to the total $\Lyc$
    budget. The global \Lyc{} emissivity drifts significantly with
    decreasing redshift towards galaxies with decreasing $\sSFRten$.}
\end{figure}

We show the correlation of $\fesc$ with the shorter-term $\sSFRten$ in
\Fig{fesc_halo_hist_sSFR10.fig} (blue curve) over six redshift
intervals. At all redshift ranges except perhaps the earliest one,
there is a strong increase in $\fesc$ with $\sSFRten$ for galaxies
experiencing the strongest starbursts, with
$\sSFRten \ga 30 \ Gyr^{-1}$. At lower $\sSFR$s, we find $\fesc$
reaching the lowest values at intermediate $\sSFR$s and rising
towards the lowest $\sSFR$s. Comparing the red curves in each panel,
we see an evolution in $\Lyc$ photons being emitted from galaxies with
decreasing $\sSFR$s with decreasing redshift.

The peak in $\fesc$ at the high $\sSFRten$-end fits the expectation
that starbursts lead to high escape fractions. But what about the high
escape fractions that we find for low $\sSFR$s? Upon examination of
randomly selected halos with low $\sSFRten$ and high $\fesc$, we find
that these are typically galaxies that have recently, \textit{but not
  very recently} (i.e. more than $\tau=10$ Myr), experienced
starbursts and then subsequent feedback episodes that ejected nearly
all the ISM gas. These violent ejections of gas have three effects: i)
$\fesc$ becomes high, because there is little gas remaining within the
halo to absorb the radiation, ii) the SFR becomes very low or even
zero, because there is no longer any gas to fuel star formation, and
iii) the \Lyc{} luminosity becomes very low because of effect ii). We
have shown two examples of low-mass halos in this quenched, high
$\fesc$ state in the rightmost two panels in the bottom row of
\Fig{map_fullbox.fig}.

Both a very high and a very low $\sSFRten$ seems to be a good
indicator of high $\fesc$, though the latter class of halos does not
contribute very significantly to reionization (purple curves in
\Fig{fesc_halo_hist_sSFR10.fig}). However, as is evident from the
background 2D histograms in \Fig{fesc_halo_hist_sSFR10.fig},
there is huge scatter in $\fesc$ at any $\sSFRten$, so spotting a
galaxy with a very high (or very low) $\sSFRten$ is by no means a
guarantee of spotting a high escape fraction.

\begin{figure}
  \centering
  \includegraphics[width=0.49\textwidth]
    {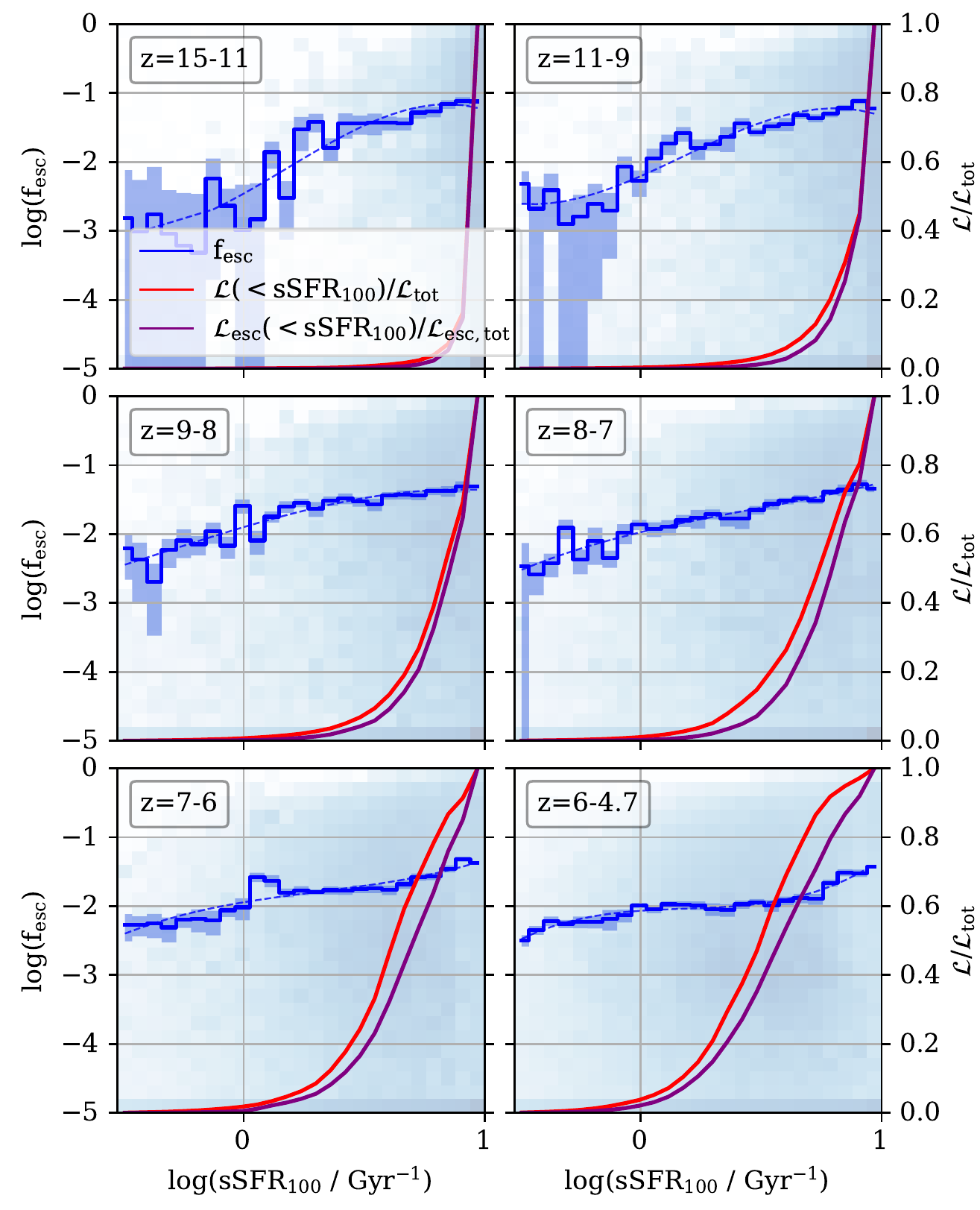}
  \caption
  {\label{fesc_halo_hist_sSFR100.fig} Correlation of
    intrinsic-luminosity-weighted mean $\fesc$ with long-term specific
    star formation rate, $\sSFRc$, in \sphinxtw at different
    redshifts. See caption of \Fig{fesc_hist_Mvir.fig} for a detailed
    explanation of the panels. Note that we have floored galaxies with
    $\sSFRc$ below the range to the leftmost $\sSFRc$ bin, so they are
    counted in the plots. The correlation of $\fesc$ with $\sSFRc$ is
    more straightforward than with $\sSFRten$, with overall decreasing
    $\fesc$ for decreasing $\sSFR$s. Like for $\sSFRten$, the global
    \Lyc{} emissivity drifts significantly with time towards galaxies
    with decreasing $\sSFRc$.  }
\end{figure}

We show in \Fig{fesc_halo_hist_sSFR100.fig} the correlation of $\fesc$
with the longer-term $\sSFRc$. As expected, the highest $\sSFRc$
galaxies have the highest escape fractions. However, the peak in
$\fesc$ is somewhat weaker than for high-$\sSFRten$ galaxies. This is
because $\sSFRc$ is not as useful as $\sSFRten$ in singling out
starbursts -- a high $\sSFRc$ can mean anything between a quick
intense starburst and a fairly robust but relatively continuous
episode of star formation with low $\fesc$, lasting over tens of Myrs
whereas the majority of \Lyc{} photons are produced within 10 Myrs
from the starburst. As for $\sSFRten$, we see an evolution of \Lyc{}
radiation being emitted from galaxies with decreasing $\sSFRc$ with
decreasing redshift.
The lack of high $\fesc$ for galaxies at the very low $\sSFRc$-end
further supports our interpretation that high $\fesc$ for low
shorter-term $\sSFRten$ in \Fig{fesc_halo_hist_sSFR10.fig} indeed
comes from highly disrupted galaxies that recently, but not too
recently, experienced starbursts and then shut down star formation,
hence having very low $\sSFRten$ but still moderate or high $\sSFRc$.

Correlations between $\fesc$ and $\sSFR$ has not been examined often
in previous simulation works predicting escape fractions, but both
\cite{Xu2016} with the Renaissance simulations, and
\cite{Paardekooper2015} with the FiBY simulations, find high escape
fractions for galaxies with the very highest short-term $\sSFR$s, with
which we are in agreement.

Several works have suggested a positive correlation between $\fesc$
and the SFR surface density $\Sigsfr$ \citep[e.g.][]{Heckman2000,
  Sharma2017, Naidu2019}.  $\Sigsfr$ is a somewhat similar quantity to
$\sSFR$, the former being SFR per area and the latter per mass. We
will analyse $\Sigsfr$ and how it correlates with $\fesc$ in \sphinx{}
in a follow-up paper, as it adds too much to the scope of the present
one.

\subsubsection{UV magnitude}

\begin{figure}
  \centering
  \includegraphics[width=0.49\textwidth]
    {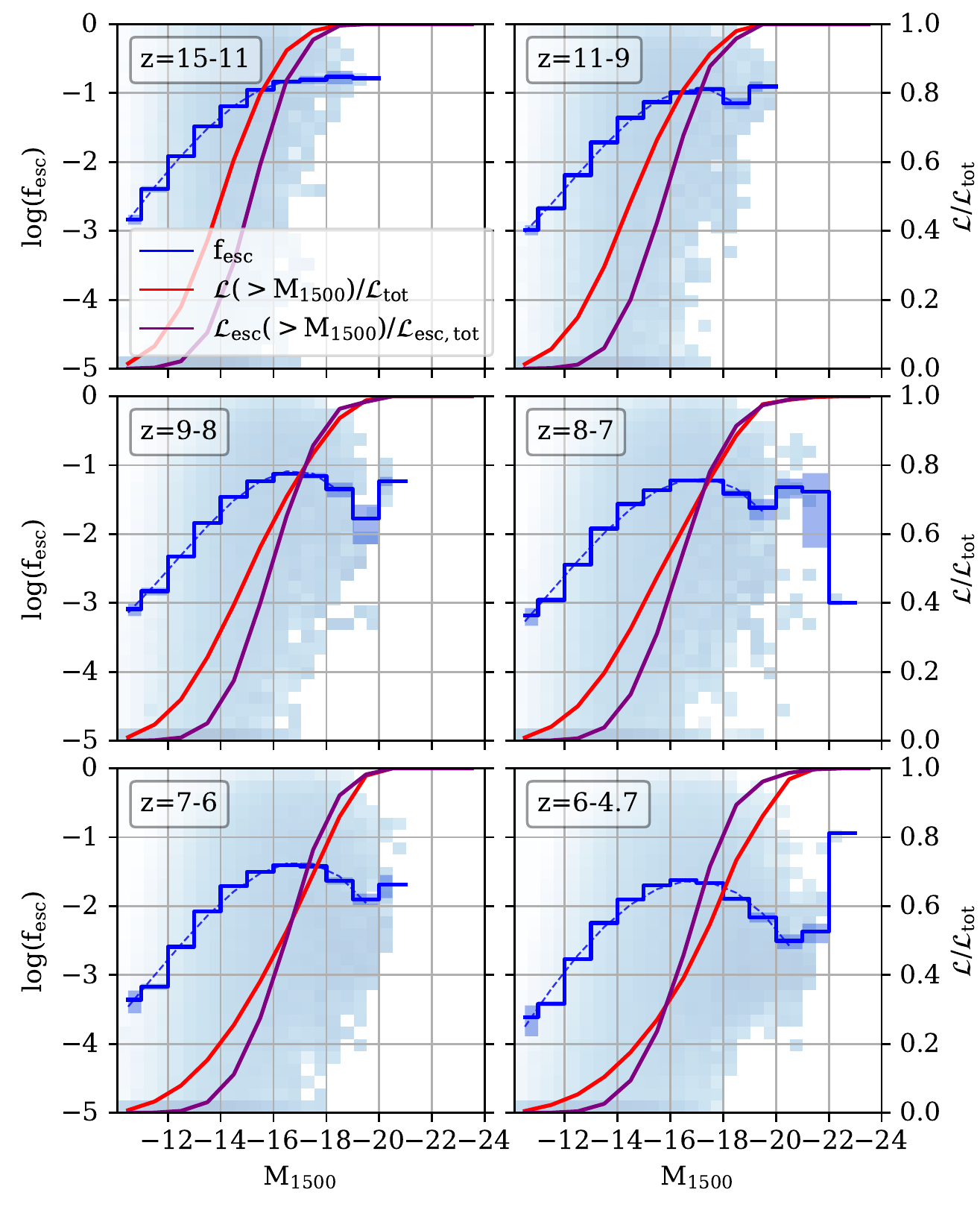}
  \caption
  {\label{fesc_halo_hist_matt_all.fig} Correlation of
    intrinsic-luminosity-weighted mean escape fraction with $1500$
    \AA{} magnitude ($\MFifteen$), in \sphinxtw at different
    redshifts. See the caption of \Fig{fesc_hist_Mvir.fig} for a
    detailed explanation of the panels. The correlation of $\fesc$
    with $\MFifteen$ more or less mirrors that of galaxy stellar mass,
    except for a conspicuous peak in $\fesc$ for the brightest
    galaxies, which is an attenuation effect.}
\end{figure}
We finally consider in \Fig{fesc_halo_hist_matt_all.fig} the
correlation of $\Lyc$ escape fractions of galaxies with their $1500$
\AA{} UV magnitudes, or $\MFifteen$. We show here attenuated UV
magnitudes, retrieved by Monte-Carlo ray-tracing from the stellar
particles and through the dusty ISM, as described in Section
\ref{fesc_rascas.sec}.

Ignoring the peaking $\fesc$ for the very brightest galaxies seen in
most panels of \Fig{fesc_halo_hist_matt_all.fig} (which we discuss in
the next paragraph), we find a similar correlation of $\fesc$ with
$\MFifteen$ as with $\Mstar$, with a peak at intermediate luminosities
where SN feedback is most disruptive and a declining $\fesc$ for both
dimmer (less massive) and brighter (more massive) galaxies. We find an
evolution with redshift of brighter galaxies dominating both the
intrinsic and escaping \Lyc{} emission (red and purple curves,
respectively).

The jump in escape fractions for the very brightest galaxies seen in
most redshift ranges is due to the small statistics at the bright end
as well as the tendency of the most attenuated galaxies at the bright
end to have low escape fractions, often leaving at the very bright end
one or a few galaxies that have small attenuation and high
$\fesc$. This noise-effect would disappear with larger volume sizes,
leaving a simple drop in $\fesc$ with increasing brightness at the
bright end.

\subsection{Contributions to the ionizing radiation
  budget}\label{dominance.sec}

\begin{figure*}
  \centering
  {\includegraphics[width=0.9\textwidth]
    {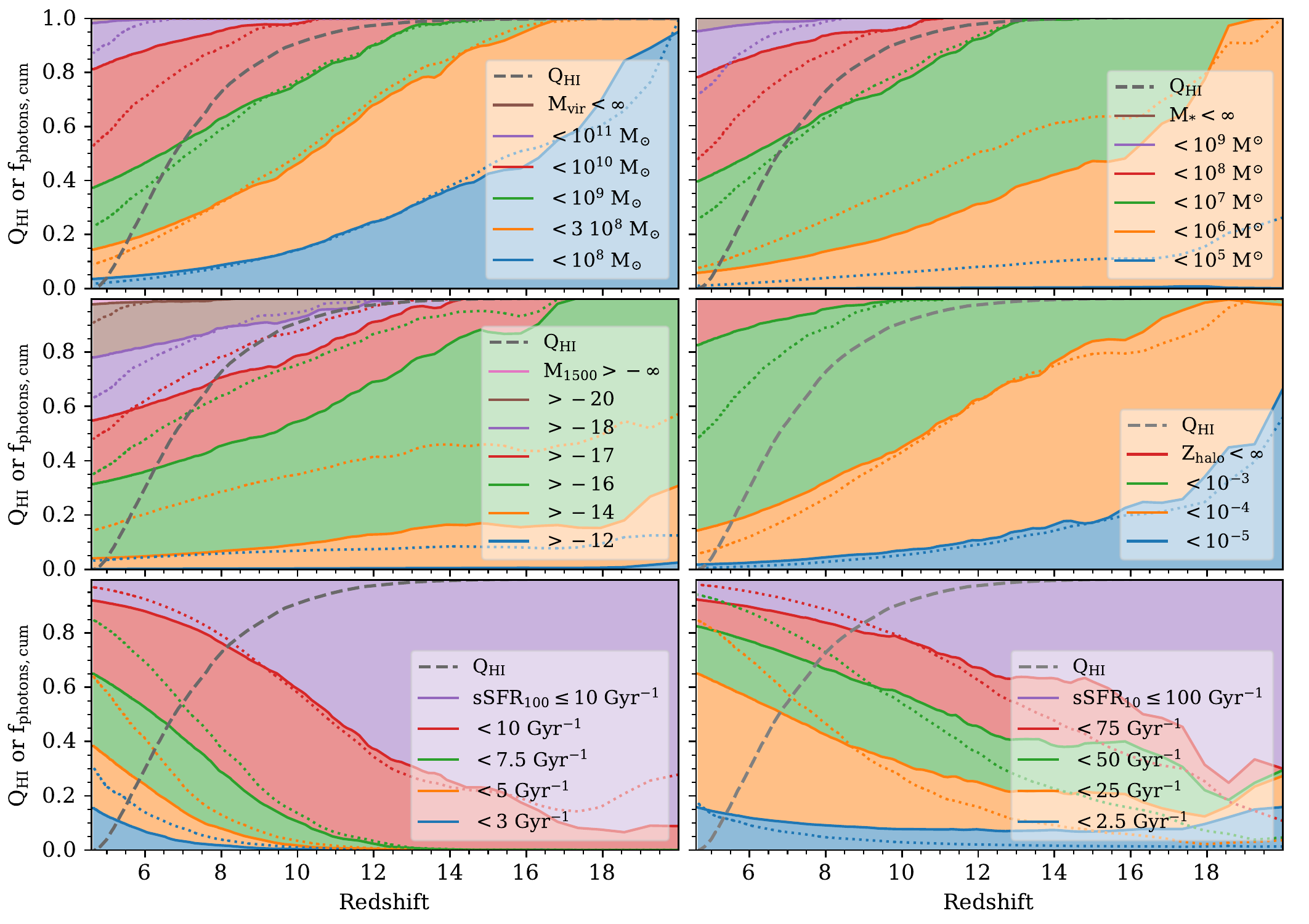}}
  \caption
  {\label{contributions.fig}Each panel shows the cumulative relative
    contributions of halos in \sphinxtw{} to the total number of
    \Lyc{} photons escaping into the IGM (emitted from stars) in solid
    (dotted) curves, and for reference we also show the redshift
    evolution of the volume-weighted neutral fraction in dashed
    curves. Colors indicate halo properties as described in the legend
    of each panel. Shaded regions describe bins in halo properties,
    for escaping luminosities only.  Clockwise from top left, the
    plots show contributions of halos with different $\Mvir$,
    $\Mstar$, \Lyc-luminosity-weighted stellar metallicity,
    $\sSFRten$, $\sSFRc$, and dust-attenuated UV magnitudes. Both the
    intrinsic and escaping production of $\Lyc{}$ photons drift over
    time towards increasingly massive halos containing galaxies with
    increasing mass, metallicity, and brightness, and decreasing
    $\sSFR$.}
\end{figure*}

We now assess in more detail how different galaxies in \sphinxtw{}
contribute to the escaping \Lyc{} budget during
reionization. \Fig{contributions.fig}, taking inspiration from
\cite{Hutter2020}, shows these contributions for different halo
properties in solid curves (and corresponding shaded regions). We
calculate this contribution as the number of escaping \Lyc{} photons
produced since the beginning of time for all halos with a given
property, divided by the total number of escaping photons that have
been produced by all halos. To demonstrate the difference between the
escape and production of \Lyc{} photons, we show in dotted curves the
\textit{intrinsic} fraction of \Lyc{} radiation produced,
i.e. assuming unity escape fraction for all halos. Finally, for
reference, the dashed curves show the reionization history in form of
the volume-weighted neutral fraction, $\Qhi{}$.

The top left panel of \Fig{contributions.fig} shows relative
contributions to the \Lyc{} budget by halo mass. Due to the
hierarchical nature of halo formation and growth of halos over time,
increasingly massive halos dominate the intrinsic production of $\Lyc$
photons with decreasing redshift. Because of the relatively low escape
fractions of the most massive halos, this evolution is not as strong for
the production of \textit{escaping} \Lyc{} photons and the most
massive halos, even if they intrinsically produce a lot of $\Lyc{}$
photons, make a disproportionally small contribution to
reionization. For example, halos with $\Mvir \ge 10^{10} \ \Msun$ are
responsible for about $47$ percent of all \Lyc{} photons produced by
the end of reionization (dotted red curve), but they account for
slightly less than $20$ percent of the \textit{escaping} \Lyc{}
photons. Looking at galaxy masses (top right panel), we find a
similar evolution in the intrinsic \Lyc{} production (dotted curves)
towards increasingly massive galaxies with time. As galaxies in the
mass range $10^{8} \ \Msun > \Mstar > 10^6 \ \Msun$ have the highest
mean escape fractions, this mass range dominates the production of
escaping \Lyc{} radiation by the end of reionization, being
responsible for about $70$ percent of the escaping budget but only
$40$ percent of the intrinsic one. This translates similarly to UV
magnitude, shown in the middle left panel. Intermediate-brightness
halos with $\MFifteen$ in the range $[-14,-18]$ are responsible for
$\approx 75$ percent of the escaping $\Lyc{}$ photons by the end of
reionization, while intrinsically producing only $\approx 50$ of the
budget.

Taking sensitivity limits of current EoR surveys at
$\MFifteen\approx -17$, where discrepancies in the UV luminosity
function between different works become larger than order-of-magnitude
at $z\lesssim 6$ \citep{Bouwens2017}, we can estimate from the middle left
panel in \Fig{contributions.fig} that the constrained part of the
high-z luminosity function contains galaxies responsible for about
$45$ ($52$) percent of the \Lyc{} photons emitted into the IGM
(produced in the ISM) up to the end of reionization. Upcoming James
Webb Space Telescope (JWST) surveys will have an sensitivity limit of
$\MFifteen \approx -16$ for $z \approx 7$ \citep[see][]{Hutter2020},
so they should, according to the same plot, capture galaxies
responsible for about $68$ percent of the \Lyc{} photons contributing
to reionization.

\Lyc{} photons are produced by increasingly metal-rich halos over time
as seen in the middle right panel of \Fig{contributions.fig}. The
escaping $\Lyc{}$ budget is shifted from the intrinsic one towards
metal-poor halos, which have higher escape fractions. Relatively
metal-rich halos with $\Zhalo\ge10^{-3}$ intrinsically produce
slightly more than $50$ percent of all $\Lyc{}$ photons by the end of
reionization, but since their escape fractions are low, they are
responsible for only $\approx 17$ percent of the \textit{escaping}
photons.

Finally, in the bottom panels, we see clearly a drift in the intrinsic
production of \Lyc{} radiation with time towards galaxies with lower
specific star formation rates, i.e. less intense star formation. Since
halos with more intense star formation tend to have higher escape
fractions, this evolution is not as strong in the sense of escaping
\Lyc{} production.

\subsubsection{Comparison with literature}

\cite{Katz2018b} used a \sphinx{}-like simulation to study the
contributions of halos in different mass ranges to reionization, using
the same code and in most respects similar methods as us, but a
smaller volume (10 cMpc wide), lower resolution (25 times more massive
DM particles and 12 times wider ISM cells), and the older BPASS SED
version that we use in \sphinxtenf{}. Their halo contributions to
reionization are fairly similar to both \sphinxtw{} and \sphinxtenf{}
(see \App{bpass.app}) in terms of halo mass. Their $\Mvir < 10^{9}$
and $< 10^{10} \ \Msun$ halos produce $\approx 20$ and $\approx 70$
percent of the \Lyc{} photons reionizing the IGM, respectively, while
our corresponding numbers are $\approx 37$ and $\approx 80$
percent. ``Intermediate'' mass halos of
$\Mvir = 10^{9} - 10^{10} \ \Msun$ are hence significantly more
dominant than in \sphinxtw{} at the cost of lower-mass halos which are
not represented in \cite{Katz2018b} due to lower resolution. Using the
CoDa II large-volume ($100^3$ cMpc$^3$) simulations, \citet{Lewis2020}
found similar halo contributions to \cite{Katz2018b}, with
$\Mvir < 10^{9}$ and $< 10^{10} \ \Msun$ halos producing $\approx 20$
and $\approx 80$ percent of the \Lyc{} photons escaping into the IGM
at $z=7$, when their volume is $50$ percent reionized.  Here the
somewhat stronger contribution of massive halos compared to \sphinx{}
is not due to the DM resolution, which is similar to
\sphinx{}. Instead, it is likely due to the lower physical resolution,
\citet{Lewis2020} having a co-moving uniform cell width of 23 kpc or
3.3 kpc at $z=6$ (300 times wider than in \sphinx{}), leading to
little star formation in low-mass halos. Yet it is quite remarkable
that these three suites of simulations find fairly similar
contributions of halo masses to reionization despite their large
differences in resolution and volume size.

\cite{Hutter2020} studied how four different semi-analytic models
affect reionization, and specifically investigated contributions from
halos with different masses to reionization as in our top left panel
in \Fig{contributions.fig}.  Their model that agrees best with our
halo-mass contribution to reionization is ``early heating'' which
assumes a halo mass-dependent escape fraction, decreasing with halo
mass. Their other models assume fixed escape fractions and have
significantly larger contribution from massive and UV-bright halos
than \sphinx{}. This is not surprising, since the ``early heating''
assumption matches qualitatively the $\fesc$ to halo mass correlation
found in \sphinx{}, as well as \citet{Lewis2020} and several other
simulation works. However, the ``early heating'' model in
\cite{Hutter2020} has reionization being predominantly driven by very
dim galaxies with $\MFifteen \ga -14$. This is far from what we find
in the middle left panel in \Fig{contributions.fig}, so the scenario
is not fully compatible with \sphinx{}. The reason for the discrepancy
is that low-mass halos in \cite{Hutter2020} are assumed to have much
higher escape fractions ($\fesc\approx 60\%$ for
$\Mvir\lesssim 10^9 \ \Msun$) than we find in \sphinx{}.

Similarly, \citet{Finkelstein2019} conclude, using the well-known
differential equation from \citet{Madau1999} describing the
competition of photo-ionization and recombination of the IGM, that
reionization is predominantly driven by significantly dimmer galaxies
than in \sphinx{}, or with $\MFifteen > -15$. Here again the
discrepancy is likely mostly due to differences in escape fractions
for low-mass galaxies. They assume escape fractions, taken from the
cosmological simulations of \citet{Paardekooper2015}, that are
strongly dependent on halo mass, being fairly high for very low-mass
halos and essentially zero for $\Mvir \ga 10^9 \ \Msun$, i.e. their
escape fraction drops with increasing mass as in \sphinx{} but starts
dropping at much lower masses. We can only guess at the reason for
this difference between \citet{Paardekooper2015} and \sphinx{}: it
could be due to their use of the non-binary Starburst99 SED model,
too-late or inefficient SN feedback, or their lack of massive halos.

\cite{Naidu2019}, using a semi-analytic model, argue that reionization
is driven primarily by the most massive “Oligarch” galaxies, with
$\Mvir\ga10^{10} \ \Msun$ halos responsible for about 95 percent of
the \Lyc{} photons reionizing the IGM. This result is in strong
contrast with \sphinx{} and rests on two assumptions: a positive
correlation $\fesc \propto \Sigma_{\rm SFR}^{0.4}$ between the escape
fraction and SFR surface density and a positive correlation between
halo mass and $\Sigma_{\rm SFR}$. Together, these assumptions lead to
a positive correlation between $\fesc$ and halo mass, i.e. the most
massive halos get the highest $\fesc$, which is contrary to \sphinx{}
and almost all other galaxy formation simulations predicting $\fesc$.
Therefore, either of the assumptions must hold false in \sphinx{} in
order for the most massive halos to have the lowest escape
fractions. Assessing this is beyond the scope of the current paper but
we will study $\Sigma_{\rm SFR}$ in \sphinx{} and how it correlates
with both $\fesc$ and halo mass in a follow-up paper.

\subsection{What regulates the global escape
  fraction?}\label{fesc_regulation.sec}

In \Fig{fesc_global.fig} we have seen that the global \Lyc{} escape
fraction from galaxies decreases significantly with decreasing
redshift. Theoretical models of reionization and for the redshift
evolution of the intergalactic UV background
\citep[e.g.][]{Haardt2012, Khaire2018, Puchwein2019,
  FaucherGiguere2020} require qualitatively the same evolution of the
global escape fraction in order to produce a sufficient amount of
ionizing radiation in the IGM during the EoR, to not produce an overly
strong low-z UV background, and match the general observational
constraint of low escape fractions at low redshift
\citep[e.g.][]{Matthee2016, Grazian2017, Rutkowski2017}. Those
theoretical models vary significantly in the exact form of the assumed
evolution of the global $\fesc$ with redshift, but the general trend
is always there and is qualitatively similar to the evolution
of $\fesc$ that naturally comes out of the \sphinx{} simulations, as
we show by including in \Fig{fesc_global.fig} the escape fraction
evolution in \cite{Haardt2012} and
\cite{FaucherGiguere2020}. Therefore, \sphinx{} not only confirms such
an evolution but also provides an opportunity to investigate what
drives it.

We have already seen in Figures
\ref{fesc_hist_Mvir.fig}-\ref{fesc_halo_hist_sSFR100.fig} that
\Lyc{}-weighted mean escape fractions for individual galaxies in
\sphinx{} decrease with increasing halo mass, increasing galaxy mass,
increasing metallicity, and decreasing $\sSFR$ (except for the
turn-over at very low $\sSFRten$). The same figures also show, with
red curves, that with decreasing redshift the intrinsic production of
\Lyc{} radiation is increasingly dominated by these massive,
metal-rich, and mildly star-forming galaxies which have low $\fesc$,
i.e. the red curve slides to the right with decreasing redshift for
$\Mvir$, $\Mstar$, and $\Zhalo$, and to the left for
$\sSFRten$. Therefore it seems plausible that the decrease of the
global $\fesc$ with redshift is dictated by this shift of intrinsic
\Lyc{} production to galaxies with low $\fesc$.

Yet, the very same figures also suggest that none of those factors
alone (i.e. mass, metallicity, or $\sSFR$) regulates the decreasing
$\fesc$. Taking for example \Fig{fesc_hist_Mvir.fig} showing the
correlation of $\fesc$ with halo mass, a comparison of the different
panels reveals that $\fesc$ decreases with redshift for any fixed
$\Mvir$. \textit{If the mass evolution predominantly drove the
  decrease in $\fesc$ with redshift, $\fesc$ would remain fixed with
  redshift for any fixed $\Mvir$, and this is clearly not the case.}
Hence there must be some other factor than just halo mass evolution
significantly contributing to the decrease of the global $\fesc$ with
time. The same is true for all the halo properties explored in Figures
\ref{fesc_hist_Mvir.fig}-\ref{fesc_halo_hist_matt_all.fig},
i.e. $\fesc$ decreases with decreasing redshift for any fixed galaxy
mass, metallicity, $\sSFR$, and magnitude.

It is possible that two or several factors regulate the escape
fraction together, e.g. an increase in mean metallicity causing a drop
in $\fesc$ for a fixed stellar mass and similarly an increase in mean
stellar mass causing a drop in $\fesc$ for a fixed
metallicity. However, we find from two-dimensional histograms,
i.e. plotting escape fractions against two halo properties together
(not shown), that this is not the case for any two properties from
Figures \ref{fesc_hist_Mvir.fig}-\ref{fesc_halo_hist_matt_all.fig}
combined.

However, even if we do not find a drop in $\fesc$ with redshift for
fixed $\sSFR$, this does not close the case on the possibility that a
decreasing intensity of star formation in the Universe with redshift,
and hence a decrease in the intensity of stellar feedback, drives a
decrease in \Lyc{} escape fractions from galaxies. As touched upon in
\Sec{ssfr.sec}, the flaw in $\sSFRten$ is that it does not capture the
long-term effect of starbursts and the ensuing feedback that can last
tens of Myrs. When a galaxy is severely disrupted by stellar feedback
and its $\fesc$ becomes high, it can remain high for a long time while
$\sSFRten$ becomes very low, since the gas fuelling star formation has
been largely ejected. Therefore, $\sSFRten$ does not capture galaxies
that have high $\fesc$ and remaining non-negligible \Lyc{}
luminosities, due to recent but quenched starbursts. These galaxies
quickly become \Lyc{}-dim so most of them do not weigh very much in
(the $\Lyc$-luminosity-weighted) $\fesc$, but some of them indeed do
for a short time. $\sSFRc$ is a more long-term measure of the specific
star formation rate, but this is also not a good measure as it does
not discriminate efficiently between starbursts (with high $\fesc$)
and steady but fairly high rates of star formation (with low $\fesc$).

As a more useful measure of current \textit{and recent} star formation
activity, we therefore take for each halo its maximum $\sSFRten$ over
the last $50$ Myrs. We calculate this in the following way for each
halo: in a given simulation snapshot, we take all the stars in the
halo and compute
\begin{align}
  \sSFRmax= max\left(
  \frac{\frac{\Mstar(X \le age < X+10 \ Myr)}{10 \ Myr}}
  {\Mstar(age\ge X)} \right), \\
  {\rm for \ X \ in \ (0, 1, 2, ..., 50) \ Myr.} \nonumber
\end{align}
The $50$ Myr lookback time is arbitrary and the results that follow
are insensitive to it as long as it is significantly larger than $10$
Myrs and $\lesssim 100$ Myrs.

\begin{figure}
  \centering
  \includegraphics[width=0.49\textwidth]
    {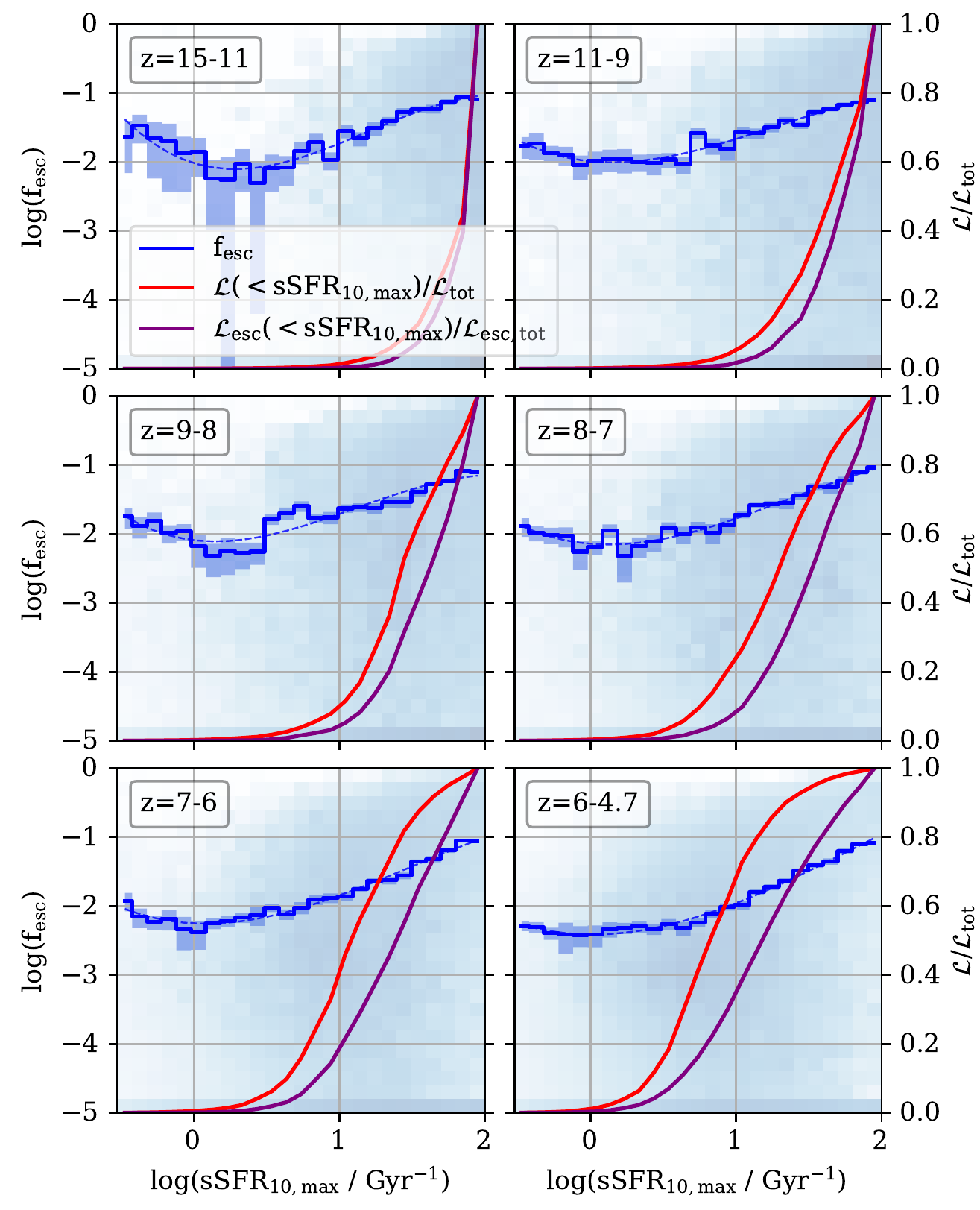}
  \caption
  {\label{fesc_halo_hist_sSFR10max.fig} Correlation of intrinsic-\Lyc
    -luminosity weighted $\fesc$ with $\sSFRmax$, the maximum
    $\sSFRten$ in the last 50 Myrs, in \sphinxtw{} at different
    redshift ranges.  See the caption of \Fig{fesc_hist_Mvir.fig} for
    a detailed explanation of the panels. Note that we have floored
    galaxies with $\sSFRmax$ below the range to the leftmost
    $\sSFRmax$ bin, so they are counted in the plots. The correlation
    of $\fesc$ is more monotonous here than with the \textit{current}
    $\sSFRten$, shown in \Fig{fesc_halo_hist_sSFR10.fig} and the
    correlation also varies much less with redshift. }
\end{figure}

We show in \Fig{fesc_halo_hist_sSFR10max.fig} how $\fesc$ correlates
with $\sSFRmax$ in the usual six redshift ranges. We find a quite
different correlation here than with $\sSFRten$, shown in
\Fig{fesc_halo_hist_sSFR10.fig}. For the high-$\sSFR$ end, the two
correlations are fairly similar, but they are very different for
$\sSFRten\lesssim 10 \ \Gyrm$, with a fairly flat and low $\fesc\approx0.01$
in the case of $\sSFRmax$ whereas there is a rise in $\fesc$ for the
lowest-$\sSFRten$ halos. The reason for this is that $\sSFRmax$ keeps
memory of recent starbursts with high $\fesc$, i.e. a galaxy that
recently experienced a starburst and is quenched but still exhibits a
high $\fesc$ will typically have a high $\sSFRmax$ but low
$\sSFRten$. Another striking difference between the correlations of
$\fesc$ with $\sSFRten$ on one hand and $\sSFRmax$ on the other is
that in the latter case there is much less evolution of the
correlation with redshift. For any fixed $\sSFRmax$ and especially for
$\sSFRmax\ga10 \ \Gyrm$, the escape fraction does not drop
significantly with redshift. Furthermore we do see in
\Fig{fesc_halo_hist_sSFR10max.fig}, with the sliding red curve towards
the left from high to low redshift, that \Lyc{} photons are produced
in galaxies with decreasing $\sSFRmax$ with decreasing redshift.

\begin{figure}
  \centering
  \includegraphics[width=0.5\textwidth]
    {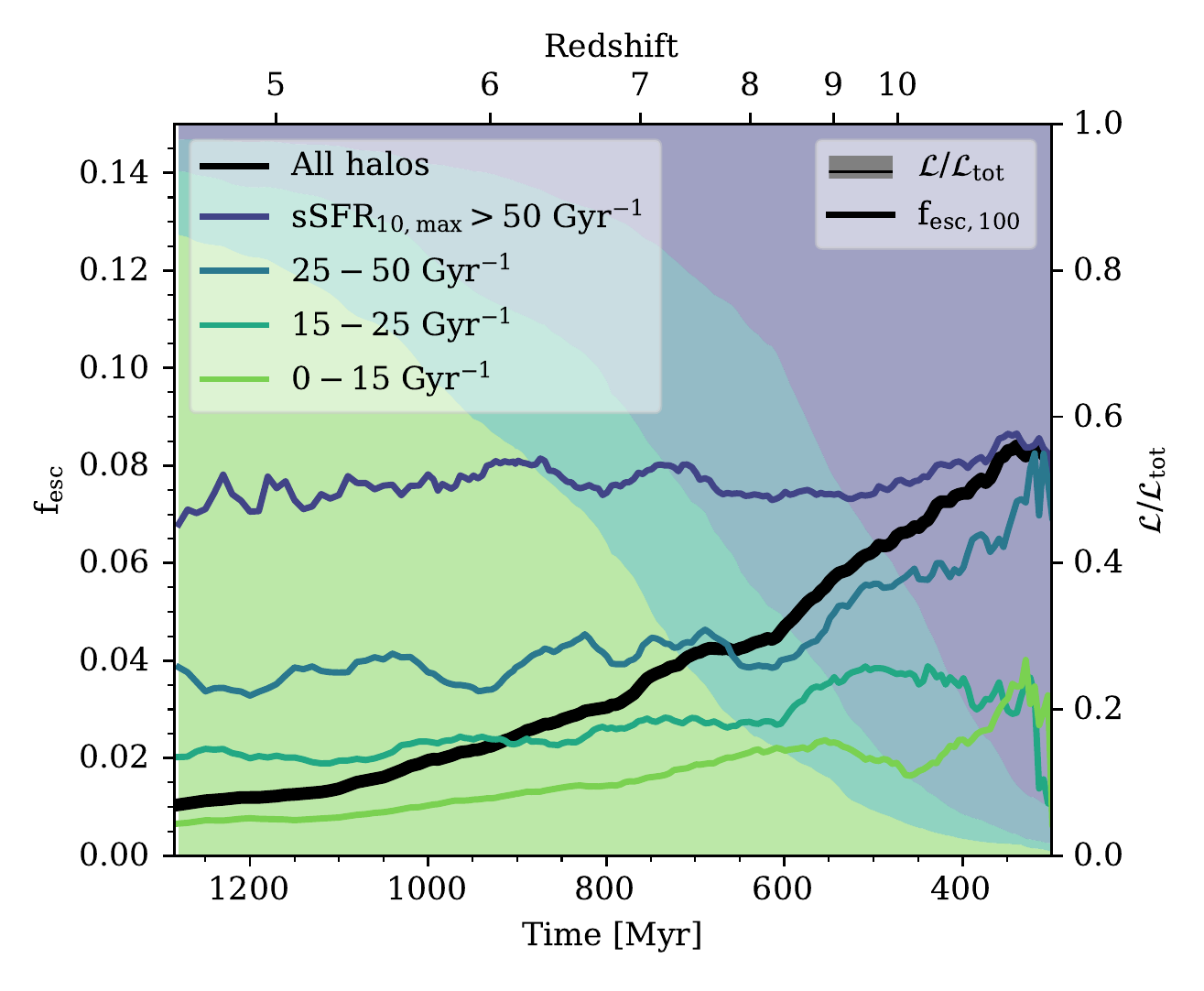}
  \caption
  {\label{fesc_fixedsSFRmax.fig}Global escape fraction averaged over
    100 Myrs ($\feschundred$) versus redshift, in solid curves, for
    all halos in \sphinxtw{} (thick black curve) and at fixed bins in
    $\sSFRmax$ (thinner coloured curves). The shaded regions indicate
    the fractions of the total \Lyc{} radiation intrinsically produced
    by halos in these fixed $\sSFRmax$ bins, and show that the \Lyc{}
    emission is dominated by decreasing $\sSFRmax$ with decreasing
    redshift. For any fixed $\sSFRmax$, halos have a fairly constant
    escape fraction. We conclude that the global decrease in $\fesc$
    with redshift is due to a decreasing efficiency of star formation
    in the Universe.}
\end{figure}

It therefore seems plausible that an evolving $\sSFRmax$ with redshift
is a driver of the decreasing global $\fesc$. To confirm that this is
the case we examine in \Fig{fesc_fixedsSFRmax.fig} the
redshift-evolution of the luminosity-weighted mean $\fesc$ for halos
in bins of fixed $\sSFRmax$. The idea is that if the global $\fesc$ is
regulated by an evolving $\sSFRmax$, \textit{it should not vary with
  redshift when $\sSFRmax$ is fixed}, while the relative contribution
to the total \Lyc{} emission should change and halos with low
$\sSFRmax$, corresponding to low $\fesc$, should increasingly dominate
the intrinsic production of \Lyc{} photons. For each bin in
$\sSFRmax$, as indicated in the legend, a solid curve shows the
\Lyc{}-luminosity-weighted mean escape fraction in \sphinxtw{}
averaged over 100 Myrs, $\feschundred$\footnote{We prefer to average
  the escape fraction over time here rather than show the
  instantaneous $\fesc$, to reduce the noise, which increases when we
  narrow the set of halos included in the analysis by fixing halo
  properties.}, as a function of redshift, for all halos in a given
$\sSFRmax$ range, and a corresponding shaded region shows the fraction
of intrinsically emitted \Lyc{} photons for the same halos.

We indeed find that a fixed burstiness, measured with $\sSFRmax$,
yields a fairly non-evolving mean $\fesc$. The figure therefore shows
that the evolution in the global $\fesc$ of \Lyc{} radiation from all
galaxies (black thick curve) is mostly driven by the \Lyc{} radiation
being emitted from decreasingly bursty galaxies (i.e. going from dark
blue to light green). There is significant noise in the curves,
especially at $z\ga8$ for the bins with lower $\sSFRmax$ and at lower
redshift for the bins with higher $\sSFRmax$, due to small numbers of
halos populating those $\sSFRmax$-redshift combinations. For the lower
redshifts shown, there is some evolution of the escape fraction for
the lowest $\sSFRmax$-bin, decreasing by more than $50\%$ from $z=9$
to $z=5$. This may be partly driven by the \Lyc{} radiation being
produced by galaxies with increasing masses or metallicities, but it
is probably predominantly an artificial effect of decreasing physical
resolution with redshift, as we will discuss in \Sec{resolution.sec}.

\begin{figure}
  \centering
  \includegraphics[width=0.45\textwidth]
    {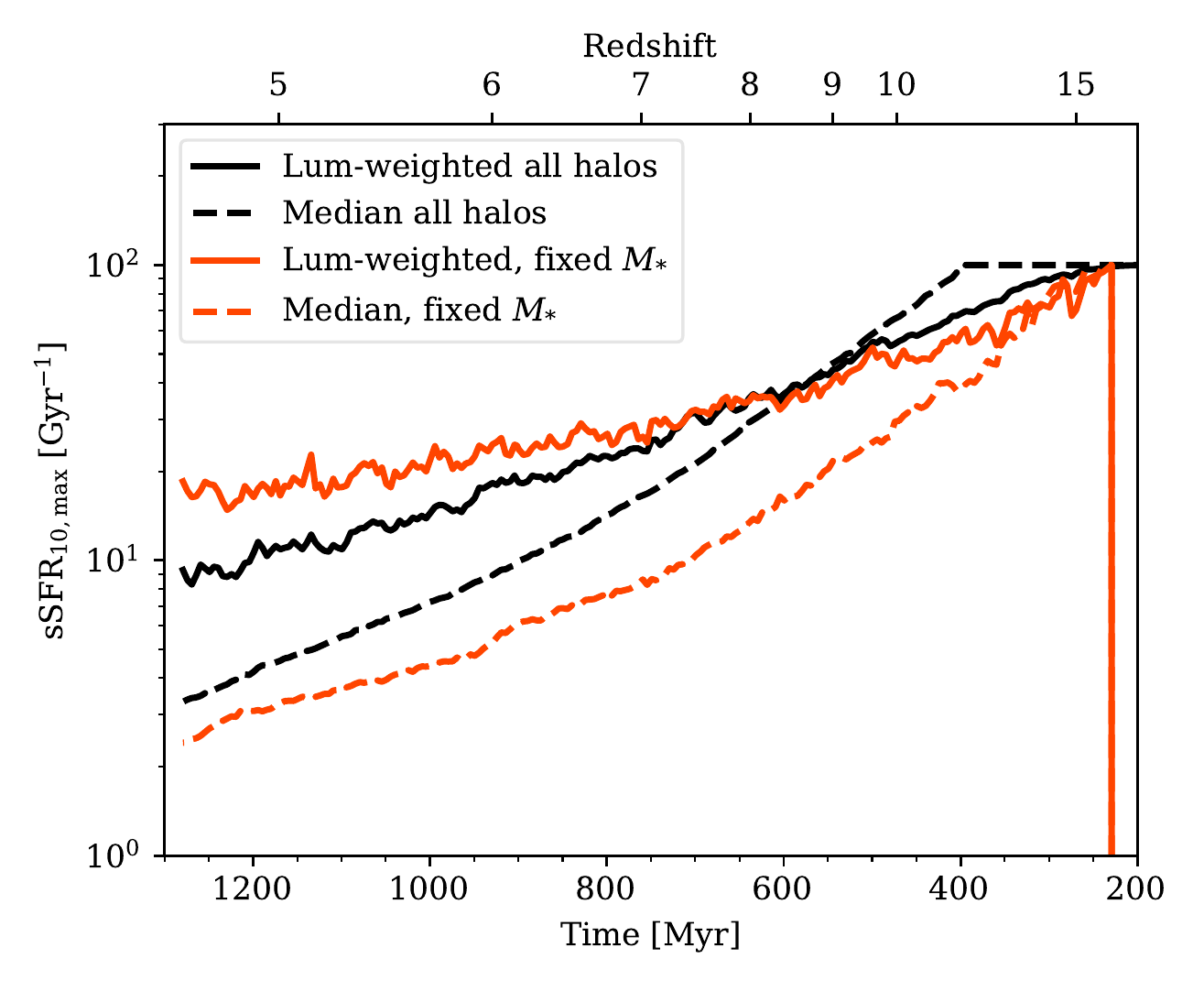}
  \caption
  {\label{sSFR1max0-z.fig} Redshift evolution of $\sSFRmax$ in
    \sphinxtw. The dashed black curve shows the median $\sSFRmax$ for
    all halos containing stars, while the solid black curve shows the
    intrinsic-\Lyc{}-luminosity-weighted mean for the same halos,
    showing better the evolution in $\sSFRmax$ for \Lyc{}-emitting
    galaxies. The dashed and solid red curves show the same, but for
    the subset of halos that at each given redshift are in the stellar
    mass range $3\times 10^6 \ \Msun < \Mstar < 3\times 10^7 \
    \Msun$. Even for a fixed galaxy mass, the specific star formation
    rate decreases significantly with redshift.}
\end{figure}

We conclude from this analysis that the redshift-evolution of the
global $\fesc$ is driven mostly by an increasing fraction of galaxies
with low $\sSFRmax$ over time. \Fig{sSFR1max0-z.fig} shows clearly
that $\sSFRmax$ decreases globally with decreasing redshift, whether
computed as the median of all halos or the intrinsic \Lyc{}-luminosity
weighted mean. To demonstrate that this evolution does not simply
reflect the general evolution towards increasingly massive galaxies,
for which extreme specific star formation rates cannot be maintained,
we also show the evolution of $\sSFRmax$ for a fixed range of stellar
mass, roughly corresponding to the peak in $\fesc$ seen in
\Fig{fesc_hist_Mstar_all.fig}. For these fixed galaxy masses,
$\sSFRmax$ also drops with redshift. The decrease in $\sSFRmax$ is
thus not simply due to galaxies becoming more massive, but is rather
due to less dense environments, less frequent mergers, and subsiding
gas accretion \citep[e.g.][]{Dekel2009, Fakhouri2010, Tillson2011}.
This evolution, with or without fixed galaxy mass, is also true for
$\sSFRten$ and $\sSFRc$ (not shown). Such an evolution of $\sSFR$ is
plausible and found, at least qualitatively, by many observational and
theoretical works \citep[][and references therein]{Lehnert2015,
  Fernandez2018, Stefanon2021}.

\section{Discussion} \label{Discussion.sec}

\subsection{On the sensitivity of escape fractions to the SED model}
\label{sed_model.sec}

\begin{figure}
  \centering
  \includegraphics[width=0.45\textwidth]
    {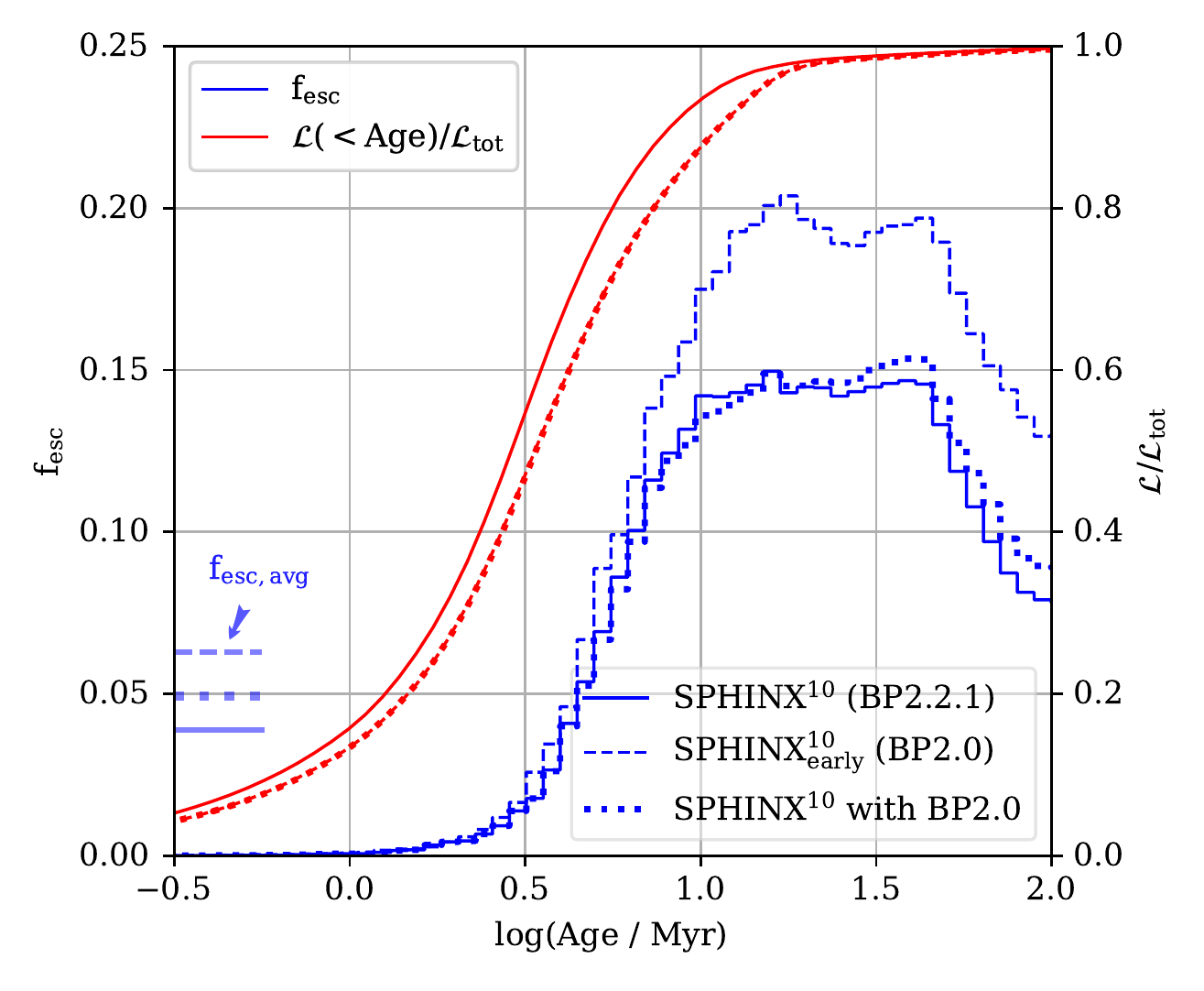}
  \caption
  {\label{fesc_hist_age_age.fig}Escape fractions versus stellar
    particle age for all 34 snapshots between redshifts $7$ and $6$ in
    the 10 cMpc volumes with BPASS 2.2.1 (solid) and with BPASS 2.0
    (dashed). Blue curves show the \Lyc{}-luminosity-weighted mean
    $\fesc$ as a function of stellar age. Red curves show the fraction
    of total intrinsic \Lyc{} emissivity for all stellar particles
    younger than the given age. Dotted curves show results for the
    \sphinxten{} simulation post-processed with the BPASS 2.0 model
    from \sphinxtenf. Horizontal lines on the left side of the plot
    show the global $\fesc$ for the three cases. The higher overall
    $\fesc$ in \sphinxtenf{} compared to \sphinxten{} is roughly
    equally due to the shift of the cumulative luminosity to older
    ages (i.e. effect of timing, isolated in the dotted curves) and
    the overall higher stellar luminosities (i.e. photo-ionization
    feedback).}
\end{figure}

We now briefly address why escape fractions (\Fig{fesc_global.fig}),
and hence reionization (\Fig{xhi.fig}), are so sensitive to rather
subtle variations in the SED model used. \Fig{fesc_hist_age_age.fig}
shows escape fractions and cumulative \Lyc{} luminosities binned by
stellar particle age in our two 10 cMpc \sphinx{} volumes, stacked
over 34 snapshots in the redshift range $z=7-6$. The fiducial BPASS
version 2.2.1 (\sphinxten) is shown in solid curves and the older
version 2.0 (\sphinxtenf) in dashed curves.  The blue curves show the
mean escape fraction as a function of age and demonstrate clearly how
sensitive the escape fraction is to stellar age. With both BPASS
versions, the escape fraction is very low for new-born stellar
populations. Between $\approx 3$ and $10$ Myrs the escape fraction
rises very steeply with age and reaches a peak. As several authors
before us \citep[e.g.][]{Kimm2014, Ma2016, Trebitsch2017}, we
interpret this sensitivity of $\fesc$ to stellar age to be due to
regulation by SN feedback: at age 3 Myrs, stellar particles start
undergoing SN explosions, dispersing surrounding gas and clearing way
for the radiation to escape from the exploding particles as well as
their neighbouring particles. The escape fraction drops again beyond
an age of $50$ Myrs, which is the time at which stellar particles
cease undergoing SN explosions. The escape fraction then stabilises at
$\sim 10$ percent, with the old stellar particles scattered over the
ISM and not especially correlated with the cold and neutral gas which
most efficiently absorbs their \Lyc{} radiation.

The overall \Lyc{}-luminosity-weighted escape fraction for all stellar
ages, indicated by horizontal lines in the left side of the plot, is
about $60$ percent higher in \sphinxtenf{} than in \sphinxten. We
attribute this difference to a combination of timing and the overall
emission of \Lyc{} photons. The former is seen by comparing the solid
and dashed red curves, which show the fraction of total intrinsic
\Lyc{} emissivity below a given age. This is shifted towards higher
ages with BPASS 2.0, as this model has significantly higher \Lyc{}
luminosities at old ages compared to BPASS 2.2.1. Stellar populations
older than 10 Myrs, with high escape fractions, emit about 10
percent of the intrinsic \Lyc{} radiation with the older BPASS model,
but only about 5 percent of it with the newer model. Since the escape
fraction is so sensitive to stellar age -- which we remind is due to
SN feedback -- the small shift in the intrinsic emissivity with age
leads to a significant difference in the overall escape fraction. The
second effect, which is due to overall \Lyc{} luminosities rather than
timing, is clear from the difference in the escape fraction versus age
for the two SED models, i.e. the solid and dashed blue curves. For a
given stellar population age, $\fesc$ is consistently higher with
BPASS 2.0 than with BPASS 2.2.1, by about $30$ percent for age $>10$
Myrs. The two simulations shown here are identical in setup except for
the \Lyc{} luminosities of the stellar particles as a function of
their age and metallicity, and we stress that the timing and energy
release by SN explosions is on average the same in both
simulations. We are therefore forced to conclude that
\textit{radiation feedback} provides an additional regulation of
escape fractions, by ionising gas and slowly dispersing it away from
stars. This happens more efficiently with BPASS 2.0, which produces
about $30$ percent more \Lyc{} photons per unit stellar mass for
metal-poor stellar populations than version 2.2.1 (see
\Fig{SEDs.fig}).

The overall higher escape fractions with BPASS 2.0 are due both to
timing -- i.e. a larger fraction of \Lyc{} radiation remaining to be
emitted at the onset of SNe as well -- and the larger number of \Lyc{}
photons emitted overall. The former effect has already been pointed
out by \cite{Ma2016, Rosdahl2018}, but the latter has largely been
ignored. To distinguish their importance, we separate the two effects
in the dotted curves in \Fig{fesc_hist_age_age.fig}. Here we have
post-processed the \sphinxten{} simulation outputs with the older
BPASS 2.0 model, i.e. we compute escape fractions in the BPASS 2.2.1
simulation assuming BPASS 2.0. This isolates the timing effect from
that of radiation feedback, since the timing is changed from the newer
BPASS version but not the feedback. As expected, the escape fraction
versus age (blue) is close to that of the \sphinxten{} simulation with
BPASS 2.2.1 while the cumulative intrinsic \Lyc{} luminosity (red) is
almost identical to the \sphinxtenf{} simulation with BPASS 2.0. The
global escape fraction from this experiment is shown in the horizontal
dotted curve on the left side. It sits right between the two previous
escape fractions, though slightly closer to that for \sphinxten{}
(solid). From this we conclude that timing and radiation feedback each
contribute similarly to boosting the \Lyc{} escape fraction with the
increased and delayed \Lyc{} luminosities in BPASS 2.0, with perhaps a
slightly stronger effect from radiation feedback. This highlights the
need for self-consistent RHD simulations to correctly predict escape
fractions. Simply post-processing a non-RHD simulation, or
post-processing an RHD simulation with an SED model different from the
on-the-fly one, will yield inconsistent and wrong \Lyc{} escape
fractions.

\subsection{Convergence of results with volume size and SED
  model} \label{volume_sed.sec}

Due to its large volume, we have focused our analysis on the
\sphinxtw{} simulation, despite the late reionization history it
produces. We have verified that the general trends and conclusions on
the dependence of escape fractions on galaxy properties, contributions
of galaxies to reionization, and evolution of escape fractions with
redshift hold for the smaller volume and brighter SED model. For
reference we present this analysis in Appendix \ref{bpass.app}. To
summarise, the higher global escape fraction in the smaller volume
\sphinxten{} simulation is mostly due to the cut-off in the halo mass
function being at lower mass than in \sphinxtw{}. Since \sphinxtw{}
has more of those massive halos, which we have shown to have low
$\fesc$, its global escape fraction is lower than in the smaller
volumes. However, we do also see a hint of escape fractions being
generally lower for galaxies of all masses, metallicities, and
magnitudes in \sphinxten{} than in \sphinxtw{}. This can be due to
cosmic variance or even simply noise, but it is possible that
environmental effects play a minor role, i.e. less massive
environments may have slightly higher escape fractions. This effect is
out of the scope of the current paper and will be explored in future
work. Since the \sphinxten{} simulation is shifted to lower halo
masses than \sphinxtw{}, the halo contributions to the \Lyc{}
luminosity budget during reionization is very similar in the two
simulations, except it is more noisy in \sphinxten{} and slightly
shifted to less massive, dimmer, and more metal-poor galaxies, whereas
the contributions in terms of $\sSFR$ is almost identical between the
two.

The only difference between \sphinxten{} and \sphinxtenf{} is the
assumed BPASS version, so we can safely conclude that the
significantly higher global escape fraction in the latter is purely
due to the different SED model. The comparison in Appendix
\ref{bpass.app} reveals much the same correlations of $\fesc$ with
halo properties in the two simulations, the $\fesc$ curve simply being
monotonously shifted up for \sphinxtenf{} compared to that of
\sphinxten{}, though perhaps slightly increasingly with halo
mass. Between \sphinxten{} and \sphinxtenf{}, the halo contributions
to the intrinsically emitted \Lyc{} radiation are very similar, while
the escaping contributions are slightly shifted towards more massive
and metal-rich halos.

Our trends of $\fesc$ with halo properties and contributions of halos
to reionization are therefore insensitive to variations in volume size
and assumed SED model. We furthermore demonstrate in Appendix
\ref{bpass.app} that although the data are very noisy in the
\sphinxtenf{} simulation due to the smaller volume, the global
evolution in $\fesc$ is driven there by a decreasing $\sSFRmax$, as in
\sphinxtw{}. Hence all our conclusions hold for reionization histories
that bracket main-stream theoretical and observational constraints of
reionization, from late in \sphinxtw{} ($z=4.64$) to early in
\sphinxtenf{} ($z=6.5$).

\subsection{Resolution convergence} \label{resolution.sec} We finally
investigate the convergence of our main results with resolution. We
focus on the simulation cell width, $\Delta x$, which directly affects
the porosity of the ISM through which \Lyc{} radiation propagates and
therefore can be expected to affect \Lyc{} escape fractions. We recall
that our minimum cell width is $\Dxmin = 10.9$ pc at $z=6$ and that
$\Dxmin$ is co-moving, i.e. the cell width scales directly with the
cosmological expansion factor $a$ and therefore the physical
resolution degrades with decreasing redshift, doubling, for example,
over the redshift interval $z=11-5$ (or $a=0.083 - 0.167$).

\begin{figure}
  \centering
  \includegraphics[width=0.45\textwidth]
    {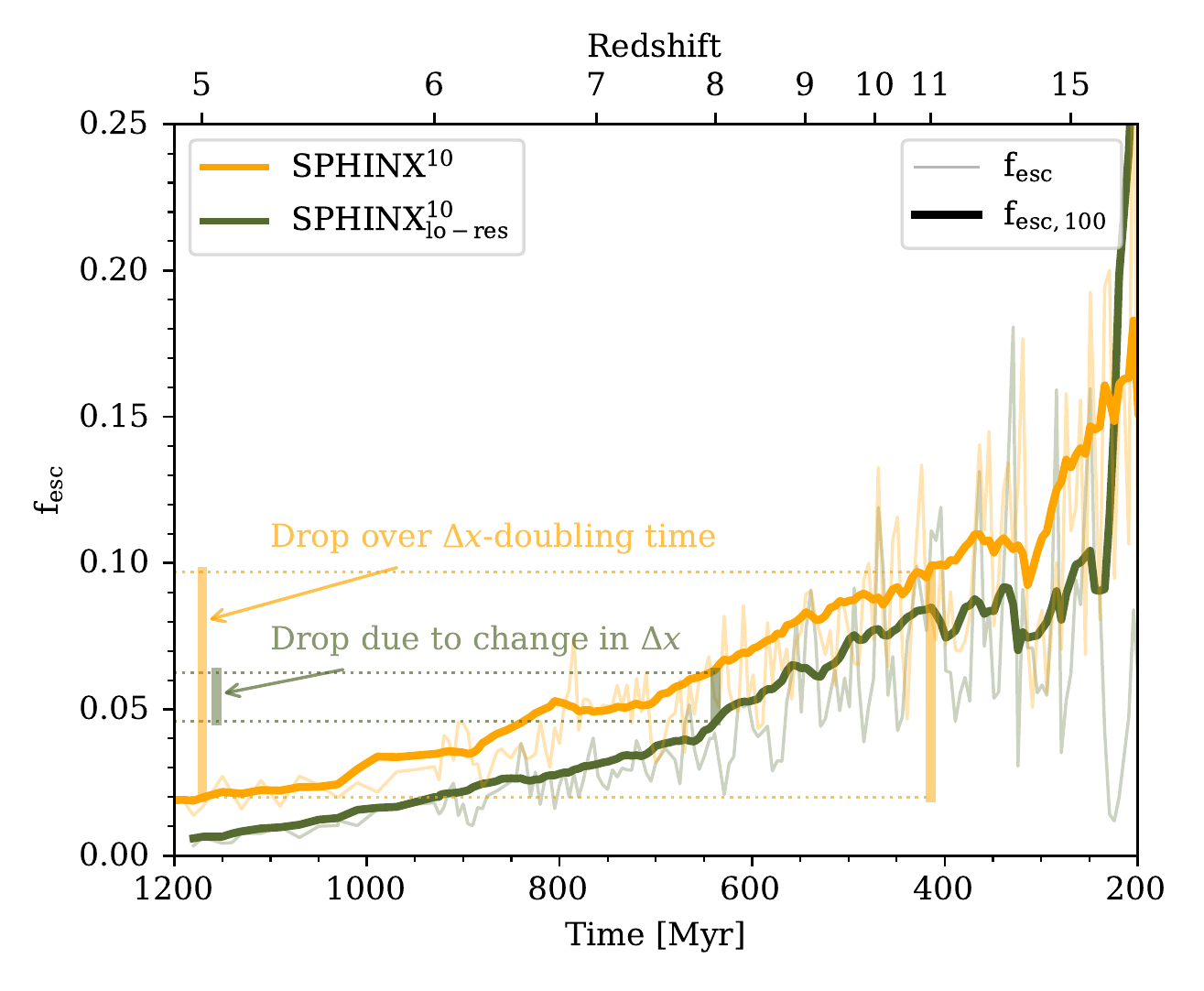}
  \caption
  {\label{fesc_global_10cmpc_resolution.fig} Global escape fraction
    with fiducial and degraded physical resolution (twice the minimum
    cell width), in yellow and green curves, respectively. Thin
    transparent curves show the instantaneous $\fesc$ while thick
    curves show sliding luminosity-weighted averages over the previous
    100 Myrs. Escape fractions are not converged with resolution, with
    lower resolution giving lower $\fesc$. The decrease in the global
    $\fesc$ with redshift is therefore partly due to decreasing
    resolution, since the cell width $\Dxmin$ decreases linearly with
    the expansion factor in our constant co-moving
    resolution. However, the drop due to the evolving $\Dxmin$ is
    sub-dominant, as indicated by the relative sizes of the yellow and
    green vertical bars, which indicate the total drop in $\fesc$ over
    the time it takes for the physical $\Dxmin$ to double and the drop
    in $\fesc$ due to directly doubling $\Dxmin$, respectively.}
\end{figure}

To study the convergence of escape fractions with resolution, we use a
$10$ cMpc wide simulation, identical to \sphinxten{} except with one
level lower maximum refinement, i.e. the physical width of the highest
resolution cells it at all times double that of \sphinxten{}. We show
in \Fig{fesc_global_10cmpc_resolution.fig} the evolution of the
intrinsic \Lyc{} emissivity and global escape fraction in \sphinxten{}
and its lower resolution counterpart. The intrinsic \Lyc{} emissivity
is very similar between the two runs before $z=6$ but is somewhat
enhanced in the lower-resolution run after that. We note (though not
shown) that the UV luminosity function is very well converged in the
two runs and matches well the observational constraints at the
redshifts considered in \Fig{flum.fig},
i.e. $z=5,6,7,8,9,10$. However, the global escape fraction, best
compared via the 100-Myr average $\feschundred$, is systematically
lower with lower resolution by a fixed $1-2$ percentage points (actual
difference, i.e. not relative). Due to the lower escape fraction, and
the star formation rate being largely unaffected by resolution, as
indicated by the thin solid lines in
\Fig{fesc_global_10cmpc_resolution.fig} showing the intrinsic rate of
\Lyc{} production per volume, the low-resolution simulation reionizes
slower than \sphinxten{} and is not fully reionized by the end of the
run at $z \approx 5$. This is somewhat expected and a similar trend of
decreasing $\fesc$ with decreasing resolution has already been
reported in the simulations of \cite{Ma2020}. We compare in detail the
correlations of $\fesc$ with halo properties for the fiducial and
lower resolution in \App{resolution.app}. To summarise, we find that
all correlations we have studied between $\fesc$ and halo properties
hold, with the lower-resolution $\fesc$ normalisation more or less
flatly scaled down, except that that $\fesc$ tends to be somewhat more
suppressed with lower resolution in galaxies with low and intermediate
sSFRs than in the galaxies with the highest sSFRs.

This non-convergence of $\fesc$ with resolution, combined with the
good convergence we find in terms of star formation, indicates that
higher-resolution counterpart simulations would likely reionize too
early due to their high escape fractions, even too early compared to
observational constraints. However, changing the resolution seems to
have similar effects on escape fractions as varying the SED model for
stellar \Lyc{} luminosities within quite reasonable limits
(e.g. reducing slightly the impact of binary stars), and therefore
these two factors are somewhat degenerate. And we do find that
changing either leaves our conclusions fairly intact on the
correlation of $\fesc$ with halo properties and how halos with
different properties contribute to reionization.

\begin{figure}
  \centering
  \includegraphics[width=0.45\textwidth]
    {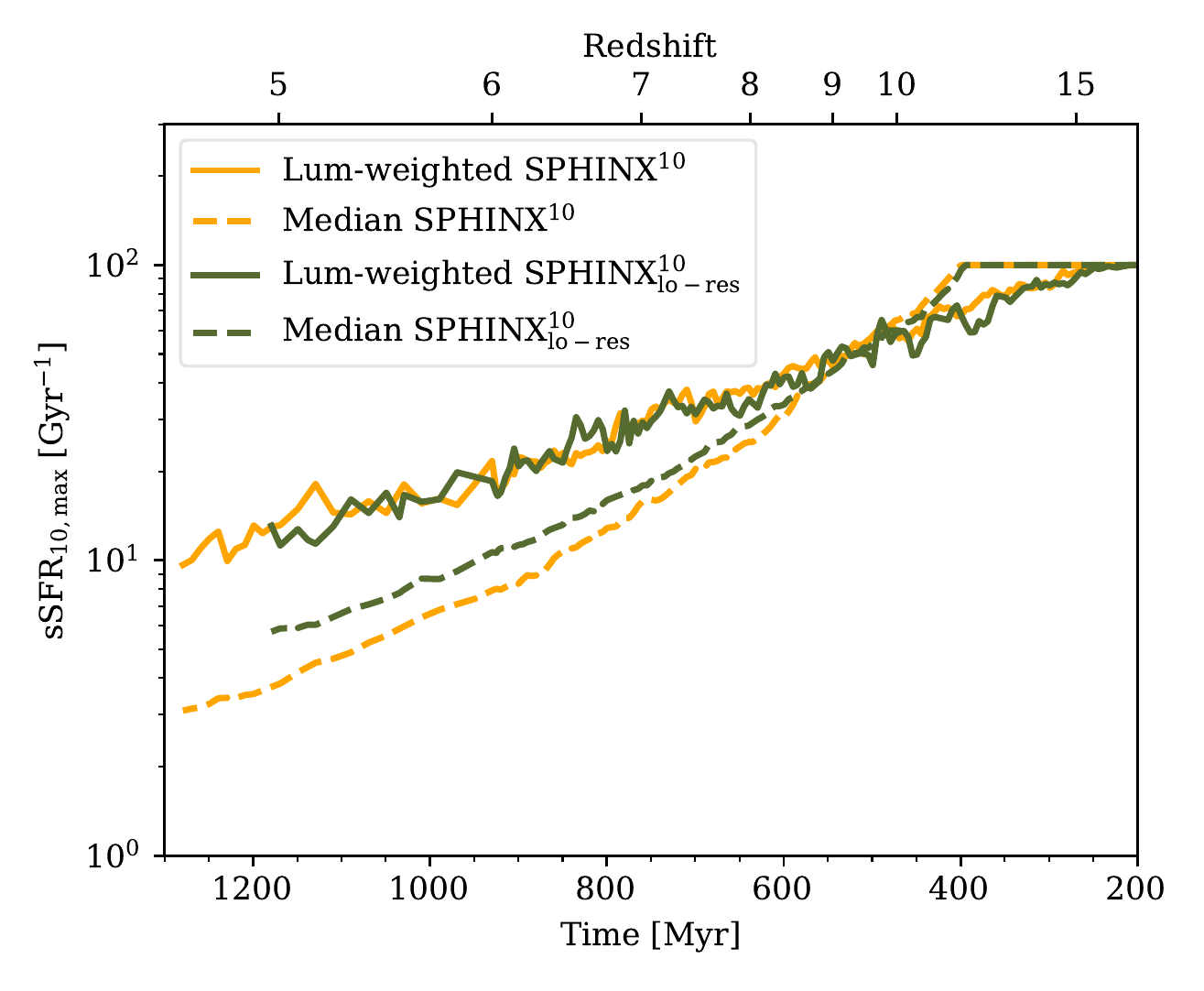}
  \caption
  {\label{sSFR1max0-z_resolution.fig} Redshift evolution of $\sSFRmax$
    in \sphinxten{} and its lower-resolution counterpart. The dashed
    curves show the median $\sSFRmax$ for all halos containing
    stars. Here the evolution with redshift is slightly weaker in the
    lower-resolution case. The solid curves show the
    intrinsic-\Lyc{}-luminosity-weighted mean for the same halos,
    showing better the evolution in $\sSFRmax$ for \Lyc{}-emitting
    galaxies. This is barely affected by the lower resolution.}
\end{figure}

Since $\Dxmin$ varies linearly with the expansion of the Universe and
$\fesc$ decreases with $\Dxmin$, we must also conclude that the
decrease we find with redshift of the global $\fesc$ must be in some
part due to the redshift-degradation of resolution. To determine how
much the decrease in $\fesc$ is due to this, we compare with vertical
bars in \Fig{fesc_global_10cmpc_resolution.fig} the overall drop in
yellow in the global $\fesc$ in \sphinxten{} from $z=11-5$, during
which the expansion factor, and therefore also $\Dxmin$, doubles, and
in green, the drop in $\fesc$ by directly doubling $\Dxmin$ in the
simulation. We find that the drop due to directly doubling $\Dxmin$ is
about $20$ percent of the overall drop in $\fesc$ with redshift over
the $\Dxmin$-doubling time. The effect of the redshift-degrading
resolution is therefore sub-dominant but non-negligible, and we
maintain our conclusion that the drop in $\fesc$ is predominantly
dictated by an overall decrease in sSFR with redshift. We finally note
that the global sSFR and its decrease with redshift is well converged
with resolution, as shown in \Fig{sSFR1max0-z_resolution.fig}.

\section{Conclusions} \label{Conclusions.sec} We use the \sphinx{}
suite of high redshift cosmological radiation-hydrodynamics
simulations to predict the escape of ionizing \Lyc{} radiation from
galaxies in the epoch of reionization. We focus our analysis on our
largest $20$ cMpc wide \sphinxtw{} volume which finishes reionizing
late, or at $z=4.64$ but complement our analysis with a $10$ cMpc wide
volume using a slightly more luminous SED model which leads to an
early reionization, finishing at $z=6.5$.  We find that our
simulations agree well with the observed UV luminosity function for
redshifts $5-10$, implying that the intrinsic volume-emissivity of
\Lyc{} photons is approximately correct. The reasonable reionization
histories produced in turn imply that the mean escape fraction is also
sensible. Our conclusions for the escape of \Lyc{} radiation from
galaxies are as follows.

\begin{itemize}
\item As found previously by several authors, the escape fraction,
  $\fesc$, fluctuates quickly and strongly over time for individual
  galaxies. This is due to the regulation of $\fesc$ by stellar
  feedback and leads to a ``disco'' reionization, where the emission
  of \Lyc{} photons into the IGM is at any given time dominated by a
  small fraction of galaxies, which are replaced over a few Myrs by a
  fresh small subset of galaxies dominating the escaping \Lyc{}
  emission (\Fig{christmas.fig}).

\item The escape fraction per galaxy decreases strongly for massive
  halos and massive, metal-rich, bright galaxies, largely in
  agreement with most previous studies (Figures
  \ref{fesc_hist_Mvir.fig}, \ref{fesc_hist_Mstar_all.fig},
  \ref{fesc_hist_Z.fig}, and \ref{fesc_halo_hist_matt_all.fig},
  respectively). We find a peak in $\fesc$ for intermediate-mass
  galaxies of $\Mstar \approx 10^7 \ \Msun$, with $\fesc$ dropping
  strongly for both less and more massive galaxies. The same applies
  for brightness, with a peak at $\MFifteen\approx -17$.
  We find that $\fesc$ correlates positively with both the short-term
  (\Fig{fesc_halo_hist_sSFR10.fig}) and long-term
  (\Fig{fesc_halo_hist_sSFR100.fig}) specific star formation rate
  (sSFR), through we also find high mean $\fesc$ for low short-term
  sSFR, which we find is due to galaxies that have recently
  experienced high sSFR and subsequent catastrophic feedback events
  which caused severe disruptions of the ISM and, hence, high $\fesc$

\item We find that escaping \Lyc{} radiation during the EoR comes from
  increasingly massive, metal-rich, and bright galaxies with
  decreasing redshift. This evolution is weaker though than for
  intrinsic emission, due to these massive, metal-rich, and bright
  galaxies having low escape fractions. The emission (both intrinsic
  and escaping) also shifts to galaxies with decreasing sSFR with
  redshift, due to an overall decrease in sSFR. Taking a sensitivity
  limit of $\MFifteen=-17$ for existing high-z surveys, we find that
  the bright and constrained part of the high-z luminosity function
  contains galaxies responsible for about $45$ ($52$) percent of the
  \Lyc{} photons emitted into the IGM (produced in the ISM) up to the
  end of reionization (\Fig{contributions.fig}). The JWST, with a
  sensitivity limit of $\MFifteen=-16$, will account for about $68$
  percent of the \Lyc{} radiation contributing to reionization.

\item The global $\fesc$, naturally and without any calibration,
  decreases with redshift in our simulations, as assumed by UV
  background models and low-redshift observations
  (\Fig{fesc_global.fig}). We find that this evolution is driven
  almost exclusively by a decreasing intensity of star formation (and
  hence feedback), measured via the maximum $\sSFR$ of a galaxy over
  the last 50 Myrs (\Fig{fesc_fixedsSFRmax.fig}). The current $\sSFR$
  does not show as clean a correlation with $\fesc$, because a burst
  in star formation can lead to an increase in $\fesc$ that lasts tens
  of Myrs after the burst has ceased and while the galaxy is
  effectively quenched.

\item Whereas previous works have attributed the regulation of $\fesc$
  primarily to SN feedback, we find that \Lyc{} radiation feedback
  also plays an important role (\Fig{fesc_hist_age_age.fig}).

\end{itemize}

Our fiducial late reionization simulation has lower overall escape
fractions and therefore a lower global escape fraction than the early
reionization simulation, but otherwise the conclusions above hold for
both simulations, in terms of the dependencies of $\fesc$ on halo
properties, the halos contributing to the \Lyc{} budget during
reionization, the evolution of $\fesc$ with redshift, and what drives
it. Lower resolution tends to produce overall lower escape fractions,
but our conclusions on the correlation of $\fesc$ with halo properties
and the evolution of $\fesc$ with redshift hold.

\section*{Acknowledgements}
We thank the referee, Nick Gnedin, for an insightful review that
helped improve our manuscript. We are also grateful to JJ Eldridge and
Elizabeth Stanway for help in interpreting the \bpass{} models.  This
work was supported by the National Research Foundation of Korea (NRF)
grant funded by the Korea government (MSIT) (No. 2019K2A9A1A06091377
and No. 2020R1C1C1007079). Support was also provided jointly by CNRS
and NRF via the ``projets de recherche conjoints'' (PRC) grant titled
RUBGY in 2020-2021.  TG acknowledges support from ERC starting grant
ERC-757258-TRIPLE. LCK was supported by the European Union’s Horizon
2020 research and innovation programme under the Marie
Skłodowska-Curie grant agreement No. 885990.  Support by ERC Advanced
Grant 320596 ``The Emergence of Structure during the Epoch of
reionization'' is gratefully acknowledged. Computing time for this
work was provided by the Partnership for Advanced Computing in Europe
(PRACE) as part of the ``First luminous objects and reionization with
SPHINX (cont.)''  (2016153539, 2018184362, 2019215124) project. We
thank Philipp Otte and Filipe Guimaraes for helpful support throughout
the project and for the extra storage they provided us. We also thank
GENCI for providing additional computing resources under GENCI grant
A0070410560.  Preparations and tests were also performed at the Common
Computing Facility (CCF) of the LABEX Lyon Institute of Origins
(ANR-10-LABX-0066) and PSMN (Pôle Scientifique de Modélisation
Numérique) at ENS de Lyon.

\section*{Data availability}
The data underlying this article will be shared on reasonable request
to the corresponding author.

\bibliography{/Users/joki/astro/latex/papers/Library.bib}

\begin{thebibliography}{}
\makeatletter
\relax
\def\mn@urlcharsother{\let\do\@makeother \do\$\do\&\do\#\do\^\do\_\do\%\do\~}
\def\mn@doi{\begingroup\mn@urlcharsother \@ifnextchar [ {\mn@doi@}
  {\mn@doi@[]}}
\def\mn@doi@[#1]#2{\def\@tempa{#1}\ifx\@tempa\@empty \href
  {http://dx.doi.org/#2} {doi:#2}\else \href {http://dx.doi.org/#2} {#1}\fi
  \endgroup}
\def\mn@eprint#1#2{\mn@eprint@#1:#2::\@nil}
\def\mn@eprint@arXiv#1{\href {http://arxiv.org/abs/#1} {{\tt arXiv:#1}}}
\def\mn@eprint@dblp#1{\href {http://dblp.uni-trier.de/rec/bibtex/#1.xml}
  {dblp:#1}}
\def\mn@eprint@#1:#2:#3:#4\@nil{\def\@tempa {#1}\def\@tempb {#2}\def\@tempc
  {#3}\ifx \@tempc \@empty \let \@tempc \@tempb \let \@tempb \@tempa \fi \ifx
  \@tempb \@empty \def\@tempb {arXiv}\fi \@ifundefined
  {mn@eprint@\@tempb}{\@tempb:\@tempc}{\expandafter \expandafter \csname
  mn@eprint@\@tempb\endcsname \expandafter{\@tempc}}}

\bibitem[\protect\citeauthoryear{Agertz et~al.,}{Agertz
  et~al.}{2020}]{Agertz2020}
Agertz O.,  et~al., 2020, \mn@doi [MNRAS] {10.1093/mnras/stz3053}, 491, 1656

\bibitem[\protect\citeauthoryear{Alvarez, Finlator  \& Trenti}{Alvarez
  et~al.}{2012}]{Alvarez2012}
Alvarez M.~A.,  Finlator K.,   Trenti M.,  2012, \mn@doi [ApJ]
  {10.1088/2041-8205/759/2/L38}, 759, L38

\bibitem[\protect\citeauthoryear{Atek et~al.,}{Atek et~al.}{2015}]{Atek2015}
Atek H.,  et~al., 2015, \mn@doi [ApJ] {10.1088/0004-637X/814/1/69}, 814, 69

\bibitem[\protect\citeauthoryear{Atek, Richard, Kneib, Schaerer, Atek, Richard,
  Kneib  \& Schaerer}{Atek et~al.}{2018}]{Atek2018}
Atek H.,  Richard J.,  Kneib J.-P.,  Schaerer D.,  Atek H.,  Richard J.,  Kneib
  J.-P.,   Schaerer D.,  2018, \mn@doi [MNRAS] {10.1093/mnras/sty1820}, 479,
  5184

\bibitem[\protect\citeauthoryear{Aubert, Pichon  \& Colombi}{Aubert
  et~al.}{2004}]{Aubert2004}
Aubert D.,  Pichon C.,   Colombi S.,  2004, MNRAS Letters, 352, 376

\bibitem[\protect\citeauthoryear{Ba{\~{n}}ados et~al.,}{Ba{\~{n}}ados
  et~al.}{2017}]{Banados2017}
Ba{\~{n}}ados E.,  et~al., 2017, \mn@doi [Nature] {10.1038/nature25180}, 553,
  473

\bibitem[\protect\citeauthoryear{Becker, D'Aloisio, Christenson, Zhu, Worseck
  \& Bolton}{Becker et~al.}{2021}]{Becker2021}
Becker G.~D.,  D'Aloisio A.,  Christenson H.~M.,  Zhu Y.,  Worseck G.,   Bolton
  J.~S.,  2021, \mn@doi [MNRAS] {10.1093/mnras/stab2696}, 508, 1853

\bibitem[\protect\citeauthoryear{Behroozi, Wechsler, Hearin  \&
  Conroy}{Behroozi et~al.}{2019}]{Behroozi2019}
Behroozi P.,  Wechsler R.~H.,  Hearin A.~P.,   Conroy C.,  2019, \mn@doi
  [MNRAS] {10.1093/mnras/stz1182}, 488, 3143

\bibitem[\protect\citeauthoryear{Bouwens et~al.,}{Bouwens
  et~al.}{2015}]{Bouwens2015}
Bouwens R.~J.,  et~al., 2015, \mn@doi [ApJ] {10.1088/0004-637X/803/1/34}, 803,
  34

\bibitem[\protect\citeauthoryear{Bouwens, Oesch, Illingworth, Ellis  \&
  Stefanon}{Bouwens et~al.}{2017}]{Bouwens2017}
Bouwens R.~J.,  Oesch P.~A.,  Illingworth G.~D.,  Ellis R.~S.,   Stefanon M.,
  2017, \mn@doi [ApJ] {10.3847/1538-4357/aa70a4}, 843, 129

\bibitem[\protect\citeauthoryear{Ceverino, Glover, Klessen, Ceverino, Glover
  \& Klessen}{Ceverino et~al.}{2017}]{Ceverino2017}
Ceverino D.,  Glover S.,  Klessen R.,  Ceverino D.,  Glover S. C.~O.,   Klessen
  R.~S.,  2017, \mn@doi [MNRAS] {10.1093/mnras/stx1386}, 470, 2791

\bibitem[\protect\citeauthoryear{Davies et~al.,}{Davies
  et~al.}{2018}]{Davies2018}
Davies F.~B.,  et~al., 2018, \mn@doi [ApJ] {10.3847/1538-4357/aad6dc}, 864, 142

\bibitem[\protect\citeauthoryear{Dayal, Ferrara, Dayal  \& Ferrara}{Dayal
  et~al.}{2019}]{Dayal2019}
Dayal P.,  Ferrara A.,  Dayal P.,   Ferrara A.,  2019, \mn@doi [Proc. Int.
  Astron. Union] {10.1017/S1743921320001106}, 15, 43

\bibitem[\protect\citeauthoryear{Dekel et~al.,}{Dekel et~al.}{2009}]{Dekel2009}
Dekel A.,  et~al., 2009, Nature, 457, 451

\bibitem[\protect\citeauthoryear{Eldridge, Izzard  \& Tout}{Eldridge
  et~al.}{2007}]{Eldridge2007}
Eldridge J.~J.,  Izzard R.~G.,   Tout C.~A.,  2007, \mn@doi [MNRAS]
  {10.1111/j.1365-2966.2007.12738.x}, 384, 1109

\bibitem[\protect\citeauthoryear{Faisst, Faisst  \& L.}{Faisst
  et~al.}{2016}]{Faisst2016}
Faisst A.~L.,  Faisst  L. A.,  2016, \mn@doi [ApJ]
  {10.3847/0004-637X/829/2/99}, 829, 99

\bibitem[\protect\citeauthoryear{Fakhouri, Ma, Boylan-Kolchin, Fakhouri, Ma  \&
  Boylan-Kolchin}{Fakhouri et~al.}{2010}]{Fakhouri2010}
Fakhouri O.,  Ma C.-P.,  Boylan-Kolchin M.,  Fakhouri O.,  Ma C.-P.,
  Boylan-Kolchin M.,  2010, \mn@doi [MNRAS] {10.1111/j.1365-2966.2010.16859.x},
  406, 2267

\bibitem[\protect\citeauthoryear{Fan, Carilli  \& Keating}{Fan
  et~al.}{2006a}]{Fan2006a}
Fan X.,  Carilli C.~L.,   Keating B.,  2006a, Annu. Rev. Astron. Astrophys.,
  44, 415

\bibitem[\protect\citeauthoryear{Fan et~al.,}{Fan et~al.}{2006b}]{Fan2006}
Fan X.,  et~al., 2006b, ApJ Letters, 132, 117

\bibitem[\protect\citeauthoryear{Farcy, Rosdahl, Dubois, Blaizot  \&
  Martin-Alvarez}{Farcy et~al.}{2022}]{Farcy2022}
Farcy M.,  Rosdahl J.,  Dubois Y.,  Blaizot J.,   Martin-Alvarez S.,  2022,
  \mn@doi [MNRAS] {10.1093/mnras/stac1196}, 513, 5000

\bibitem[\protect\citeauthoryear{Faucher-Gigu{\`{e}}re}{Faucher-Gigu{\`{e}}re}{2020}]{FaucherGiguere2020}
Faucher-Gigu{\`{e}}re C.-A.,  2020, \mn@doi [MNRAS] {10.1093/mnras/staa302},
  493, 1614

\bibitem[\protect\citeauthoryear{Federrath \& Klessen}{Federrath \&
  Klessen}{2012}]{Federrath2012}
Federrath C.,  Klessen R.~S.,  2012, ApJ, 761, 156

\bibitem[\protect\citeauthoryear{Ferland, Korista, Verner, Ferguson, Kingdon
  \& Verner}{Ferland et~al.}{1998}]{Ferland1998}
Ferland G.~J.,  Korista K.~T.,  Verner D.~A.,  Ferguson J.~W.,  Kingdon J.~B.,
   Verner E.~M.,  1998, Publ. Astron. Soc. Pacific, 110, 761

\bibitem[\protect\citeauthoryear{Fern{\'{a}}ndez et~al.,}{Fern{\'{a}}ndez
  et~al.}{2018}]{Fernandez2018}
Fern{\'{a}}ndez R.~L.,  et~al., 2018, \mn@doi [A{\&}A]
  {10.1051/0004-6361/201732358}, 615, A27

\bibitem[\protect\citeauthoryear{Ferrara, Loeb, Ferrara  \& Loeb}{Ferrara
  et~al.}{2012}]{Ferrara2012}
Ferrara A.,  Loeb A.,  Ferrara A.,   Loeb A.,  2012, \mn@doi [MNRAS]
  {10.1093/mnras/stt381}, 431, 2826

\bibitem[\protect\citeauthoryear{Finkelstein et~al.,}{Finkelstein
  et~al.}{2015}]{Finkelstein2015}
Finkelstein S.~L.,  et~al., 2015, \mn@doi [ApJ] {10.1088/0004-637X/810/1/71},
  810, 71

\bibitem[\protect\citeauthoryear{Finkelstein et~al.,}{Finkelstein
  et~al.}{2019}]{Finkelstein2019}
Finkelstein S.~L.,  et~al., 2019, \mn@doi [ApJ] {10.3847/1538-4357/ab1ea8},
  879, 36

\bibitem[\protect\citeauthoryear{Garel, Blaizot, Rosdahl, Michel-Dansac,
  Haehnelt, Katz, Kimm  \& Verhamme}{Garel et~al.}{2021}]{Garel2021}
Garel T.,  Blaizot J.,  Rosdahl J.,  Michel-Dansac L.,  Haehnelt M.~G.,  Katz
  H.,  Kimm T.,   Verhamme A.,  2021, \mn@doi [MNRAS] {10.1093/mnras/stab990},
  504, 1902

\bibitem[\protect\citeauthoryear{Gnedin}{Gnedin}{2014}]{Gnedin2014}
Gnedin N.~Y.,  2014, \mn@doi [ApJ] {10.1088/0004-637X/793/1/29}, 793, 29

\bibitem[\protect\citeauthoryear{Gnedin, Kravtsov  \& Chen}{Gnedin
  et~al.}{2008}]{Gnedin2008}
Gnedin N.~Y.,  Kravtsov A.~V.,   Chen H.-W.,  2008, ApJ, 672, 765

\bibitem[\protect\citeauthoryear{Grazian et~al.,}{Grazian
  et~al.}{2017}]{Grazian2017}
Grazian A.,  et~al., 2017, \mn@doi [A{\&}A] {10.1051/0004-6361/201730447}, 602,
  A18

\bibitem[\protect\citeauthoryear{Greig, Mesinger, Haiman  \& Simcoe}{Greig
  et~al.}{2017}]{Greig2017}
Greig B.,  Mesinger A.,  Haiman Z.,   Simcoe R.~A.,  2017, \mn@doi [MNRAS]
  {10.1093/mnras/stw3351}, 466, stw3351

\bibitem[\protect\citeauthoryear{Greig, Mesinger  \& Ba{\~{n}}ados}{Greig
  et~al.}{2019}]{Greig2019}
Greig B.,  Mesinger A.,   Ba{\~{n}}ados E.,  2019, \mn@doi [MNRAS]
  {10.1093/mnras/stz230}, 484, 5094

\bibitem[\protect\citeauthoryear{Gunn, Peterson, Gunn  \& Peterson}{Gunn
  et~al.}{1965}]{Gunn1965}
Gunn J.~E.,  Peterson B.~A.,  Gunn J.~E.,   Peterson B.~A.,  1965, \mn@doi
  [ApJ] {10.1086/148444}, 142, 1633

\bibitem[\protect\citeauthoryear{Haardt \& Madau}{Haardt \&
  Madau}{2012}]{Haardt2012}
Haardt F.,  Madau P.,  2012, \mn@doi [ApJ] {10.1088/0004-637X/746/2/125}, 746,
  125

\bibitem[\protect\citeauthoryear{Hahn \& Abel}{Hahn \& Abel}{2011}]{Hahn2011}
Hahn O.,  Abel T.,  2011, \mn@doi [MNRAS] {10.1111/J.1365-2966.2011.18820.X},
  415, 2101

\bibitem[\protect\citeauthoryear{Heckman}{Heckman}{2000}]{Heckman2000}
Heckman T.~M.,  2000, \mn@doi [arXiv:astro-ph/0009075]
  {10.48550/arxiv.astro-ph/0009075}

\bibitem[\protect\citeauthoryear{Heckman et~al.,}{Heckman
  et~al.}{2011}]{Heckman2011}
Heckman T.~M.,  et~al., 2011, \mn@doi [ApJ] {10.1088/0004-637X/730/1/5}, 730, 5

\bibitem[\protect\citeauthoryear{Hutter et~al.,}{Hutter
  et~al.}{2020}]{Hutter2020}
Hutter A.,  et~al., 2020, \mn@doi [MNRAS] {10.1093/mnras/stab877}, 506, 215

\bibitem[\protect\citeauthoryear{Iliev, Mellema, Ahn, Shapiro, Mao  \&
  Pen}{Iliev et~al.}{2013}]{Iliev2013}
Iliev I.~T.,  Mellema G.,  Ahn K.,  Shapiro P.~R.,  Mao Y.,   Pen U.-L.,  2013,
  \mn@doi [MNRAS] {10.1093/mnras/stt2497}, 439, 725

\bibitem[\protect\citeauthoryear{Inoue \& Iwata}{Inoue \&
  Iwata}{2008}]{Inoue2008}
Inoue A.~K.,  Iwata I.,  2008, \mn@doi [MNRAS]
  {10.1111/j.1365-2966.2008.13350.x}, 387, 1681

\bibitem[\protect\citeauthoryear{Inoue et~al.,}{Inoue et~al.}{2018}]{Inoue2018}
Inoue A.~K.,  et~al., 2018, \mn@doi [Publ. Astron. Soc. Japan]
  {10.1093/pasj/psy048}, 70, 55

\bibitem[\protect\citeauthoryear{Ishigaki et~al.,}{Ishigaki
  et~al.}{2017}]{Ishigaki2017}
Ishigaki M.,  et~al., 2017, \mn@doi [ApJ] {10.3847/1538-4357/aaa544}, 854, 73

\bibitem[\protect\citeauthoryear{Jung et~al.,}{Jung et~al.}{2020}]{Jung2020}
Jung I.,  et~al., 2020, \mn@doi [ApJ] {10.3847/1538-4357/abbd44}, 904, 144

\bibitem[\protect\citeauthoryear{Kannan, Garaldi, Smith, Pakmor, Springel,
  Vogelsberger  \& Hernquist}{Kannan et~al.}{2021}]{Kannan2021}
Kannan R.,  Garaldi E.,  Smith A.,  Pakmor R.,  Springel V.,  Vogelsberger M.,
   Hernquist L.,  2021, \mn@doi [MNRAS] {10.1093/mnras/stab3710}, 511, 4005

\bibitem[\protect\citeauthoryear{Katz, Kimm, Sijacki  \& Haehnelt}{Katz
  et~al.}{2017}]{Katz2017}
Katz H.,  Kimm T.,  Sijacki D.,   Haehnelt M.~G.,  2017, \mn@doi [MNRAS]
  {10.1093/mnras/stx608}, 468, 4831

\bibitem[\protect\citeauthoryear{Katz, Kimm, Haehnelt, Sijacki, Rosdahl  \&
  Blaizot}{Katz et~al.}{2018}]{Katz2018b}
Katz H.,  Kimm T.,  Haehnelt M.~G.,  Sijacki D.,  Rosdahl J.,   Blaizot J.,
  2018, \mn@doi [MNRAS] {10.1093/mnras/sty3154}, 483, 1029

\bibitem[\protect\citeauthoryear{Katz et~al.,}{Katz et~al.}{2020a}]{Katz2020}
Katz H.,  et~al., 2020a, \mn@doi [MNRAS] {10.1093/mnras/staa639}, 494, 2200

\bibitem[\protect\citeauthoryear{Katz et~al.,}{Katz et~al.}{2020b}]{Katz2020a}
Katz H.,  et~al., 2020b, \mn@doi [MNRAS] {10.1093/mnras/staa2355}, 498, 164

\bibitem[\protect\citeauthoryear{Katz et~al.,}{Katz et~al.}{2021}]{Katz2021c}
Katz H.,  et~al., 2021, \mn@doi [MNRAS] {10.1093/mnras/stac028}, 510, 5603

\bibitem[\protect\citeauthoryear{Keating, Weinberger, Kulkarni, Haehnelt,
  Chardin  \& Aubert}{Keating et~al.}{2020}]{Keating2020}
Keating L.~C.,  Weinberger L.~H.,  Kulkarni G.,  Haehnelt M.~G.,  Chardin J.,
  Aubert D.,  2020, \mn@doi [MNRAS] {10.1093/mnras/stz3083}, 491, 1736

\bibitem[\protect\citeauthoryear{Kennicutt \& Evans}{Kennicutt \&
  Evans}{2012}]{Kennicutt2012}
Kennicutt R.~C.,  Evans N.~J.,  2012, \mn@doi [ARA{\&}A]
  {10.1146/annurev-astro-081811-125610}, 50, 531

\bibitem[\protect\citeauthoryear{Khaire \& Srianand}{Khaire \&
  Srianand}{2018}]{Khaire2018}
Khaire V.,  Srianand R.,  2018, \mn@doi [MNRAS] {10.1093/mnras/stz174}, 484,
  4174

\bibitem[\protect\citeauthoryear{Khaire, Srianand, Choudhury  \&
  Gaikwad}{Khaire et~al.}{2015}]{Khaire2015}
Khaire V.,  Srianand R.,  Choudhury T.~R.,   Gaikwad P.,  2015, \mn@doi [MNRAS]
  {10.1093/mnras/stw192}, 457, 4051

\bibitem[\protect\citeauthoryear{Kimm \& Cen}{Kimm \& Cen}{2014}]{Kimm2014}
Kimm T.,  Cen R.,  2014, ApJ, 788, 121

\bibitem[\protect\citeauthoryear{Kimm, Cen, Devriendt, Dubois  \& Slyz}{Kimm
  et~al.}{2015}]{Kimm2015}
Kimm T.,  Cen R.,  Devriendt J.,  Dubois Y.,   Slyz A.,  2015, MNRAS, 451, 2900

\bibitem[\protect\citeauthoryear{Kimm, Katz, Haehnelt, Rosdahl, Devriendt  \&
  Slyz}{Kimm et~al.}{2017}]{Kimm2017}
Kimm T.,  Katz H.,  Haehnelt M.,  Rosdahl J.,  Devriendt J.,   Slyz A.,  2017,
  \mn@doi [MNRAS] {10.1093/mnras/stx052}, 652, 4826

\bibitem[\protect\citeauthoryear{Kroupa}{Kroupa}{2001}]{Kroupa2001}
Kroupa P.,  2001, MNRAS Letters, 322, 231

\bibitem[\protect\citeauthoryear{Kulkarni, Keating, Haehnelt, Bosman, Puchwein,
  Chardin  \& Aubert}{Kulkarni et~al.}{2019}]{Kulkarni2019}
Kulkarni G.,  Keating L.~C.,  Haehnelt M.~G.,  Bosman S. E.~I.,  Puchwein E.,
  Chardin J.,   Aubert D.,  2019, \mn@doi [MNRAS Letters]
  {10.1093/mnrasl/slz025}, 485, L24

\bibitem[\protect\citeauthoryear{Laursen, Sommer-Larsen  \& Andersen}{Laursen
  et~al.}{2009}]{Laursen2009}
Laursen P.,  Sommer-Larsen J.,   Andersen A.~C.,  2009, \mn@doi [ApJ]
  {10.1088/0004-637X/704/2/1640}, 704, 1640

\bibitem[\protect\citeauthoryear{Lehnert et~al.,}{Lehnert
  et~al.}{2015}]{Lehnert2015}
Lehnert M.~D.,  et~al., 2015, \mn@doi [A{\&}A] {10.1051/0004-6361/201322630},
  577, A112

\bibitem[\protect\citeauthoryear{Lewis et~al.,}{Lewis et~al.}{2020}]{Lewis2020}
Lewis J. S.~W.,  et~al., 2020, \mn@doi [MNRAS] {10.1093/MNRAS/STAA1748}, 496,
  4342

\bibitem[\protect\citeauthoryear{Livermore, Finkelstein  \& Lotz}{Livermore
  et~al.}{2017}]{Livermore2017}
Livermore R.~C.,  Finkelstein S.~L.,   Lotz J.~M.,  2017, \mn@doi [ApJ]
  {10.3847/1538-4357/835/2/113}, 835, 113

\bibitem[\protect\citeauthoryear{Ma et~al.,}{Ma et~al.}{2015}]{Ma2015}
Ma X.,  et~al., 2015, \mn@doi [MNRAS] {10.1093/mnras/stv1679}, 453, 960

\bibitem[\protect\citeauthoryear{Ma, Hopkins, Kasen, Quataert, Faucher-Giguere,
  Keres, Murray  \& Strom}{Ma et~al.}{2016}]{Ma2016}
Ma X.,  Hopkins P.~F.,  Kasen D.,  Quataert E.,  Faucher-Giguere C.-A.,  Keres
  D.,  Murray N.,   Strom A.,  2016, \mn@doi [MNRAS] {10.1093/mnras/stw941},
  459, 3614

\bibitem[\protect\citeauthoryear{Ma et~al.,}{Ma et~al.}{2018}]{Ma2018}
Ma X.,  et~al., 2018, \mn@doi [MNRAS] {10.1093/mnras/sty1024}, 478, 1694

\bibitem[\protect\citeauthoryear{Ma, Quataert, Wetzel, Hopkins,
  Faucher-Gigu{\`{e}}re  \& Kere{\v{s}}}{Ma et~al.}{2020}]{Ma2020}
Ma X.,  Quataert E.,  Wetzel A.,  Hopkins P.~F.,  Faucher-Gigu{\`{e}}re C.-A.,
   Kere{\v{s}} D.,  2020, \mn@doi [MNRAS] {10.1093/mnras/staa2404}, 498, 2001

\bibitem[\protect\citeauthoryear{Madau, Haardt  \& Rees}{Madau
  et~al.}{1999}]{Madau1999}
Madau P.,  Haardt F.,   Rees M.~J.,  1999, \mn@doi [ApJ] {10.1086/306975}, 514,
  648

\bibitem[\protect\citeauthoryear{Mason et~al.,}{Mason et~al.}{2018}]{Mason2018}
Mason C.~A.,  et~al., 2018, \mn@doi [ApJ] {10.3847/1538-4357/aab0a7}, 856, 2

\bibitem[\protect\citeauthoryear{Mason et~al.,}{Mason et~al.}{2019}]{Mason2019}
Mason C.~A.,  et~al., 2019, \mn@doi [MNRAS] {10.1093/mnras/stz632}, 485, 3947

\bibitem[\protect\citeauthoryear{Matthee et~al.,}{Matthee
  et~al.}{2016}]{Matthee2016}
Matthee J.,  et~al., 2016, \mn@doi [MNRAS] {10.1093/mnras/stw2973}, 465, 3637

\bibitem[\protect\citeauthoryear{Matthee et~al.,}{Matthee
  et~al.}{2021}]{Matthee2021}
Matthee J.,  et~al., 2021, \mn@doi [MNRAS] {10.1093/mnras/stac801}

\bibitem[\protect\citeauthoryear{Mauerhofer, Verhamme, Blaizot, Garel, Kimm,
  Michel-Dansac  \& Rosdahl}{Mauerhofer et~al.}{2021}]{Mauerhofer2021}
Mauerhofer V.,  Verhamme A.,  Blaizot J.,  Garel T.,  Kimm T.,  Michel-Dansac
  L.,   Rosdahl J.,  2021, \mn@doi [A{\&}A] {10.1051/0004-6361/202039449}, 646,
  A80

\bibitem[\protect\citeauthoryear{McGreer, Mesinger  \& D'Odorico}{McGreer
  et~al.}{2015}]{McGreer2015}
McGreer I.,  Mesinger A.,   D'Odorico V.,  2015, \mn@doi [MNRAS]
  {10.1093/mnras/stu2449}, 447, 499

\bibitem[\protect\citeauthoryear{Me{\v{s}}tri{\'{c}}
  et~al.,}{Me{\v{s}}tri{\'{c}} et~al.}{2021}]{Mestric2021}
Me{\v{s}}tri{\'{c}} U.,  et~al., 2021, \mn@doi [MNRAS]
  {10.1093/mnras/stab2615}, 508, 4443

\bibitem[\protect\citeauthoryear{Michel-Dansac et~al.,}{Michel-Dansac
  et~al.}{2020}]{Michel-Dansac2020}
Michel-Dansac L.,  et~al., 2020, \mn@doi [A{\&}A]
  {10.1051/0004-6361/201834961}, 635, A154

\bibitem[\protect\citeauthoryear{Naidu et~al.,}{Naidu et~al.}{2019}]{Naidu2019}
Naidu R.~P.,  et~al., 2019, \mn@doi [ApJ] {10.3847/1538-4357/ab7cc9}, 892, 109

\bibitem[\protect\citeauthoryear{Ocvirk et~al.,}{Ocvirk
  et~al.}{2018}]{Ocvirk2018b}
Ocvirk P.,  et~al., 2018, \mn@doi [MNRAS] {10.1093/mnras/staa1266}, 496, 4087

\bibitem[\protect\citeauthoryear{Ocvirk et~al.,}{Ocvirk
  et~al.}{2021}]{Ocvirk2021}
Ocvirk P.,  et~al., 2021, \mn@doi [MNRAS] {10.1093/mnras/stab2502}, 507, 6108

\bibitem[\protect\citeauthoryear{Oesch, Bouwens, Illingworth, Labb{\'{e}}  \&
  Stefanon}{Oesch et~al.}{2018}]{Oesch2018}
Oesch P.~A.,  Bouwens R.~J.,  Illingworth G.~D.,  Labb{\'{e}} I.,   Stefanon
  M.,  2018, \mn@doi [ApJ] {10.3847/1538-4357/aab03f}, 855, 105

\bibitem[\protect\citeauthoryear{Ouchi et~al.,}{Ouchi
  et~al.}{2009}]{Ouchi2009a}
Ouchi M.,  et~al., 2009, \mn@doi [ApJ] {10.1088/0004-637X/706/2/1136}, 706,
  1136

\bibitem[\protect\citeauthoryear{Ouchi et~al.,}{Ouchi et~al.}{2017}]{Ouchi2017}
Ouchi M.,  et~al., 2017, \mn@doi [Publ. Astron. Soc. Japan]
  {10.1093/pasj/psx074}, 70, S13

\bibitem[\protect\citeauthoryear{Paardekooper, Khochfar  \&
  Vecchia}{Paardekooper et~al.}{2015}]{Paardekooper2015}
Paardekooper J.-P.,  Khochfar S.,   Vecchia C.~D.,  2015, \mn@doi [MNRAS]
  {10.1093/mnras/stv1114}, 451, 2544

\bibitem[\protect\citeauthoryear{Pahl et~al.,}{Pahl et~al.}{2021}]{Pahl2021}
Pahl A.~J.,  et~al., 2021, \mn@doi [MNRAS] {10.1093/mnras/stab1374}, 505, 2447

\bibitem[\protect\citeauthoryear{Penzias \& Wilson}{Penzias \&
  Wilson}{1965}]{Penzias1965}
Penzias A.~A.,  Wilson R.~W.,  1965, Astrophys. J. Lett. v.489, 142, 419

\bibitem[\protect\citeauthoryear{{Planck Collaboration}}{{Planck
  Collaboration}}{2014}]{Ade2014}
{Planck Collaboration} 2014, \mn@doi [A{\&}A] {10.1051/0004-6361/201321529},
  571, A1

\bibitem[\protect\citeauthoryear{{Planck Collaboration} et~al.,}{{Planck
  Collaboration} et~al.}{2018}]{PlanckCollaboration2018}
{Planck Collaboration} P.,  et~al., 2018, \mn@doi [A{\&}A]
  {10.1051/0004-6361/201833910}, 641, A6

\bibitem[\protect\citeauthoryear{Price, Trac  \& Cen}{Price
  et~al.}{2016}]{Price2016}
Price L.~C.,  Trac H.,   Cen R.,  2016, arXiv:1605.03970

\bibitem[\protect\citeauthoryear{Puchwein, Haardt, Haehnelt, Madau, Puchwein,
  Haardt, Haehnelt  \& Madau}{Puchwein et~al.}{2019}]{Puchwein2019}
Puchwein E.,  Haardt F.,  Haehnelt M.~G.,  Madau P.,  Puchwein E.,  Haardt F.,
  Haehnelt M.~G.,   Madau P.,  2019, \mn@doi [MNRAS] {10.1093/mnras/stz222},
  485, 47

\bibitem[\protect\citeauthoryear{Rasera \& Teyssier}{Rasera \&
  Teyssier}{2006}]{Rasera2006}
Rasera Y.,  Teyssier R.,  2006, A{\&}A, 445, 1

\bibitem[\protect\citeauthoryear{Razoumov \& Sommer-Larsen}{Razoumov \&
  Sommer-Larsen}{2010}]{Razoumov2010}
Razoumov A.~O.,  Sommer-Larsen J.,  2010, Astrophys. J. Lett. v.489, 710, 1239

\bibitem[\protect\citeauthoryear{Read, Iorio, Agertz  \& Fraternali}{Read
  et~al.}{2017}]{Read2017}
Read J.~I.,  Iorio G.,  Agertz O.,   Fraternali F.,  2017, \mn@doi [MNRAS]
  {10.1093/mnras/stx147}, 467, 2019

\bibitem[\protect\citeauthoryear{Robertson}{Robertson}{2021}]{Robertson2021}
Robertson B.~E.,  2021, eprint arXiv:2110.13160

\bibitem[\protect\citeauthoryear{Rosdahl \& Teyssier}{Rosdahl \&
  Teyssier}{2015}]{Rosdahl2015a}
Rosdahl J.,  Teyssier R.,  2015, \mn@doi [MNRAS] {10.1093/mnras/stv567}, 449,
  4380

\bibitem[\protect\citeauthoryear{Rosdahl, Blaizot, Aubert, Stranex  \&
  Teyssier}{Rosdahl et~al.}{2013}]{Rosdahl2013}
Rosdahl J.,  Blaizot J.,  Aubert D.,  Stranex T.,   Teyssier R.,  2013, MNRAS,
  436, 2188

\bibitem[\protect\citeauthoryear{Rosdahl et~al.,}{Rosdahl
  et~al.}{2018}]{Rosdahl2018}
Rosdahl J.,  et~al., 2018, \mn@doi [MNRAS] {10.1093/mnras/sty1655}, 479, 994

\bibitem[\protect\citeauthoryear{Rosen \& Bregman}{Rosen \&
  Bregman}{1995}]{Rosen1995}
Rosen A.,  Bregman J.~N.,  1995, ApJ, 440, 634

\bibitem[\protect\citeauthoryear{Rutkowski et~al.,}{Rutkowski
  et~al.}{2017}]{Rutkowski2017}
Rutkowski M.~J.,  et~al., 2017, \mn@doi [ApJL] {10.3847/2041-8213/aa733b}, 841,
  L27

\bibitem[\protect\citeauthoryear{Saxena et~al.,}{Saxena
  et~al.}{2021}]{Saxena2021}
Saxena A.,  et~al., 2021, \mn@doi [MNRAS] {10.1093/mnras/stab3728}, 511, 120

\bibitem[\protect\citeauthoryear{Schroeder, Mesinger  \& Haiman}{Schroeder
  et~al.}{2012}]{Schroeder2012}
Schroeder J.,  Mesinger A.,   Haiman Z.,  2012, \mn@doi [MNRAS]
  {10.1093/mnras/sts253}, 428, 3058

\bibitem[\protect\citeauthoryear{Sharma, Theuns, Frenk, Bower, Crain, Schaller
  \& Schaye}{Sharma et~al.}{2017}]{Sharma2017}
Sharma M.,  Theuns T.,  Frenk C.,  Bower R.~G.,  Crain R.~A.,  Schaller M.,
  Schaye J.,  2017, \mn@doi [MNRAS] {10.1093/mnras/stx578}, 468, 2176

\bibitem[\protect\citeauthoryear{Smith, Bryan, Somerville, Hu, Teyssier,
  Burkhart  \& Hernquist}{Smith et~al.}{2020}]{Smith2020}
Smith M.~C.,  Bryan G.~L.,  Somerville R.~S.,  Hu C.-Y.,  Teyssier R.,
  Burkhart B.,   Hernquist L.,  2020, \mn@doi [MNRAS] {10.1093/mnras/stab1896},
  506, 3882

\bibitem[\protect\citeauthoryear{Stanway \& Eldridge}{Stanway \&
  Eldridge}{2018}]{Stanway2018}
Stanway E.~R.,  Eldridge J.~J.,  2018, \mn@doi [MNRAS] {10.1093/mnras/sty1353},
  479, 75

\bibitem[\protect\citeauthoryear{Stanway, Eldridge  \& Becker}{Stanway
  et~al.}{2016}]{Stanway2016}
Stanway E.~R.,  Eldridge J.~J.,   Becker G.~D.,  2016, \mn@doi [MNRAS]
  {10.1093/mnras/stv2661}, 456, 485

\bibitem[\protect\citeauthoryear{Stefanon, Bouwens, Labb{\'{e}}, Illingworth,
  Gonzalez  \& Oesch}{Stefanon et~al.}{2021a}]{Stefanon2021a}
Stefanon M.,  Bouwens R.~J.,  Labb{\'{e}} I.,  Illingworth G.~D.,  Gonzalez V.,
    Oesch P.~A.,  2021a, \mn@doi [ApJ] {10.3847/1538-4357/ac1bb6}, 922, 29

\bibitem[\protect\citeauthoryear{Stefanon et~al.,}{Stefanon
  et~al.}{2021b}]{Stefanon2021}
Stefanon M.,  et~al., 2021b, \mn@doi [ApJ] {10.3847/1538-4357/ac3de7}, 927, 48

\bibitem[\protect\citeauthoryear{Steidel et~al.,}{Steidel
  et~al.}{2018}]{Steidel2018}
Steidel C.~C.,  et~al., 2018, \mn@doi [ApJ] {10.3847/1538-4357/aaed28}, 869,
  123

\bibitem[\protect\citeauthoryear{Sun \& Furlanetto}{Sun \&
  Furlanetto}{2015}]{Sun2015}
Sun G.,  Furlanetto S.~R.,  2015, \mn@doi [MNRAS] {10.1093/mnras/stw980}, 460,
  417

\bibitem[\protect\citeauthoryear{Tacchella, Bose, Conroy, Eisenstein  \&
  Johnson}{Tacchella et~al.}{2018}]{Tacchella2018}
Tacchella S.,  Bose S.,  Conroy C.,  Eisenstein D.~J.,   Johnson B.~D.,  2018,
  \mn@doi [ApJ] {10.3847/1538-4357/aae8e0}, 868, 92

\bibitem[\protect\citeauthoryear{Teyssier}{Teyssier}{2002}]{Teyssier2002}
Teyssier R.,  2002, A{\&}A, 385, 337

\bibitem[\protect\citeauthoryear{Tillson, Miller  \& Devriendt}{Tillson
  et~al.}{2011}]{Tillson2011}
Tillson H.,  Miller L.,   Devriendt J.,  2011, \mn@doi [MNRAS]
  {10.1111/j.1365-2966.2011.19311.x}, 417, 666

\bibitem[\protect\citeauthoryear{Trebitsch, Blaizot, Rosdahl, Devriendt  \&
  Slyz}{Trebitsch et~al.}{2017}]{Trebitsch2017}
Trebitsch M.,  Blaizot J.,  Rosdahl J.,  Devriendt J.,   Slyz A.,  2017,
  \mn@doi [MNRAS] {10.1093/mnras/stx1060}, 470, 224

\bibitem[\protect\citeauthoryear{Trebitsch et~al.,}{Trebitsch
  et~al.}{2021}]{Trebitsch2021}
Trebitsch M.,  et~al., 2021, \mn@doi [A{\&}A] {10.1051/0004-6361/202037698},
  653, 154

\bibitem[\protect\citeauthoryear{Tweed, Devriendt, Blaizot, Colombi  \&
  Slyz}{Tweed et~al.}{2009}]{Tweed2009}
Tweed D.,  Devriendt J.,  Blaizot J.,  Colombi S.,   Slyz A.,  2009, A{\&}A,
  506, 647

\bibitem[\protect\citeauthoryear{Wise \& Cen}{Wise \& Cen}{2009}]{Wise2009}
Wise J.~H.,  Cen R.,  2009, ApJ Letters, 693, 984

\bibitem[\protect\citeauthoryear{Wise, Demchenko, Halicek, Norman, Turk, Abel
  \& Smith}{Wise et~al.}{2014}]{Wise2014}
Wise J.~H.,  Demchenko V.~G.,  Halicek M.~T.,  Norman M.~L.,  Turk M.~J.,  Abel
  T.,   Smith B.~D.,  2014, MNRAS, 442, 2560

\bibitem[\protect\citeauthoryear{Xu, Wise, Norman, Ahn  \& O'Shea}{Xu
  et~al.}{2016}]{Xu2016}
Xu H.,  Wise J.~H.,  Norman M.~L.,  Ahn K.,   O'Shea B.~W.,  2016, \mn@doi
  [ApJ] {10.3847/1538-4357/833/1/84}, 833, 84

\bibitem[\protect\citeauthoryear{Yajima, Choi  \& Nagamine}{Yajima
  et~al.}{2011}]{Yajima2011}
Yajima H.,  Choi J.-H.,   Nagamine K.,  2011, MNRAS, 412, 411

\bibitem[\protect\citeauthoryear{Yajima et~al.,}{Yajima
  et~al.}{2020}]{Yajima2020}
Yajima H.,  et~al., 2020, \mn@doi [MNRAS] {10.1093/mnras/stab3092}, 509, 4037

\bibitem[\protect\citeauthoryear{Yoo, Kimm  \& Rosdahl}{Yoo
  et~al.}{2020}]{Yoo2020}
Yoo T.,  Kimm T.,   Rosdahl J.,  2020, \mn@doi [MNRAS]
  {10.1093/mnras/staa3187}, 499, 5175

\bibitem[\protect\citeauthoryear{Yung, Somerville, Finkelstein, Popping,
  Dav{\'{e}}, Venkatesan, Behroozi  \& Ferguson}{Yung et~al.}{2020}]{Yung2020}
Yung L. Y.~A.,  Somerville R.~S.,  Finkelstein S.~L.,  Popping G.,  Dav{\'{e}}
  R.,  Venkatesan A.,  Behroozi P.,   Ferguson H.~C.,  2020, \mn@doi [MNRAS]
  {10.1093/mnras/staa1800}, 496, 4574

\bibitem[\protect\citeauthoryear{Zaroubi, Zaroubi  \& Saleem}{Zaroubi
  et~al.}{2012}]{Zaroubi2012}
Zaroubi S.,  Zaroubi  Saleem 2012, \mn@doi [ASSL]
  {10.1007/978-3-642-32362-1_2}, 396, 45

\makeatother
\end{thebibliography}

\appendix

\section{Escape fractions for fixed masses, metallicities, and
  sSFR} \label{fescfixed.app}

To follow up on the discussion in \Sec{fesc_regulation.sec} on how the
evolution of the global $\fesc$ with redshift is driven by an
evolution in $\sSFRmax$, we show here how it is not driven by an
evolution in galaxy mass, metallicity, or $\sSFRten$. This is already
quite clear from the redshift evolution of escape fractions in Figures
\ref{fesc_hist_Mvir.fig}-\ref{fesc_halo_hist_sSFR100.fig} but we show
it here for completeness.

\begin{figure}
  \centering
  \includegraphics[width=0.5\textwidth]
    {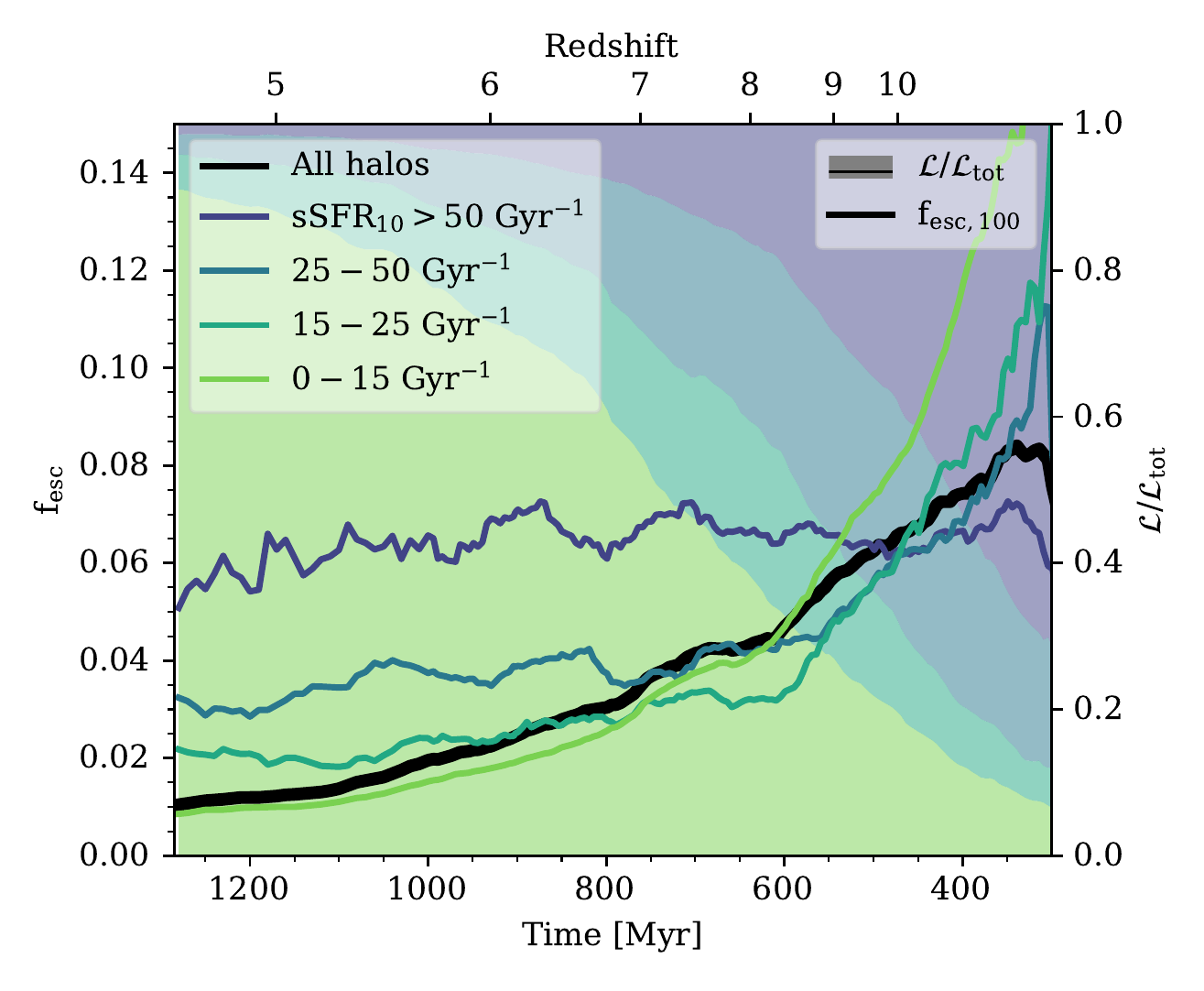}
  \caption
  {\label{fesc_fixedSFR.fig} Global escape fraction averaged over 100
    Myrs ($\feschundred$) versus redshift, in solid curves, for all
    halos in \sphinxtw{} and for bins of fixed $\sSFRten$, as
    indicated in the legend. The shaded regions indicate the fractions
    of \Lyc{} photons intrinsically produced in these fixed $\sSFRten$
    bins, and show that the \Lyc{} production is dominated by
    decreasing $\sSFRten$ with decreasing redshift. For the higher end
    of $\sSFRten$, halos with fixed $\sSFRten$ have fairly constant
    mean escape fractions, but at the lower end the escape fraction
    still decreases with redshift for fixed $\sSFRten$, indicating
    some but not quite dominant redshift-regulation of $\fesc$ via
    $\sSFRten$.}
\end{figure}

We show in \Fig{fesc_fixedSFR.fig} how the escape fraction evolves for
different bins in $\sSFRten$ (whereas we used $\sSFRmax$ in
\Fig{fesc_fixedSFR.fig} in the main text). For the highest two bins,
i.e. with $\sSFRten\ge25 \ \Gyrm$, and ignoring the noise in
$\feschundred$ due to small numbers of halos in these ranges, the
escape fraction does remain fairly constant with redshift, and there
is a significant difference of $\feschundred\approx0.06$ for the
highest $\sSFRten$ bin and $\feschundred\approx0.035$ for the
second-highest bin. Also, with decreasing redshift, the total \Lyc{}
luminosity goes from being dominated by high to lower $\sSFRten$
halos, as indicated by the shaded regions. However, for halos with
$\sSFRten < 25 \ \Gyrm$, which start to dominate the \Lyc{} emissivity
below a redshift of about $8$, the escape fraction does not stay
constant with redshift for a fixed $\sSFRten$, but rather decreases in
more or less the same way as the global escape fraction in black. A
similar plot for the longer-timescale $\sSFRc$ (not shown) shows an
even more variability in $\feschundred$ with redshift for fixed
$\sSFRc$-bins. Therefore, as we argue in the main text, $\sSFRmax$ is
a much better measure of the intensity of star formation and feedback
than $\sSFRten$ or $\sSFRc$, as it keeps memory of recent starbursts.

\begin{figure}
  \centering
  \includegraphics[width=0.5\textwidth]
    {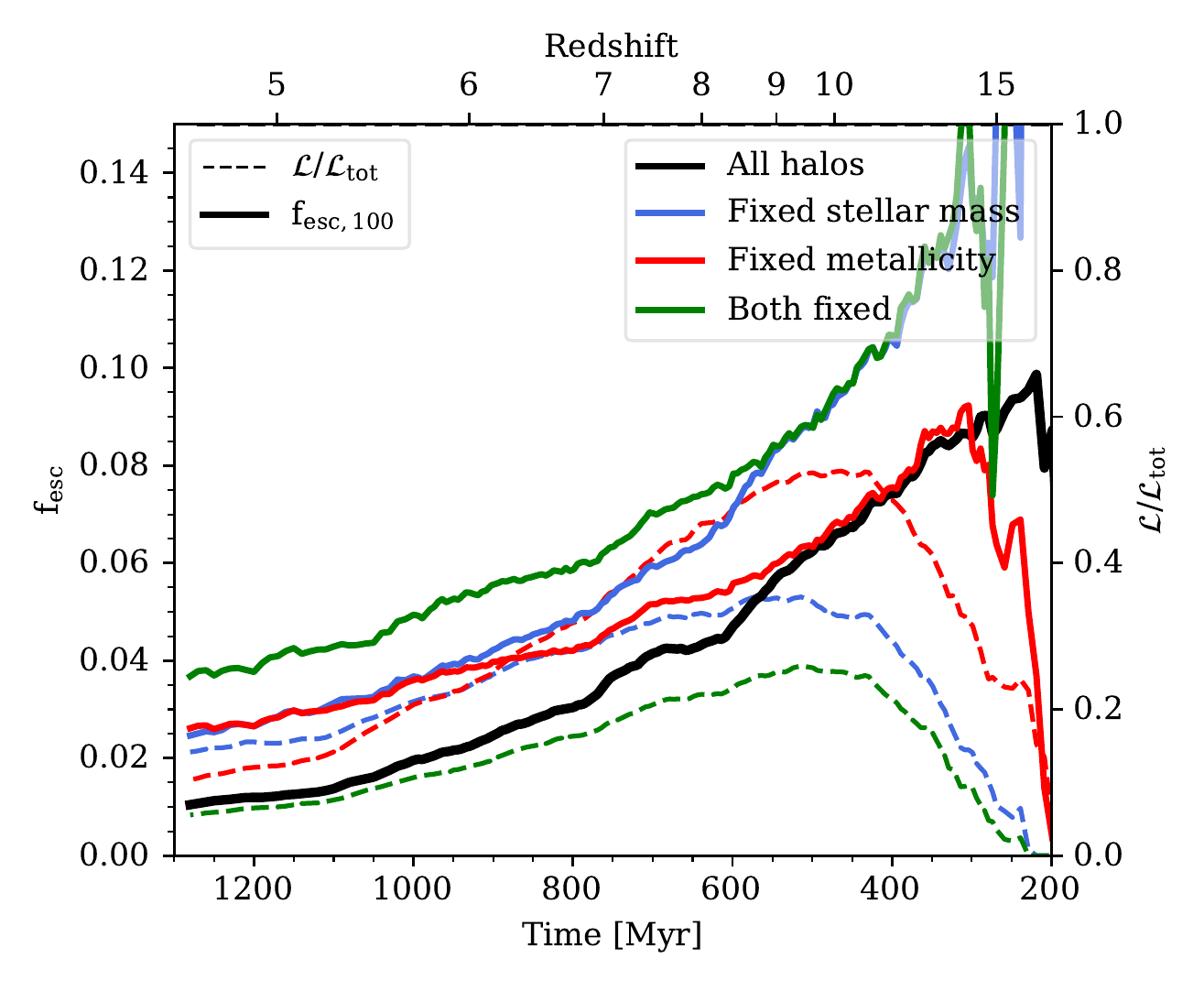}
  \caption
  {\label{fesc_fixMZ.fig} Global escape fractions in \sphinxtw{}
    averaged over 100 Myrs ($\feschundred$), in solid, for all halos
    and at fixed halo mass ($3 \times 10^6 - 3 \times 10^7 \ \Msun$)
    and/or halo (luminosity-weighted stellar) metallicity
    ($\Zhalo=10^{-4} - 4 \times 10^{-4}$), as indicated in the
    legend. The dashed curves show the fractions of the total number
    of \Lyc{} photons intrinsically produced by halos in these fixed
    mass and metallicity ranges. Fixing either halo mass or
    metallicity gives consistently larger $\feschundred$ than in the
    unfiltered case, simply due to the chosen ranges corresponding to
    high escape fractions. Fixing the metallicity reduces the
    redshift-evolution of $\feschundred$ somewhat more than fixing the
    halo mass, indicating that evolving metallicity has a stronger
    regulating effect on the escape fraction than halo mass. However,
    even for fixed metallicity the escape fraction is far from
    non-evolving with redshift, so halo mass and metallicity, alone or
    together, do not regulate the decrease of global escape fraction
    with redshift.}
\end{figure}

We then show a similar analysis in \Fig{fesc_fixMZ.fig} for fixed
galaxy mass and metallicity. The blue solid curve shows $\feschundred$
as a function of redshift for all halos in the stellar mass range
$3\times 10^6 \ \Msun < \Mstar < 3\times 10^7 \ \Msun$. The dashed
blue curve shows the relative contribution of those same halos to the
total intrinsic \Lyc{} emissivity. The escape fraction decreases with
redshift for this fixed mass-range, similarly to the the global one
for all halos (solid black curve), though the global one is always
lower owing to the ``peak'' in $\fesc$ for the selected mass range in
\Fig{fesc_hist_Mstar_all.fig}. The similar slope of the blue and black
curves demonstrates that the decreasing escape fraction with redshift
is \textit{not} regulated by an evolution towards more massive
galaxies -- rather it suggests that the mass-evolution has very little
effect on driving the evolution of the global escape
fraction. Similarly, the red solid curve shows $\feschundred$ for
halos with luminosity-weighted stellar metallicities in the range
$10^{-4} < \Zhalo < 4 \times 10^{-4}$. For this fixed metallicity
range, we also see a decrease in $\feschundred$ with
redshift. Although the drop is significant, it is not as strong as for
fixed galaxy mass, suggesting that metallicity plays a somewhat
stronger role than galaxy mass in driving a decreasing $\fesc$ with
redshift, though clearly it is not the dominant driver. The solid
green curve finally shows $\feschundred$ for halos fixed to both the
mass and metallicity ranges noted above. Once again we retrieve an
escape fraction that evolves very strongly with redshift. This means
that evolving galaxy masses or metallicities are not the dominant
factors driving a decreasing escape fraction with redshift, neither
alone nor together. We note that we have played extensively with
varying the fixed ranges in both galaxy mass and metallicity and we
always find a similar evolution of decreasing $\fesc$ with redshift,
so we are confident that neither galaxy mass nor metallicity is a
major factor in regulating the $\fesc$ evolution in \sphinx.

We note that a different conclusion was reached by \citet{Yoo2020},
studying $\fesc$ from idealised disc galaxies simulated with the same
code as \sphinx{} and similar methods. They found metallicity to be a
much stronger regulator of $\fesc$ than both mass and $\sSFR$, in
conflict with our results. However, we argue that i) the galaxies in
\citet{Yoo2020} are likely more relevant to galaxies with well defined
disks than the messy and bursty high-z galaxies found in the EoR and
ii) the $\sSFR \ (\approx 0.1-1 \,{\rm Gyr^{-1}})$ was not tested at
extreme enough values in \citet{Yoo2020} to reach the regime where
$\fesc$ becomes high in \sphinx{}.

\section{Volume size and SED variations} \label{bpass.app}

We collect here for reference a comparison of our main results for the
different \sphinx{} runs, i.e. the fiducial \sphinxtw{} run reported
on in the main text, the smaller-volume but otherwise identical
\sphinxten{} run and the smaller-volume and more luminous SED
\sphinxtenf{} run. We remind the reader that due to their smaller
volumes, the results from \sphinxten{} and \sphinxtenf{} are much
noisier than those from \sphinxtw{}. 

\begin{figure}
  \centering
  \includegraphics[width=0.43\textwidth]
    {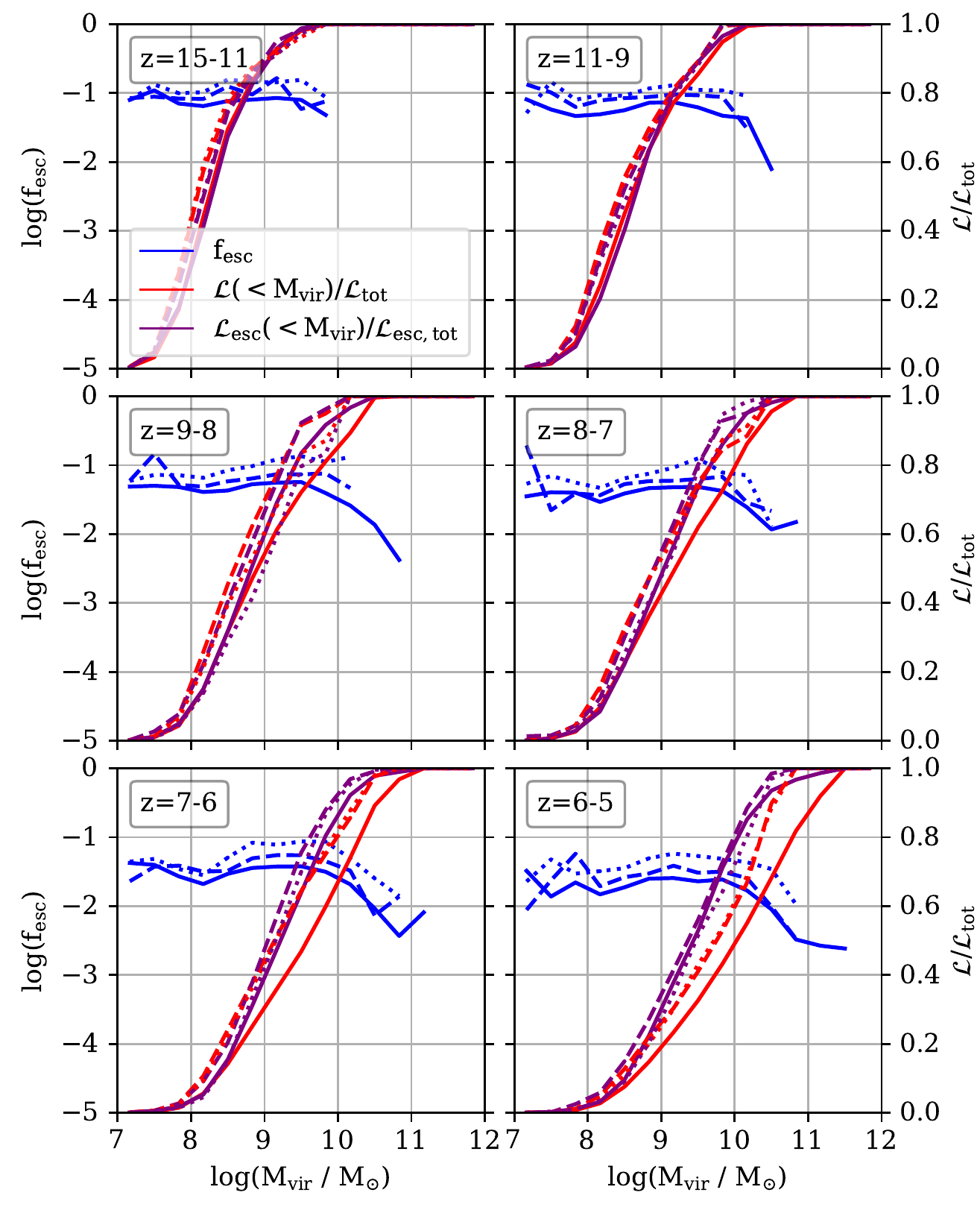}
  \caption
  {\label{fesc_hist_Mvir_all_10Mpc.fig}$\fesc$ versus $\Mvir$
    correlation, cumulative intrinsic emissivity, and cumulative
    escaping \Lyc{} emissivity (blue, red, and purple curves,
    respectively) at different redshifts, same as
    \Fig{fesc_hist_Mvir.fig} but here compared for \sphinxtw{}, the
    smaller volume \sphinxten{}, and the smaller volume plus more
    luminous SED \sphinxtenf{} simulations (solid, dashed, and dotted
    curves, respectively). The smaller volume case shows a hint of
    higher $\fesc$ than in \sphinxtw{}, especially at the highest
    redshifts. The more luminous SED increases $\fesc$ even further
    and roughly equally for any halo mass. Intrinsic luminosities vary
    significantly between the two volume sizes but escaping
    luminosities much less so due to the low $\fesc$ of
    massive halos.}
\end{figure}

\begin{figure}
  \centering
  \includegraphics[width=0.43\textwidth]
    {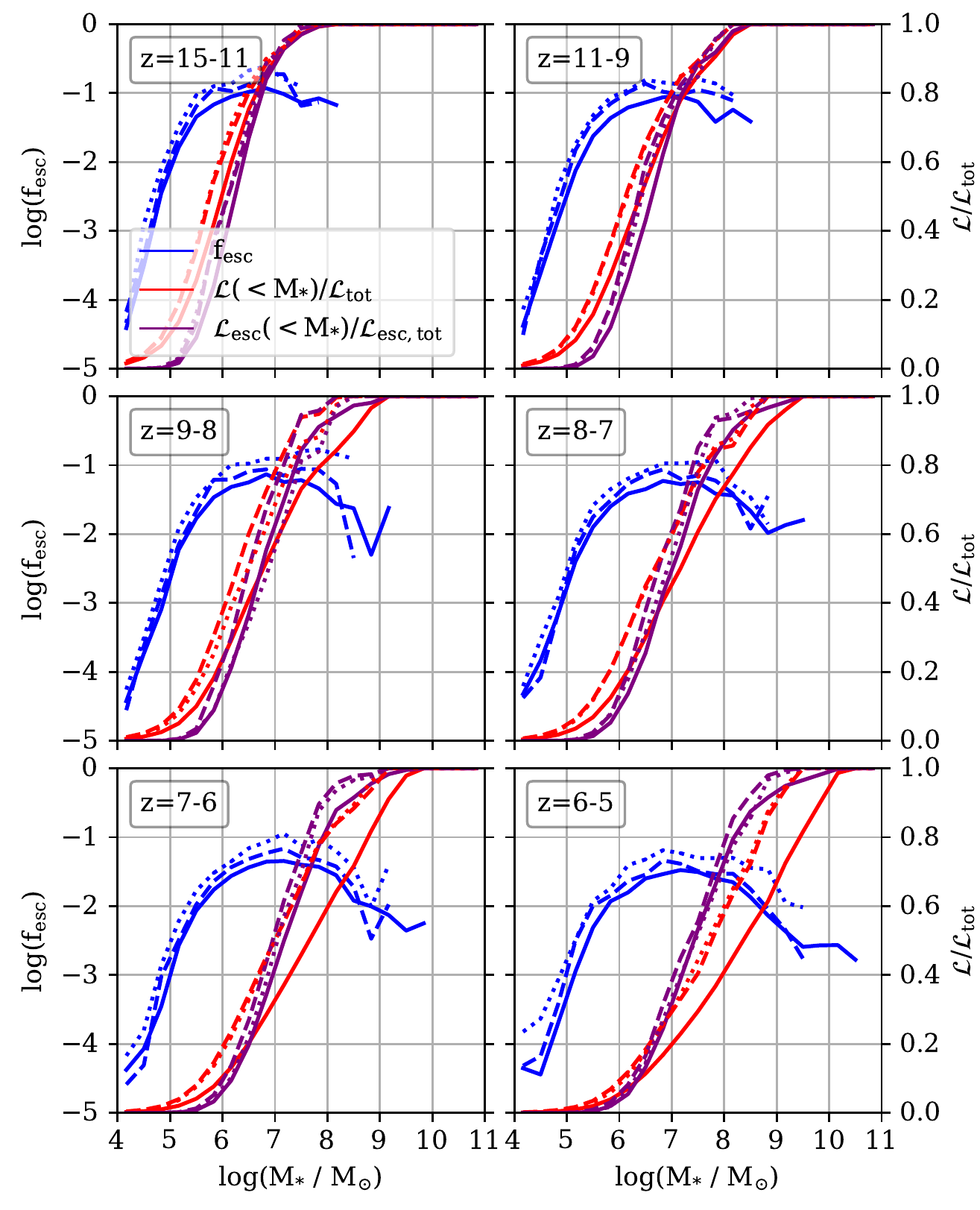}
  \caption
  {\label{fesc_hist_Mstar_all_10Mpc.fig}As
    \Fig{fesc_hist_Mvir_all_10Mpc.fig} but for galaxy mass
    $\Mstar$. The smaller volume shows higher $\fesc$ than in
    \sphinxtw{} at high-z. The more luminous SED increases $\fesc$
    overall, i.e. at any galaxy mass. Intrinsic luminosities vary
    significantly between the two volume sizes but escaping
    luminosities much less so due to the low $\fesc$ of massive
    galaxies.}
\end{figure}

\begin{figure}
  \centering
  \includegraphics[width=0.43\textwidth]
    {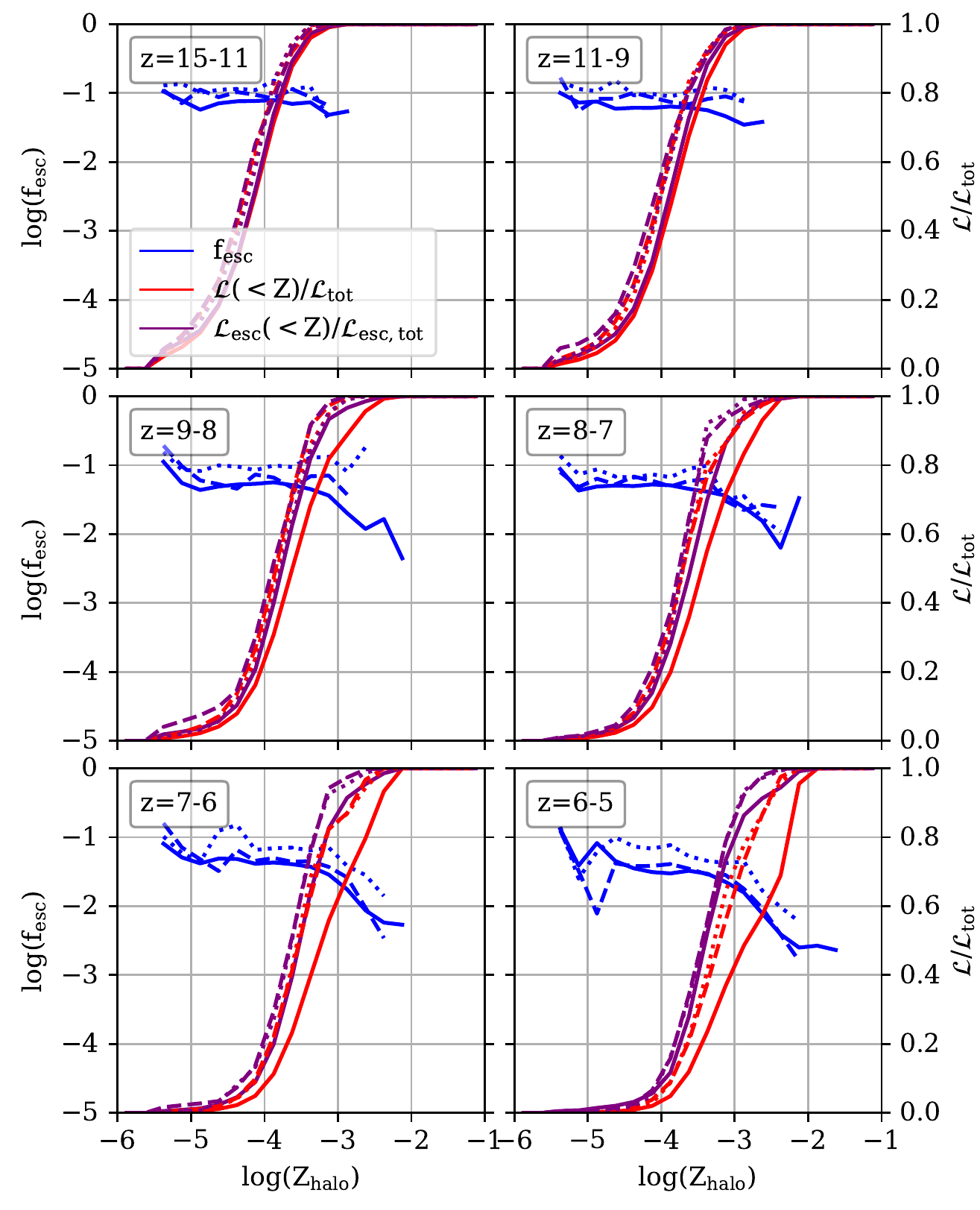}
  \caption
  {\label{fesc_hist_halozlw_all_10Mpc_lg.fig}As
    \Fig{fesc_hist_Mvir_all_10Mpc.fig} but for galaxy metallicity. The
    smaller volume shows higher $\fesc$ than in \sphinxtw{} at
    high-z. The more luminous SED increases $\fesc$ overall,
    i.e. at any metallicity. Intrinsic luminosities vary significantly
    between the two volume sizes but escaping luminosities much less
    so due to the low $\fesc$ of metal-rich galaxies.}
\end{figure}

\begin{figure}
  \centering
  \includegraphics[width=0.43\textwidth]
    {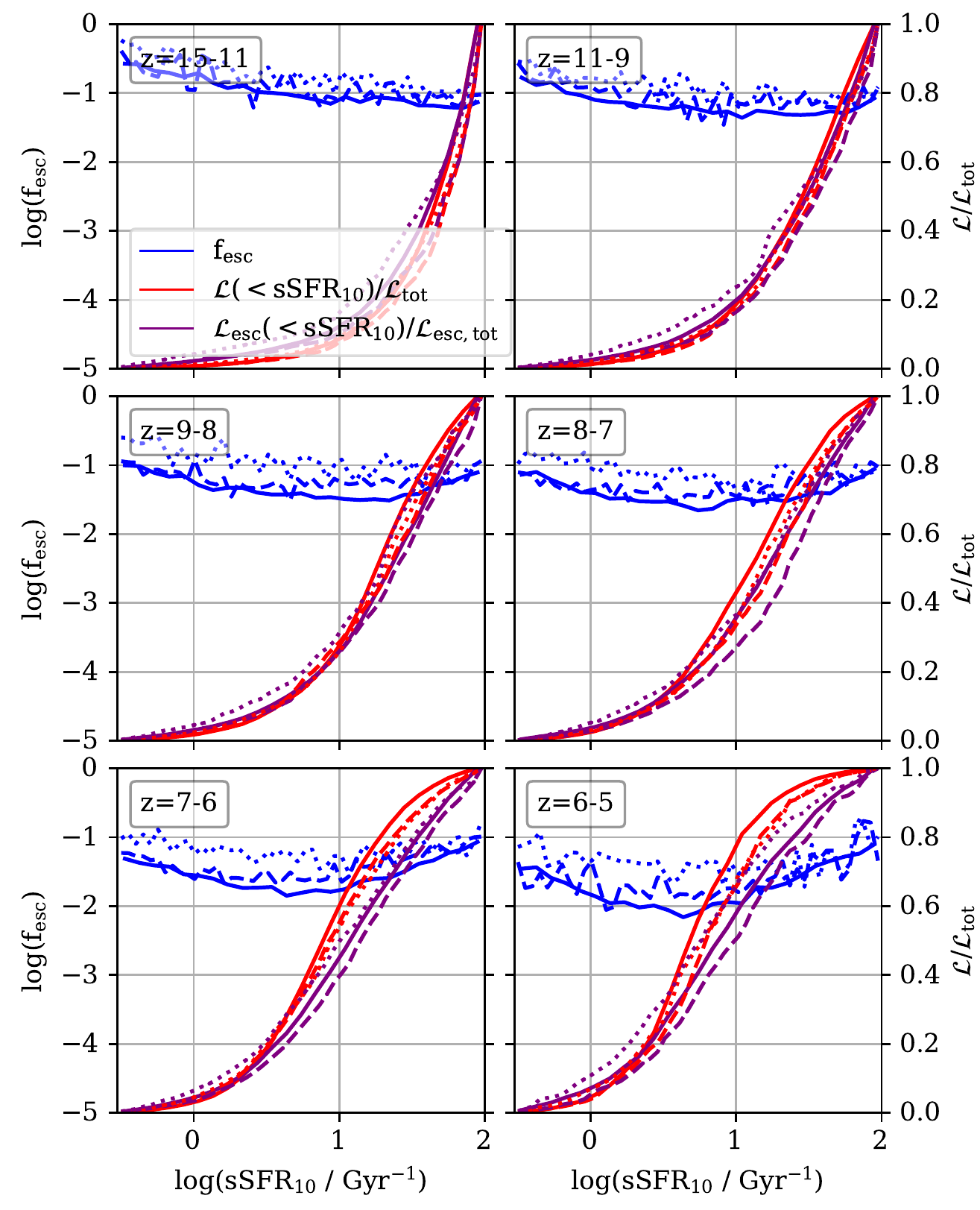}
  \caption
  {\label{fesc_hist_ssfr10_all_10cMpc_lg.fig} As
    \Fig{fesc_hist_Mvir_all_10Mpc.fig} but for $\sSFRten$. The smaller
    volume shows higher $\fesc$ than in \sphinxtw{} for intermediate
    $\sSFRten$, likely due to the galaxies being less massive. The
    more luminous SED increases $\fesc$ overall but least
    significantly at the highest $\sSFRten$. Here escaping
    luminosities vary somewhat between the runs, especially at the
    lowest-z, with the escaping luminosity notably shifted to moderate
    $\sSFRten$ in the case of the more luminous SED model.}
\end{figure}

\begin{figure}
  \centering
  \includegraphics[width=0.43\textwidth]
    {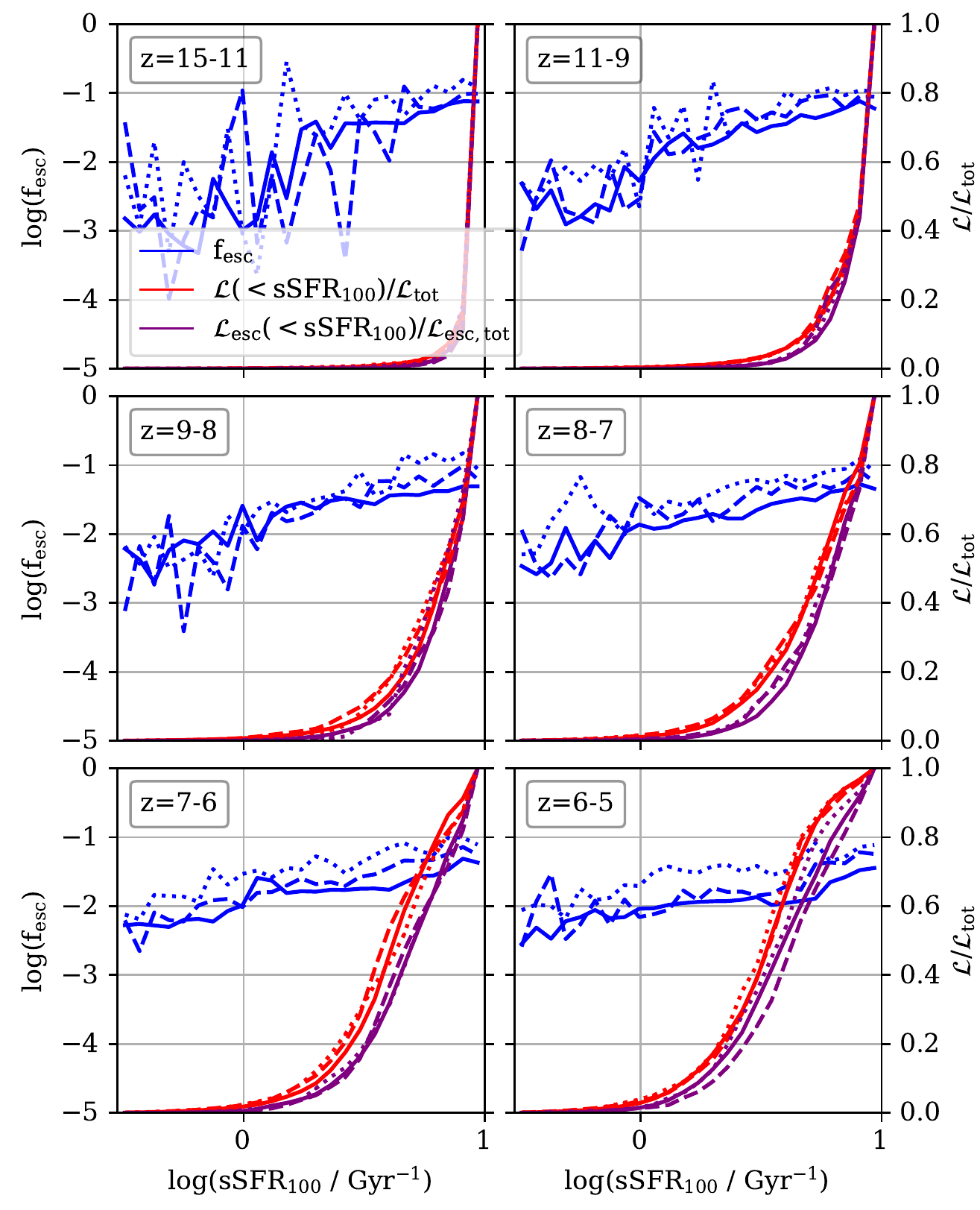}
  \caption
  {\label{fesc_hist_ssfr10_all_100cMpc_lg.fig}As
    \Fig{fesc_hist_Mvir_all_10Mpc.fig} but for $\sSFRc$. The luminous
    SED increases $\fesc$ overall, but there is small difference in
    intrinsic or escaping emissivities.}
\end{figure}

\begin{figure}
  \centering
  \includegraphics[width=0.43\textwidth]
    {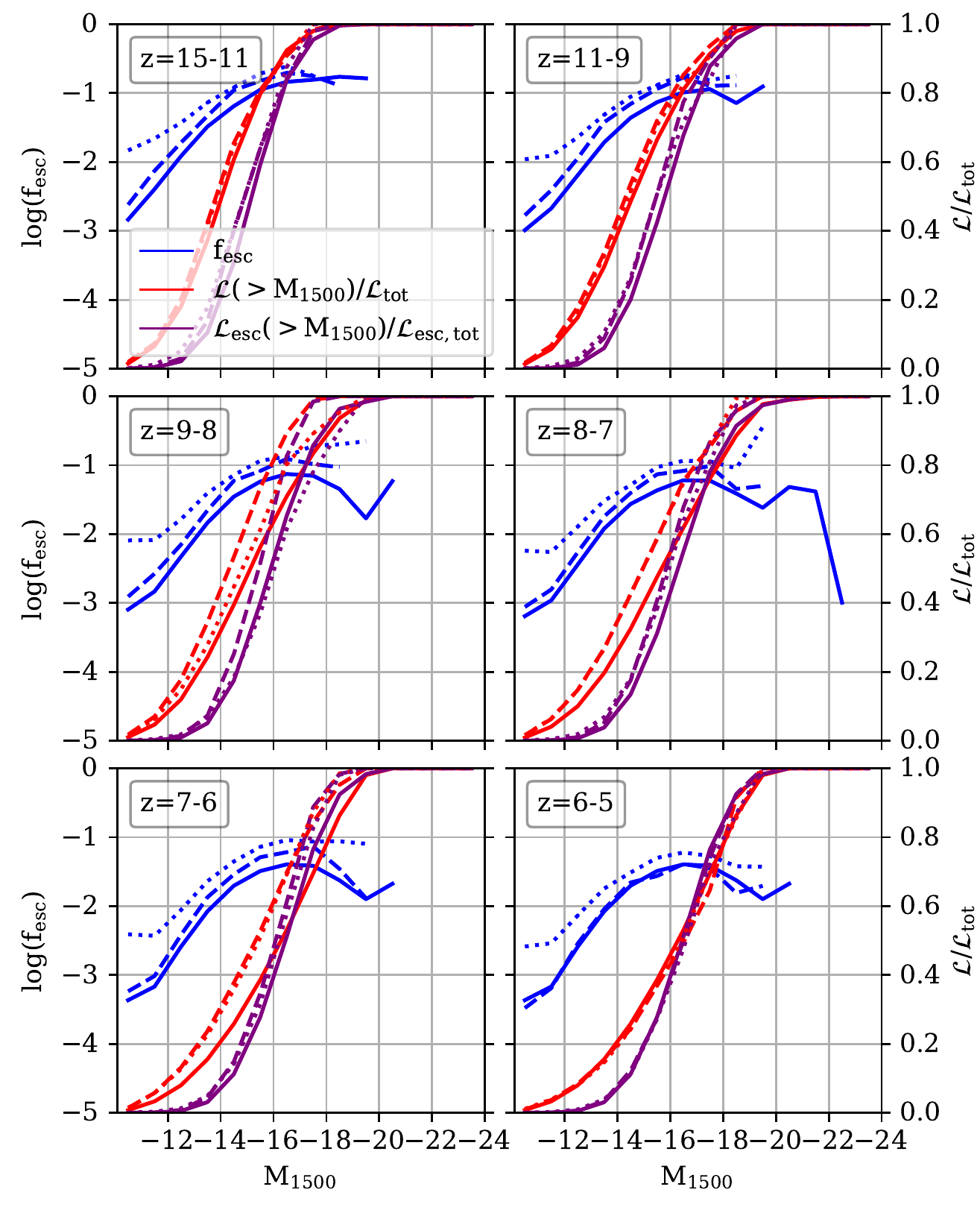}
  \caption
  {\label{fesc_hist_matt_all_10cMpc_lg.fig}As
    \Fig{fesc_hist_Mvir_all_10Mpc.fig} but for $\MFifteen$.  The
    smaller volume has higher $\fesc$ than \sphinxtw{} at
    high-z. The luminous SED increases $\fesc$ overall, but there is
    small difference in intrinsic or escaping.}
\end{figure}

\begin{figure}
  \centering
  \includegraphics[width=0.43\textwidth]
    {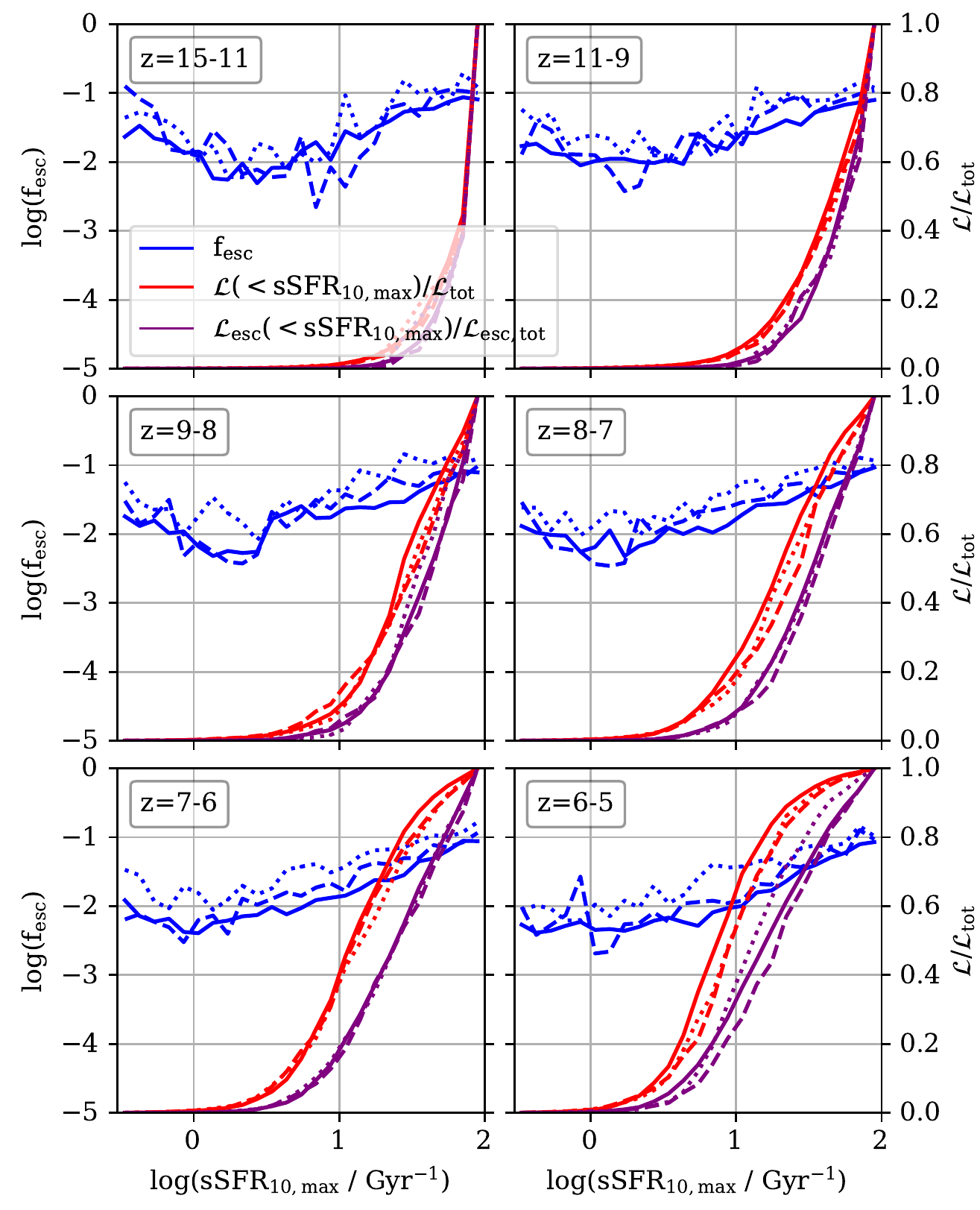}
  \caption
  {\label{fesc_hist_ssfr10max50_all_10cMpc_lg.fig} As
    \Fig{fesc_hist_Mvir_all_10Mpc.fig} but for $\sSFRmax$. The data is
    noisy for the small-volume simulations, especially at low
    $\sSFRmax$. The more luminous SED increases $\fesc$ overall. For
    all runs, the $\fesc$-$\sSFRmax$ relation is fairly non-evolving
    with redshift (except for low $\sSFRmax$ where little \Lyc{}
    radiation is emitted anyway), implying that the evolving global
    $\fesc$ is driven by the shift to lower $\sSFRmax$ in the smaller
    volumes, just as in \sphinxtw. }
\end{figure}

We first compare in Figures
\ref{fesc_hist_Mvir_all_10Mpc.fig}-\ref{fesc_hist_ssfr10max50_all_10cMpc_lg.fig}
the dependence of $\fesc$ on halo mass, galaxy mass, metallicity,
$\sSFRten$, $\sSFRc$, $\MFifteen$, and $\sSFRmax$. Here we have made
sure to use only the subset of snapshots that exist in all runs,
i.e. the same number of snapshots and at the same redshifts, so not as
to introduce any bias in the comparison.

First focusing on volume size, for mass, metallicity, and magnitude,
the escape fraction is generally slightly higher in the smaller-volume
\sphinxten{} run than in \sphinxtw{}, especially at the higher
redshifts. We are not sure why but speculate that this hints at a
small but non-negligible environmental effect on $\fesc$,
i.e. galaxies of a given mass and metallicity residing in massive
environments tend to have lower escape fractions. We will explore this
in future work. The larger difference between the volume sizes is that
the intrinsic \Lyc{} emission is shifted towards significantly more
massive and metal-rich halos. Because these halos have small $\fesc$,
this translates to a much smaller difference in the \textit{escaping}
emission, shown in purple curves. For $\sSFRten$ and $\sSFRmax$, the
intrinsic \Lyc{} emission is shifted to higher sSFRs in \sphinxten{}
compared to \sphinxtw{} and $\fesc$ is noticeably higher in
\sphinxten{} at intermediate $\sSFRten$. Both are effects of the
relative lack of massive galaxies with are luminous but have low
escape fractions. This difference is sort of smoothed out for the
longer-term $\sSFRc$, where $\fesc$ is just overall slightly higher
for \sphinxten{}.

Then focusing on the different SED models and comparing \sphinxten{}
and \sphinxtenf{}, the more luminous SED model leads to $\fesc$ being
overall higher in \sphinxtenf{} for any halo property correlation
considered, i.e. the correlation as a whole shifts to higher $\fesc$,
due to more \Lyc{} photons being emitted at ages of $\ga 5$ Myrs in
the more luminous model (see Figures \ref{SEDs.fig} and
\ref{fesc_hist_age_age.fig}). The cumulative intrinsic \Lyc{} fraction
tends to be very similar for the two models and mostly this is also
the case for the \textit{escaping} emission.

To summarise, there is no significant or surprising difference in the
dependence of $\fesc$ on halo properties when changing the volume size
or SED model, except for an unexplained slight overall increase in
$\fesc$ with a smaller volume, which we speculate may be due to
environmental effects.

\begin{figure*}
  \centering
  \includegraphics[width=0.75\textwidth]
    {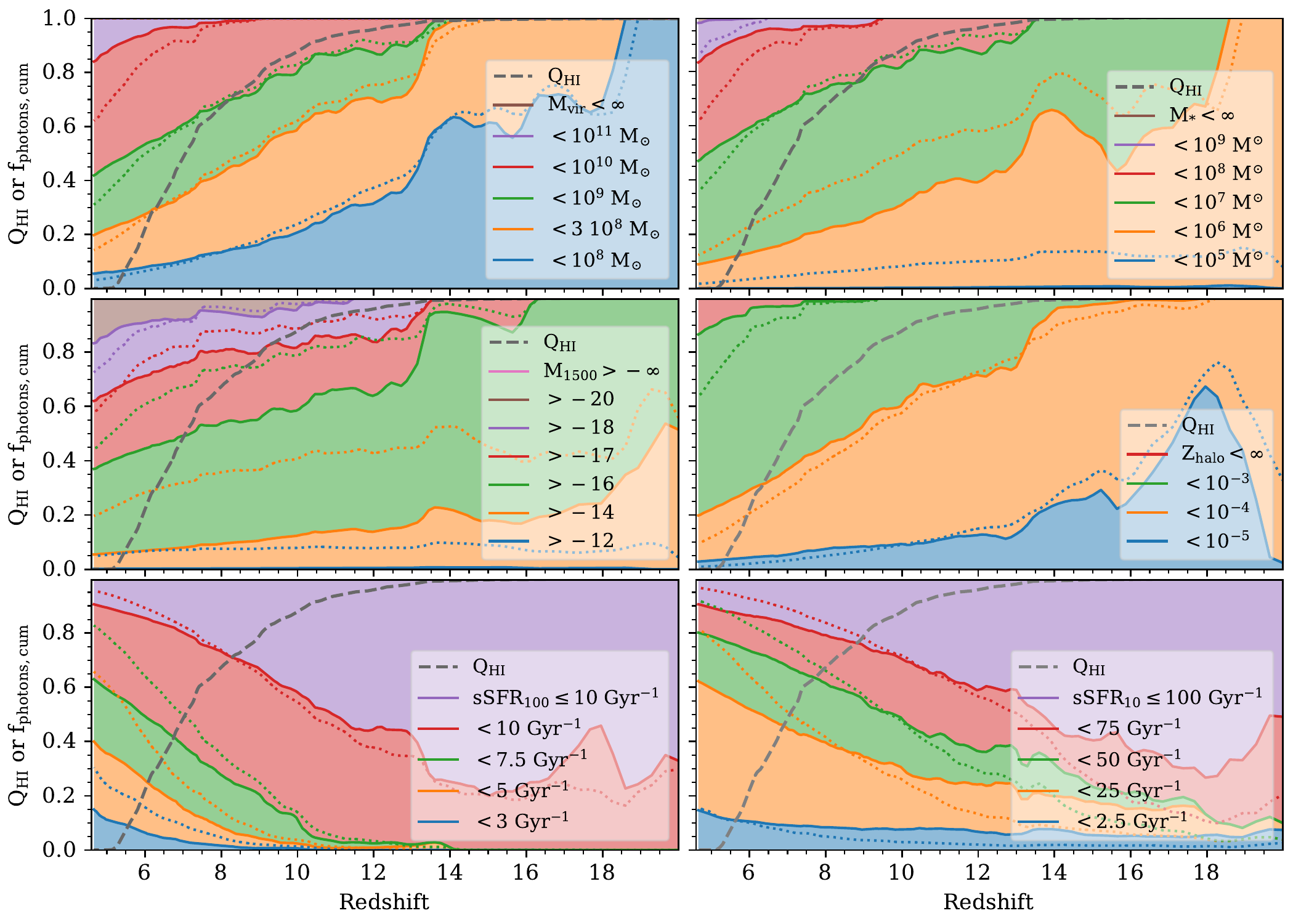}
  \caption
  {\label{contributions_0507.fig}Contributions of different halos to
    reionization in the 10 cMpc \sphinxten{} volume (with BPASS
    v2.2.1). Compared to the larger-volume but otherwise identical
    \sphinxtw{} run (\Fig{contributions.fig}), the intrinsic \Lyc{}
    emission (dotted curves) is shifted to lower masses, lower
    metallicities, and dimmer magnitudes. However, since the 'extra'
    massive, metal-rich and bright galaxies represented in \sphinxtw{}
    but not in \sphinxten{} tend to have low escape fractions, the
    escaping luminosities are significantly less shifted. For sSFRs,
    both the intrinsic and escaping emission is very similar between
    \sphinxten{} and \sphinxtw.}
\end{figure*}

\begin{figure*}
  \centering
  \includegraphics[width=0.75\textwidth]
    {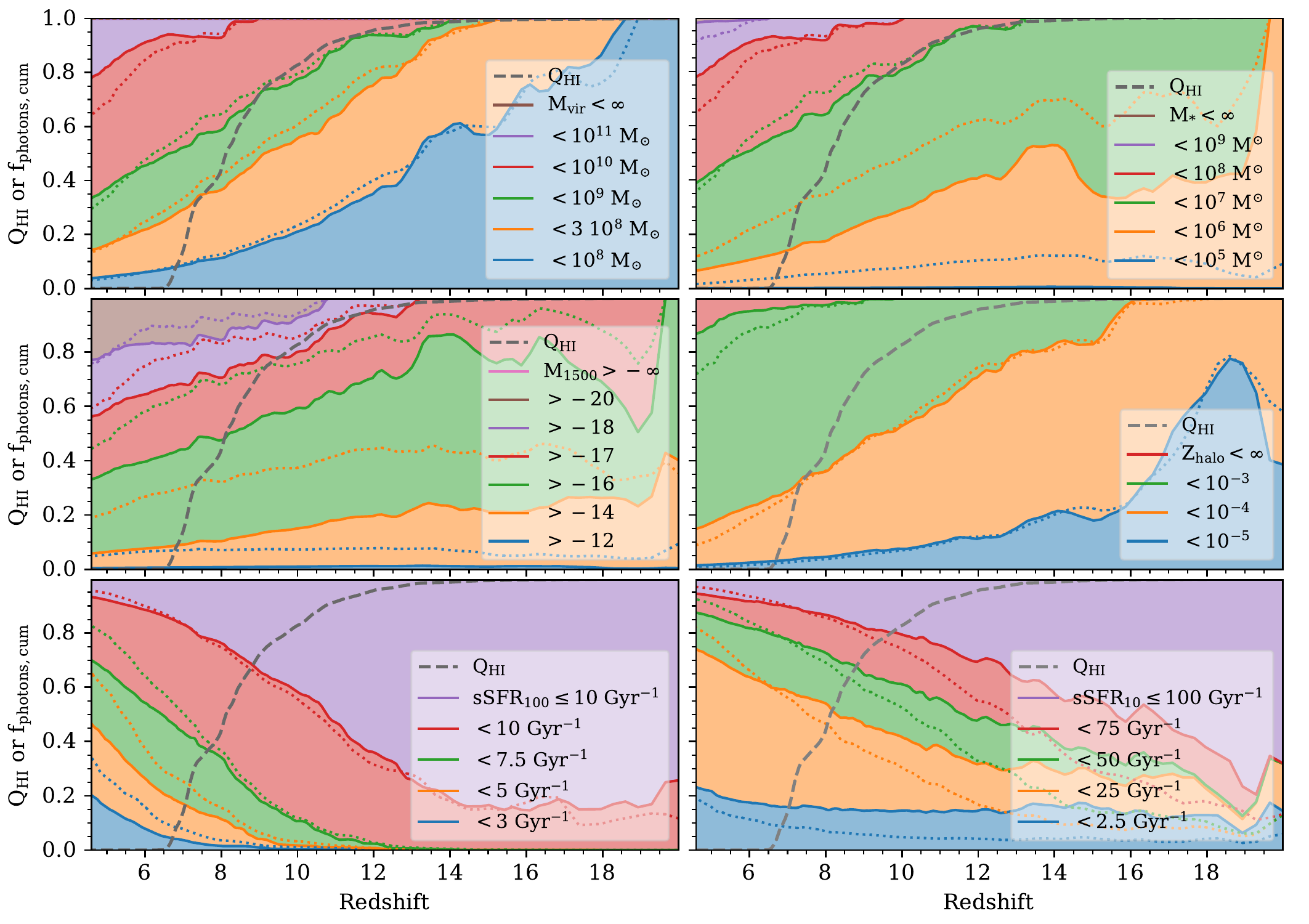}
  \caption
  {\label{contributions_0508.fig}Contributions of different halos to
    reionization in the 10 cMpc \sphinxtenf{} volume (with BPASS
    v2.0). Compared to \sphinxten{} (\Fig{contributions_0507.fig}),
    which is identical save for the SED version, both intrinsic and
    escaping contributions to the \Lyc{} budget by $z=5$ are shifted
    somewhat to more massive, brighter, more metal-rich, and less
    star-forming galaxies. However, taking into account that
    reionization finishes $\approx 1.5$ dex earlier in \sphinxtenf{},
    halo contributions by the end of reionization ($z=5.1$ in
    \sphinxten{}, $z=6.5$ in \sphinxtenf{}) are actually for the most
    part shifted very slightly in the opposite direction.}
\end{figure*}

We show in Figures \ref{contributions_0507.fig} and
\ref{contributions_0508.fig} the contributions to emitted and escaping
\Lyc{} radiation of halos of different properties in the
smaller-volume \sphinxten{} run and luminous SED \sphinxtenf{} run,
respectively. Qualitatively they are in both cases similar to
\sphinxtw{}. Keep in mind, again, that the data is significantly more
noisy than in \sphinxtw{}, especially at high redshift. Due to the
smaller volume, the intrinsic contribution in \sphinxten{} is shifted
a somewhat towards objects that are less massive, less metal-rich, and
dimmer. However, due to the low escape fractions of the massive,
bright, and metal-rich galaxies that are ``missing'' in \sphinxten{},
the shift is much smaller for escaping \Lyc{} contributions. The
effect of the more luminous SED model in \sphinxtenf{} is to shift
contributions slightly back to higher masses, metallicities, and
brighter galaxies, compared to \sphinxten{}, if taking the total
contribution to the \Lyc{} budget at $z\approx5$. However, if we
consider the contribution at the end of reionization and factor in that
reionization finishes at $z=5.1$ in \sphinxten{} but a dex and a half
earlier in \sphinxtenf{}, the contributions shift back again and
become fairly similar between the two simulations. Hence we conclude
that the contributions of halos to reionization are very similar
between the three simulations considered and for all the halo
properties considered.

\begin{figure}
  \centering
  \includegraphics[width=0.45\textwidth]
    {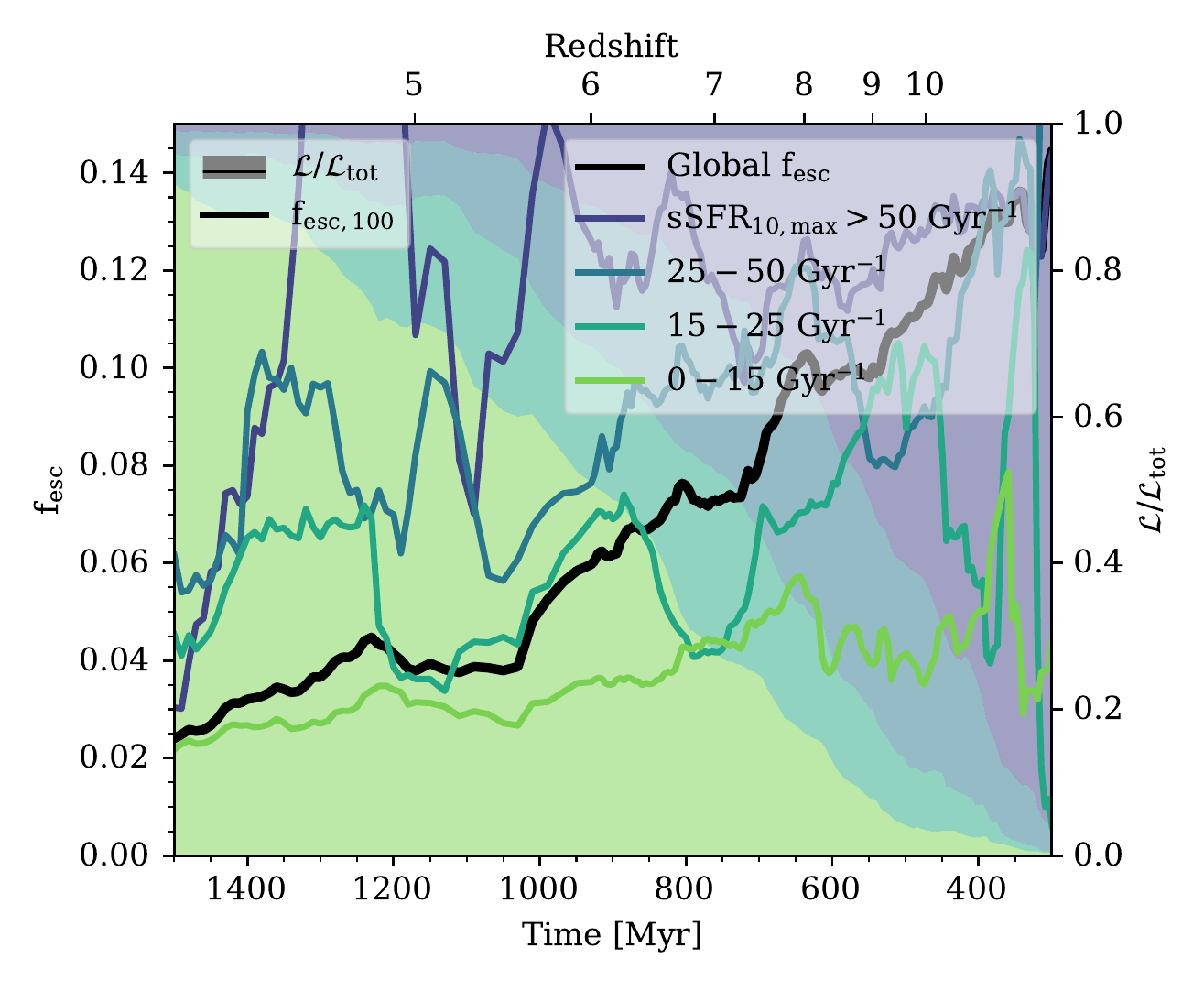}
  \caption
  {\label{fesc_global_halos_sSFR10maxfix_0508.fig} Global escape
    fractions in the \sphinxtenf{} run averaged over 100 Myrs
    ($\feschundred$) versus redshift, in solid, for all halos and at
    fixed bins in $\sSFRmax$, as indicated in the legend. The
    shaded regions indicate the fractions of the total \Lyc{}
    radiation intrinsically produced by halos in these fixed
    $\sSFRmax$ bins. This is the same as \Fig{fesc_fixedsSFRmax.fig}
    but for a smaller volume and the more \Lyc{}-luminous SED
    model. The data is much noisier than in
    \Fig{fesc_fixedsSFRmax.fig} (due to the small volume), but we
    still find a clear difference in $\fesc$ between the
    $\sSFRmax$-bins and the global evolution in $\fesc$ still appears
    driven mostly by this difference and the evolution in $\sSFRmax$.}
\end{figure}

We finally demonstrate that the evolution of the global escape
fraction in the \sphinxtw{} simulation is driven by decreasing
$\sSFRmax$, just as it is in \sphinxtw{}. We show in
\Fig{fesc_global_halos_sSFR10maxfix_0508.fig} a dissection of the
evolving global escape fraction in \sphinxtenf{} for halos with
different $\sSFRmax$, just like in \Fig{fesc_fixedsSFRmax.fig} for
\sphinxtw{}. Once again there is much more noise in the data for this
smaller volume simulation, especially for halo categories that contain
relatively few halos. Despite large fluctuations it is still fairly
clear that the decrease with redshift in the global $\fesc$ is largely
driven by an evolution where the \Lyc{} radiation is emitted from
halos with decreasing $\sSFRmax$ and hence decreasing $\fesc$. Our
conclusion -- that the decreasing $\fesc$ with redshift is driven by
an increasing fraction of galaxies with low $\sSFRmax$ with redshift
in the expanding Universe -- holds for both our simulations bracketing
the observational constraints on the reionization history of the
Universe.

\section{Resolution convergence} \label{resolution.app} 
\begin{figure}
  \centering
  \includegraphics[width=0.43\textwidth]
    {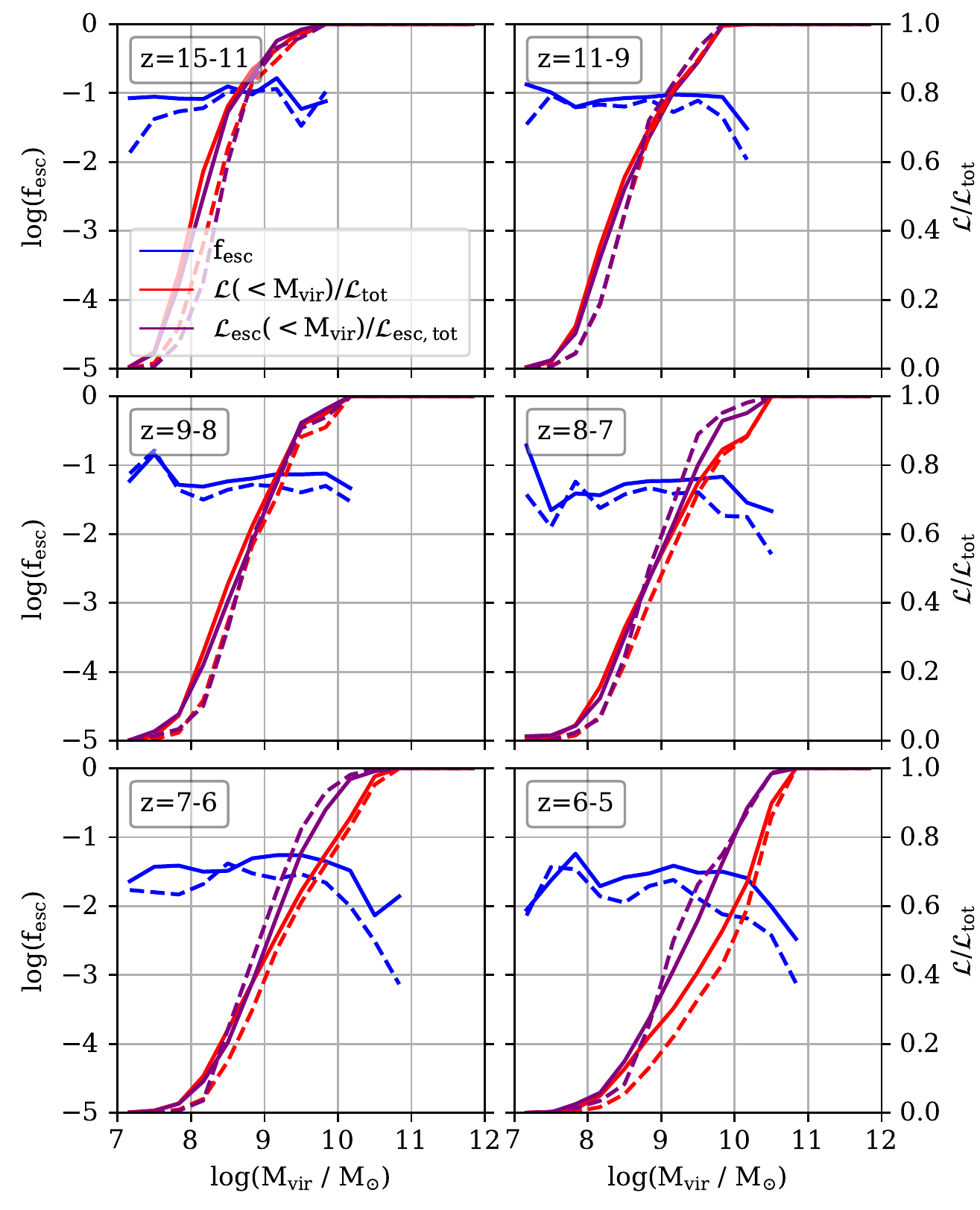}
  \caption
  {\label{fesc_hist_Mvir_all_resolution_lg.fig}Correlation of $\fesc$
    with $\Mvir$ , cumulative intrinsic emissivity, and cumulative
    escaping \Lyc{} emissivity (blue, red, and purple curves,
    respectively) at different redshifts, same as
    \Fig{fesc_hist_Mvir.fig} but here compared for \sphinxten{}, in
    solid curves, and an identical lower resolution run with double
    the minimum cell width at any point, shown in dashed curves.}
\end{figure}

\begin{figure}
  \centering
  \includegraphics[width=0.43\textwidth]
    {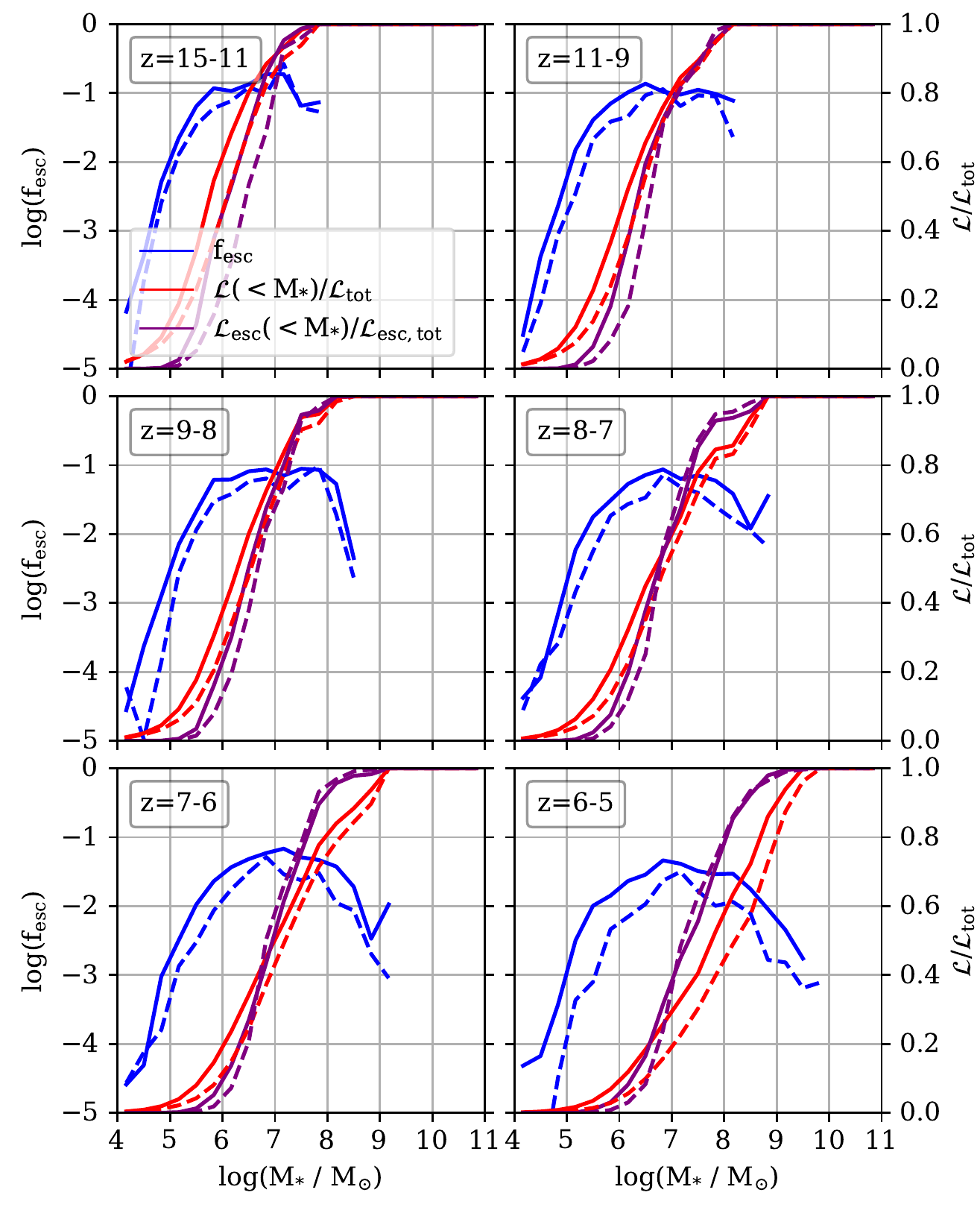}
  \caption
  {\label{fesc_hist_Mstar_all_resolution_lg.fig}As
    \Fig{fesc_hist_Mvir_all_resolution_lg.fig} but for $\fesc$ versus galaxy mass
    $\Mstar$.}
\end{figure}

\begin{figure}
  \centering
  \includegraphics[width=0.43\textwidth]
    {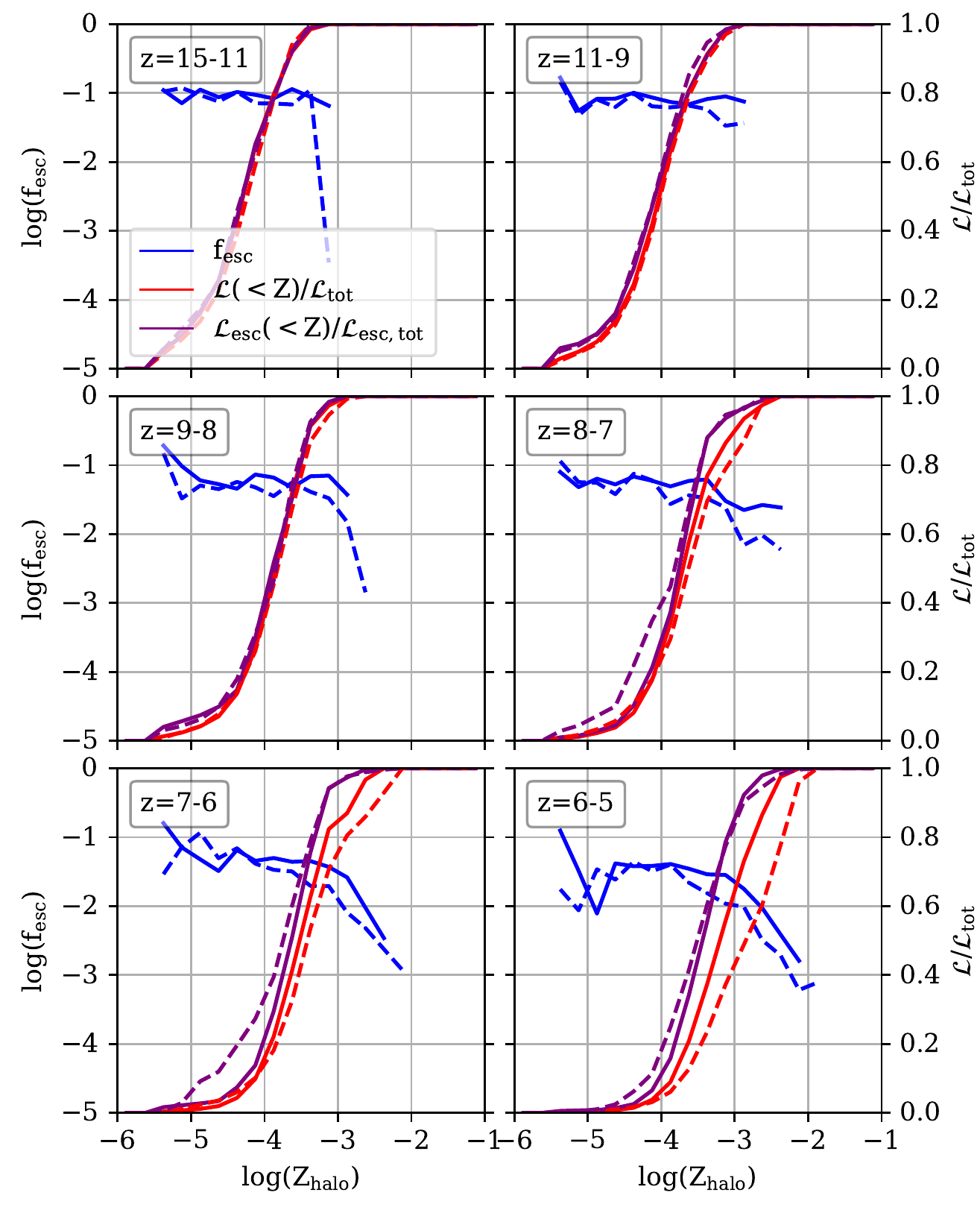}
  \caption
  {\label{fesc_hist_halozlw_all_resolution_lg.fig}As
    \Fig{fesc_hist_Mvir_all_resolution_lg.fig} but for $\fesc$ versus galaxy metallicity.}
\end{figure}

\begin{figure}
  \centering
  \includegraphics[width=0.43\textwidth]
    {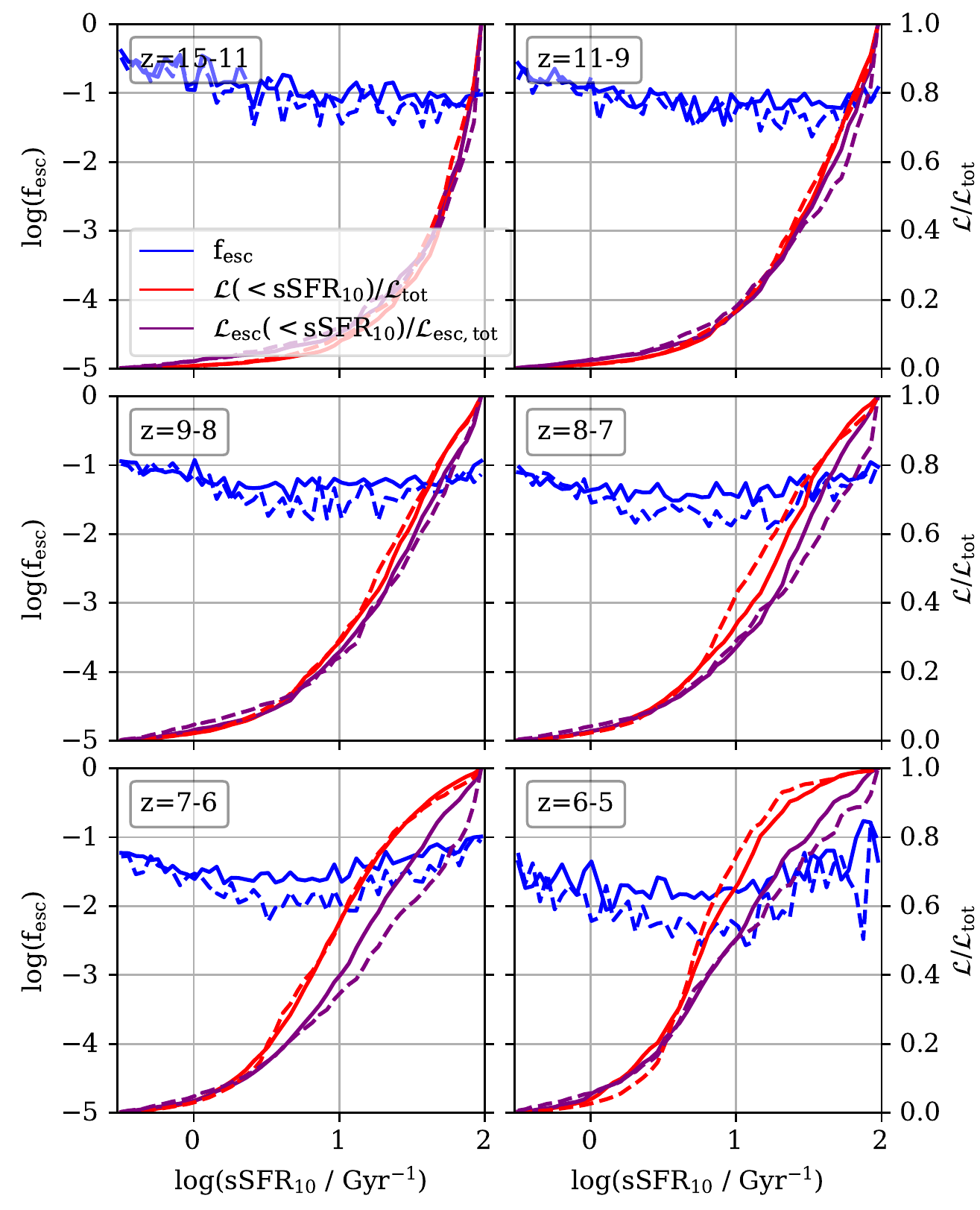}
  \caption
  {\label{fesc_hist_ssfr10_all_resolution_lg.fig} As
    \Fig{fesc_hist_Mvir_all_resolution_lg.fig} but for $\fesc$ versus $\sSFRten$.}
\end{figure}

\begin{figure}
  \centering
  \includegraphics[width=0.43\textwidth]
    {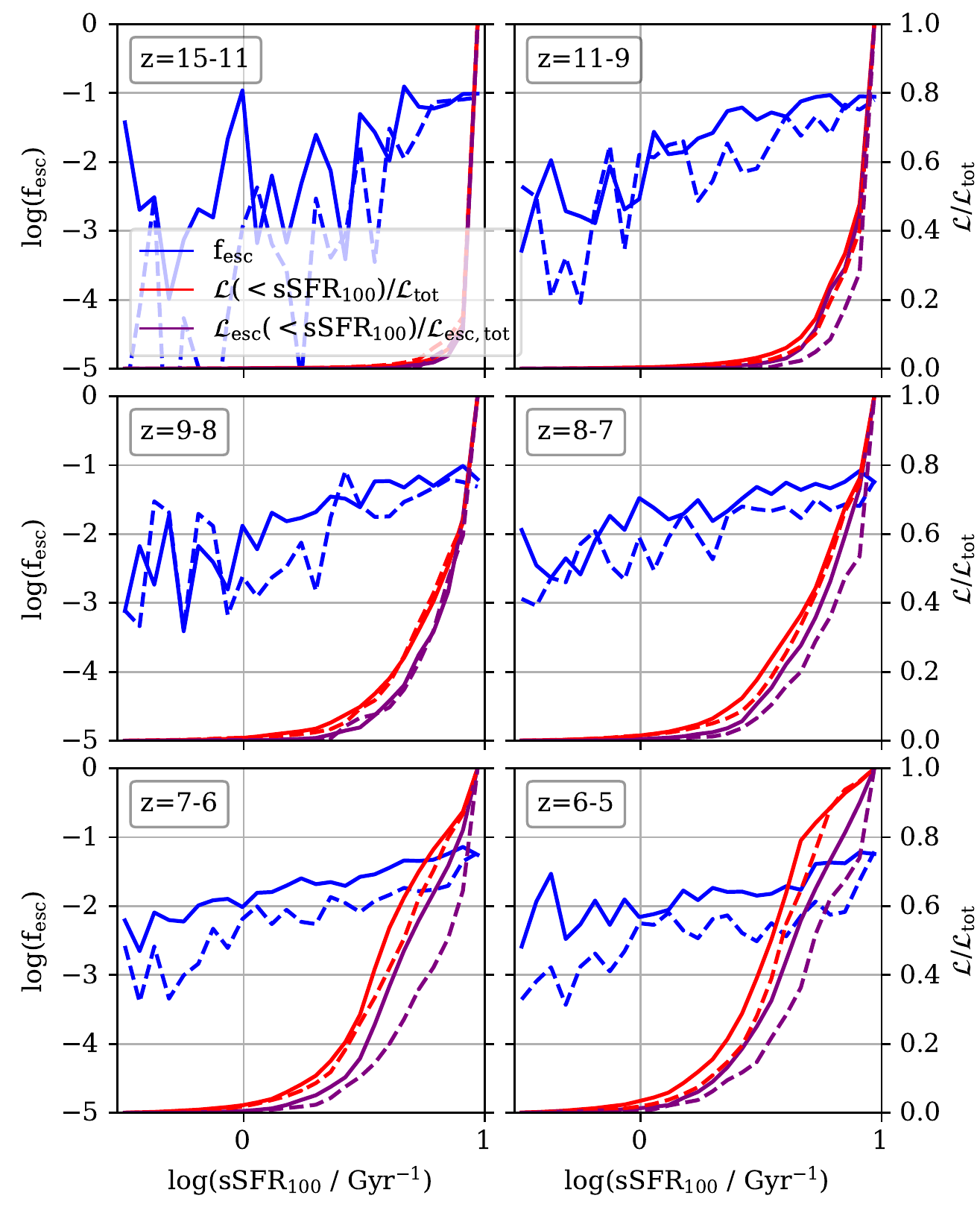}
  \caption
  {\label{fesc_hist_ssfr100_all_resolution_lg.fig}As
    \Fig{fesc_hist_Mvir_all_resolution_lg.fig} but for $\fesc$ versus $\sSFRc$.}
\end{figure}

\begin{figure}
  \centering
  \includegraphics[width=0.43\textwidth]
    {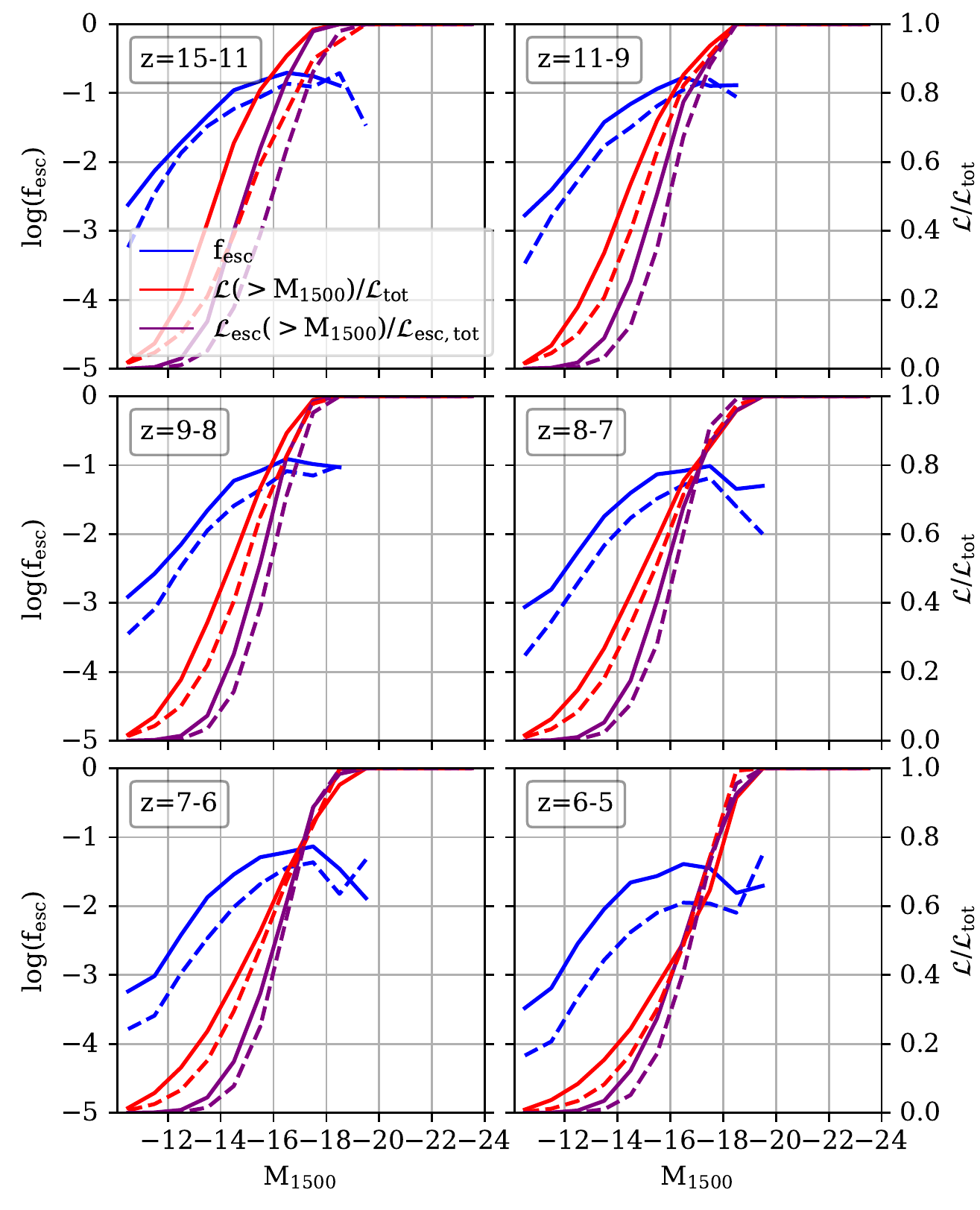}
  \caption
  {\label{fesc_hist_matt_all_resolution_lg.fig}As
    \Fig{fesc_hist_Mvir_all_resolution_lg.fig} but for $\fesc$ versus $\MFifteen$.}
\end{figure}

\begin{figure}
  \centering
  \includegraphics[width=0.43\textwidth]
    {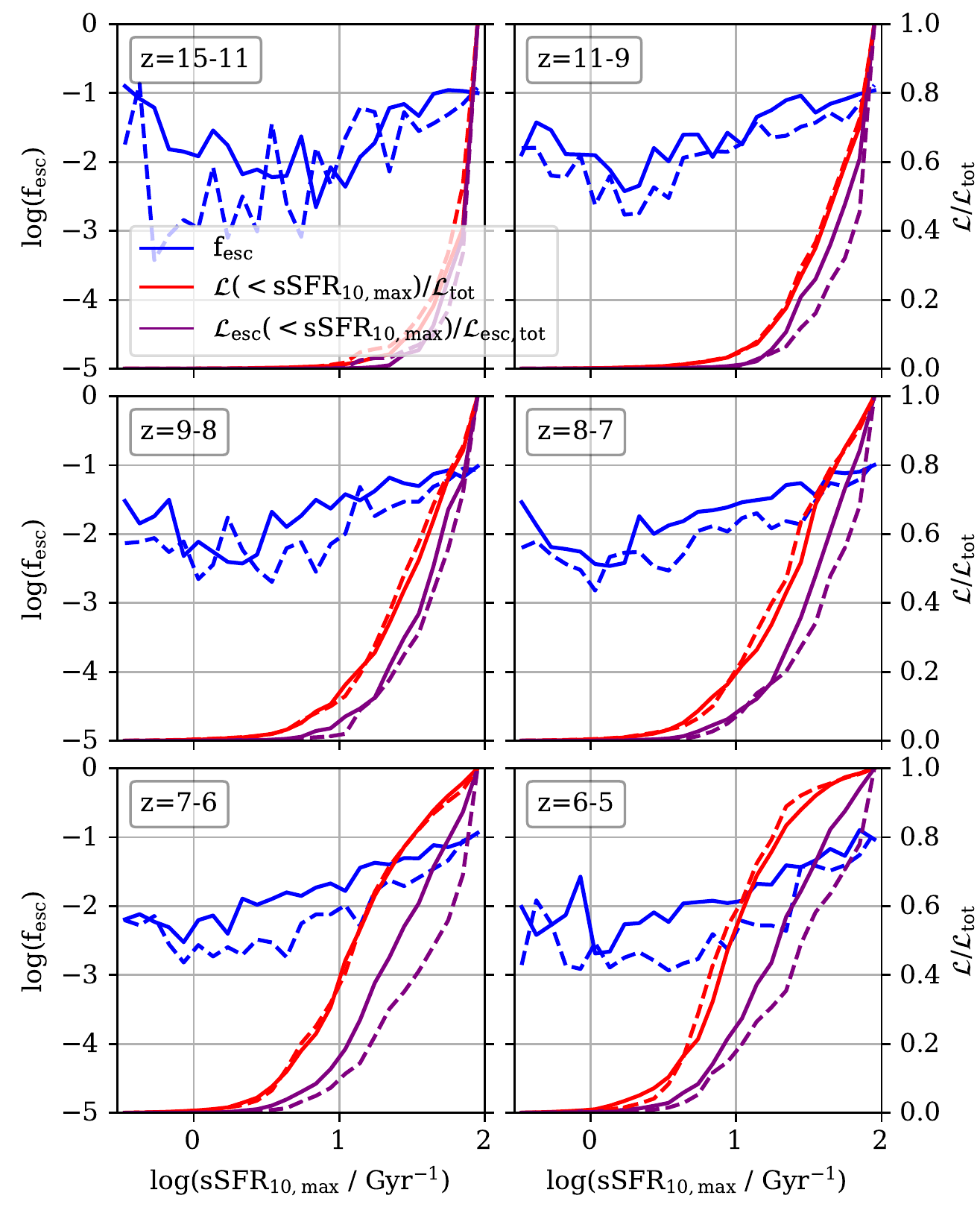}
  \caption
  {\label{fesc_hist_ssfr10max50_all_resolution_lg.fig} As
    \Fig{fesc_hist_Mvir_all_resolution_lg.fig} but for $\fesc$ versus
    $\sSFRmax$.}
\end{figure}

We
repeat the exercise of the previous appendix, now comparing our
fiducial resolution \sphinxten{} run to an identical run with one
level lower maximum resolution, i.e. the minimum cell width is twice
as large at any point in the simulation. The resulting correlations of
$\fesc$ versus halo properties are shown in Figures
\ref{fesc_hist_Mvir_all_resolution_lg.fig}-\ref{fesc_hist_ssfr10max50_all_resolution_lg.fig}
with \sphinxten{} in solid curves and its lower-resolution counterpart
in dashed curves. Generally the correlations of $\fesc$ with the halo
properties considered are unaffected, with $\fesc$ being simply
overall lower with lower resolution, but with a somewhat enhanced
difference for the highest metallicities and intermediate-to-high
$\sSFRten$ and $\sSFRmax$.

\section{Fits for escape fractions} \label{fits.app} In case the
escape fractions measured in \sphinx{} may be useful for analytic
models of reionization or to compare to other simulation works, we
provide here our fits for the correlations of $\fesc$ with the various
halo properties plotted in Figures \ref{fesc_hist_Mvir.fig} -
\ref{fesc_halo_hist_matt_all.fig} and
\Fig{fesc_halo_hist_sSFR10max.fig}. The fits are shown in each of
those figures in transparent blue curves. The fitted function is in
all cases the logarithm of a third degree polynomial, i.e.
\begin{align}
  \fesc (x) = 10^{a + b x + c x^2 + d x^3},
\end{align}
where x is (typically the base-10 logarithm of) the halo property
under consideration. We use the \pyth{scipy.optimize.curve_fit}
fitting function to derive the best-fit values for the polynomial
coefficients $a,b,c,d$ for each halo property and at each redshift
range considered. We use the default parameters for the
\pyth{curve_fit} function, i.e. we use the ``least squares'' fitting
method, we do not provide any guess for the best fit, we do not weigh
by uncertainties, and we do not assume any bounds. We do, however,
exclude any $\fesc$ data from the simulations represented by less than
$100$ halos. This is why the fits sometimes cover a shorter range than
the histograms in Figures \ref{fesc_hist_Mvir.fig} -
\ref{fesc_halo_hist_matt_all.fig} and
\Fig{fesc_halo_hist_sSFR10max.fig}.

We provide the best-fit values for the polynomial coefficients
$a,b,c,d$ in Table \ref{fits.tbl}, which also includes the ranges for
which our fits are valid.

\begin{table*}
  \begin{center}
  \caption
  {Our polynomial coefficient fits for the correlation of $\fesc$ with
    the halo properties considered in this paper at various redshift
    ranges, using the third-degree polynomial
    $\fesc (x) = 10^{a + b x + c x^2 + d x^3}$. The first row gives
    the name of the halo property which goes into the polynomial as
    the $x$ parameter. The second column gives the relevant redshift
    range. The third column gives the range in the halo property $x$
    for which the fit is valid, and the last four columns give the
    best fit coefficients.}
  \label{fits.tbl}
  \begin{tabular}{l|l|l|cccc}
    \toprule
    Halo property x & redshift range 
    & x-range
    & a & b & c & d \\
    \midrule
    $\log(\Mvir / \Msun)$ & $z=15-11$ & $\log(\Mvir / \Msun) = [7, 9.7]$
    & $5.3570$ & $-1.9448$ & $0.1874$ & $-0.0057$      \\
               & $z=11-9$ & $\log(\Mvir / \Msun) = [7, 10]$
    & $104.5197$ & $-37.6354$ & $4.4391$ & $-0.1735$   \\
               & $z=9-8$ & $\log(\Mvir / \Msun) = [7, 10.3]$
    & $54.3631$ & $-19.9959$ & $2.3789$ & $-0.0938$    \\
               & $z=8-7$ & $\log(\Mvir / \Msun) = [7, 10.7]$
    & $60.6799$ & $-22.3502$ & $2.6610$ & $-0.1048$    \\
               & $z=7-6$ & $\log(\Mvir / \Msun) = [7, 11]$
    & $77.7443$ & $-27.8650$ & $3.2448$ & $-0.1252$    \\
               & $z=6-4.7$ & $\log(\Mvir / \Msun) = [7, 11.7]$
    & $24.4699$ & $-9.2316$ & $1.0853$ & $-0.0425$    \\
    \midrule
    $\log(\Mstar / \Msun)$ & $z=15-11$ & $\log(\Mstar / \Msun) = [4, 7.7]$
    & $-53.4495$ & $22.1841$ & $-3.0984$ & $0.1428$     \\
               & $z=11-9$ & $\log(\Mstar / \Msun) = [4, 8.3]$
    & $-39.8329$ & $15.2711$ & $-1.9774$ & $0.0838$  \\
               & $z=9-8$ & $\log(\Mstar / \Msun) = [4, 8.7]$
    & $-35.3369$ & $12.9054$ & $-1.5841$ & $0.0626$   \\
               & $z=8-7$ & $\log(\Mstar / \Msun) = [4, 9]$
    & $-28.2177$ & $9.4517$ & $-1.0492$ & $0.0358$   \\
               & $z=7-6$ & $\log(\Mstar / \Msun) = [4, 9.3]$
    & $-27.5989$ & $9.0776$ & $-0.9955$ & $0.0335$   \\
               & $z=6-4.7$ & $\log(\Mstar / \Msun) = [4, 10.3]$
    & $-27.2628$ & $9.0356$ & $-1.0212$ & $0.0364$   \\
    \midrule
    $\log(\Zhalo)$ & $z=15-11$ & $\log(\Zhalo) = [-5.3 , -3]$
    & $-18.2132$ & $-12.3055$ & $-2.9206$ & $-0.2284$    \\
               & $z=11-9$ & $\log(\Zhalo) = [-5.3, -2.7]$
    & $-10.5256$ & $-6.7514$ & $-1.6252$ & $-0.1300$ \\
               & $z=9-8$ & $\log(\Zhalo) = [-5.3, -2.7]$
    & $-20.9345$ & $-14.4295$ & $-3.4955$ & $-0.2793$  \\
               & $z=8-7$ & $\log(\Zhalo) = [-5.3, -2.3]$
    & $-13.1098$ & $-8.5654$ & $-2.0575$ & $-0.1635$  \\
               & $z=7-6$ & $\log(\Zhalo) = [-5.3, -2]$
    & $-7.5847$ & $-4.0109$ & $-0.8745$ & $-0.0649$  \\
               & $z=6-4.7$ & $\log(\Zhalo) = [-5.3, -1.7]$
    & $-7.8124$ & $-4.1390$ & $-0.9282$ & $-0.0722$  \\
    \midrule
    $\log(\sSFRten/Gyr^{-1})$ & $z=15-11$ & $\log (\sSFRten/Gyr^{-1}) = [-0.45 , 2]$
    & $-0.7918$ & $-0.4428$ & $0.1651$ & $-0.0200$  \\
               & $z=11-9$ & $\log( \sSFRten/Gyr^{-1}) = [-0.45, 2]$
    & $-1.0398$ & $-0.4359$ & $0.1510$ & $0.0218$\\
               & $z=9-8$ & $\log( \sSFRten/Gyr^{-1}) = [-0.45, 2]$
    & $-1.2243$ & $-0.4452$ & $0.0820$ & $0.0914$ \\
               & $z=8-7$ & $\log( \sSFRten/Gyr^{-1}) = [-0.45, 2]$
    & $-1.3863$ & $-0.4905$ & $0.2776$ & $0.0252$ \\
               & $z=7-6$ & $\log( \sSFRten/Gyr^{-1}) = [-0.45, 2]$
    & $-1.5840$ & $-0.5096$ & $0.3093$ & $0.0504$ \\
               & $z=6-4.7$ & $\log( \sSFRten/Gyr^{-1}) = [-0.45, 2]$
    & $-1.9525$ & $-0.6187$ & $0.8167$ & $-0.1412$ \\
    \midrule
    $\log(\sSFRc/Gyr^{-1})$ & $z=15-11$ & $\log (\sSFRc/Gyr^{-1}) = [-0.45 , 2]$
    & $-2.4597$ & $1.9507$ & $0.8506$ & $-1.5870$ \\
               & $z=11-9$ & $\log( \sSFRc/Gyr^{-1}) = [-0.5, 1]$
    & $-2.2077$ & $1.5319$ & $0.7636$ & $-1.4199$ \\
               & $z=9-8$ & $\log( \sSFRc/Gyr^{-1}) = [-0.5, 1]$
    & $-1.8990$ & $1.0351$ & $-0.2655$ & $-0.2329$ \\
               & $z=8-7$ & $\log( \sSFRc/Gyr^{-1}) = [-0.5, 1]$
    & $-1.9739$ & $0.8402$ & $-0.4007$ & $0.2843$ \\
               & $z=7-6$ & $\log( \sSFRc/Gyr^{-1}) = [-0.5, 1]$
    & $-1.9435$ & $0.5844$ & $-0.4538$ & $0.4819$ \\
               & $z=6-4.7$ & $\log( \sSFRc/Gyr^{-1}) = [-0.5, 1]$
    & $-2.0670$ & $0.2429$ & $-0.6316$ & $1.1331$ \\
    \midrule
    $\MFifteen$ & $z=15-11$ & $\MFifteen = [-10.5 , -17.5]$
    & $-5.2932$ & $0.3721$ & $0.0902$ & $0.0031$ \\
               & $z=11-9$ & $\MFifteen = [-10.5, -18.5]$
    & $3.0704$ & $2.2120$ & $0.2186$ & $0.0060$ \\
               & $z=9-8$ & $\MFifteen = [-10.5, -18.5]$
    & $5.0293$ & $2.6969$ & $0.2557$ & $0.0070$ \\
               & $z=8-7$ & $\MFifteen = [-10.5, -19.5]$
    & $-1.6847$ & $1.2212$ & $0.1476$ & $0.0044$ \\
               & $z=7-6$ & $\MFifteen = [-10.5, -19.5]$
    & $-1.5486$ & $1.3384$ & $0.1596$ & $0.0047$ \\
               & $z=6-4.7$ & $\MFifteen = [-10.5, -20.5]$
    & $-10.3763$ & $-0.4500$ & $0.0386$ & $0.0020$ \\
    \midrule
    $\log(\sSFRmax/Gyr^{-1})$ & $z=15-11$ & $\log (\sSFRmax/Gyr^{-1}) = [-0.45 , 2]$
    & $-2.0148$ & $-0.6498$ & $1.2940$ & $-0.3616$  \\
               & $z=11-9$ & $\log( \sSFRmax/Gyr^{-1}) = [-0.45, 2]$
    & $-1.9890$ & $-0.2057$ & $0.7240$ & $-0.1960$ \\
               & $z=9-8$ & $\log( \sSFRmax/Gyr^{-1}) = [-0.45, 2]$
    & $-2.0879$ & $-0.2526$ & $0.9023$ & $-0.2701$ \\
               & $z=8-7$ & $\log( \sSFRmax/Gyr^{-1}) = [-0.45, 2]$
    & $-2.1417$ & $-0.1700$ & $0.7125$ & $-0.1745$ \\
               & $z=7-6$ & $\log( \sSFRmax/Gyr^{-1}) = [-0.45, 2]$
    & $-2.2493$ & $-0.1025$ & $0.6510$ & $-0.1438$ \\
               & $z=6-4.7$ & $\log( \sSFRmax/Gyr^{-1}) = [-0.45, 2]$
    & $-2.4119$ & $-0.0168$ & $0.5724$ & $-0.1002$ \\
    \bottomrule
  \end{tabular}
  \end{center}
\end{table*}

\bsp	
\label{lastpage}
\end{document}